\documentclass[letterpaper]{JHEP3}
\usepackage{amsmath}
\usepackage{amssymb, amsfonts}
\usepackage{wasysym}
\usepackage{graphicx,subfig}

\usepackage[usenames,dvipsnames]{color}
\let\normalcolor\relax

\global\long\def\usp#1{\mathfrak{#1}}

\makeatletter

\providecommand{\tabularnewline}{\\}

\title{ \Large  The Veneziano Limit  of ${\cal N} = 2$ Superconformal QCD:   \\
\smallskip
\smallskip
Towards the String Dual of ${\cal N}=2$ $ SU(N_c)$ SYM with $N_f=2N_c$  
}

\preprint{ YITP-SB-09-48}

\author{Abhijit Gadde\footnote{Email: abhijit.gadde@stonybrook.edu }$\,$, Elli Pomoni\footnote{Email: elli.pomoni@stonybrook.edu } $\,$  and Leonardo Rastelli\footnote{Email: leonardo.rastelli@stonybrook.edu}
\\ \\ \\
\it  C.N. Yang Institute for Theoretical Physics,\\
\it Stony Brook University, \\
\it Stony Brook, NY 11794-3840, USA}

\abstract{ 

\medskip

We attack the long-standing problem of finding the AdS dual of ${\cal N} = 2$ superconformal QCD, the ${\cal N}=2$ super Yang Mills theory with gauge group $SU(N_c)$ and
$N_f = 2 N_c$ fundamental hyper multiplets.
The theory admits a Veneziano expansion of large $N_c$ and large $N_f$, with $N_f/N_c$ and $\lambda = g^2_{YM} N_c$ kept fixed. 
The topological structure of   large $N$ diagrams motivates a general conjecture: the {\it flavor-singlet sector} of a gauge theory in the Veneziano limit
is dual to a closed string theory; single closed string states correspond to ``generalized single-trace'' operators, where adjoint letters and flavor-contracted fundamental/antifundamental pairs
are stringed together in a closed chain.
We look for the string dual of ${\cal N} = 2$ superconformal QCD  from two  fronts. 
From the bottom-up, 
 we perform a systematic analysis of the protected spectrum using superconformal representation theory. We also evaluate
 the one-loop dilation operator in the scalar sector, finding a novel spin chain.
From the top-down, we consider the decoupling limit of known brane constructions.
In both approaches, more insight is gained by viewing the theory as the degenerate limit of the ${\cal N} = 2$ $\mathbb{Z}_2$ orbifold
of ${\cal N} = 4$ SYM,   as one of the two gauge couplings is tuned to zero.
A consistent picture emerges. We conclude that the string dual is a sub-critical background with seven ``geometric'' dimensions,
 containing both an $AdS_5$ and an $S^1$ factor.   The supergravity approximation is never entirely valid, even for large $\lambda$,
 indeed  the field theory  has an exponential degeneracy of  exactly protected states with higher spin, 
 which must be dual to  a sector of light string states.

\bigskip

}

\renewcommand{\[}{\begin{equation}}
\renewcommand{\]}{\end{equation}}

\makeatother

\global\long\global\long\def\m{\mu}

\global\long\global\long\def\n{\nu}

\global\long\global\long\def\La{\Lambda}

\global\long\global\long\def\s{\sigma}

\global\long\global\long\def\f{\phi}

\global\long\global\long\def\e{\epsilon}

\global\long\global\long\def\del{\partial}

\global\long\global\long\def\D{\Delta}

\global\long\global\long\def\al{\alpha}

\global\long\global\long\def\ad{\dot{\alpha}}

\global\long\global\long\def\bd{\dot{\beta}}

\global\long\global\long\def\la{\lambda}

\global\long\global\long\def\ra{\rightarrow}

\global\long\global\long\def\fbar{\bar{\phi}}

\global\long\global\long\def\p{\partial}

\global\long\global\long\def\bA{{\bf A}}

\global\long\global\long\def\OO{\mathcal{O}}

\global\long\global\long\def\II{\mathcal{I}}

\global\long\global\long\def\JJ{\mathcal{J}}

\global\long\global\long\def\KK{\mathcal{K}}

\global\long\global\long\def\LL{\mathcal{L}}

\global\long\global\long\def\TT{\mathcal{T}}

\global\long\global\long\def\NN{\mathcal{N}}

\global\long\global\long\def\MM{\mathcal{M}}

\global\long\global\long\def\PP{\mathcal{P}}

\global\long\global\long\def\nn{ \mathfrak{n} }

\global\long\global\long\def\qq{ \mathfrak{q} }

\global\long\global\long\def\mm{ \mathfrak{m} }

\global\long\global\long\def\pp{ \mathfrak{p}}

\global\long\global\long\def\Tr{\mbox{Tr}}

\global\long\global\long\def\Q{\mathcal{Q}}

\global\long\global\long\def\TT{\mathcal{T}}

\global\long\global\long\def\SS{\mathcal{S}}

\global\long\global\long\def\RR{\mathcal{R}}

\global\long\global\long\def\TpT{\mathcal{T}^{\prime}}

\global\long\global\long\def\IIh{\hat{\mathcal{I}}}

\global\long\global\long\def\JJh{\hat{\mathcal{J}}}

\global\long\global\long\def\KKh{\hat{\mathcal{K}}}

\global\long\global\long\def\LLh{\hat{\mathcal{L}}}

\global\long\global\long\def\SSh{\hat{\mathcal{S}}}

\global\long\global\long\def\RRh{\hat{\mathcal{R}}}

\global\long\global\long\def\dprime{\prime\prime}

\global\long\global\long\def\topp#1{\check{#1}}

\global\long\def\fh{\topp{\f}}

\global\long\def\QQ{\mathcal{Q}}

\global\long\def\AA{\mathcal{A}}

\global\long\def\BB{\mathcal{B}}

\global\long\def\CC{\mathcal{C}}

\global\long\def\DD{\mathcal{D}}

\global\long\def\EE{\mathcal{E}}

\newcommand{\q}{\mathfrak{Q}}

\global\long\def\ind#1{\mathtt{#1}}

\def\bea{\begin{eqnarray}}
\def\eea{\end{eqnarray}}
\def\be{\begin{equation}}
\def\ee{\end{equation}}
\def\ea{\end{align}}
\def\bse{\begin{subequations}}
\def\ese{\end{subequations}}
\def\1F1{{}_1\!F_1}
\def\2F0{{}_2\!F_0}

\newcommand{\Torb}{{\cal T}}

\begin{document}

\section{Motivation}

How general is the gauge/string correspondence? 't Hooft's topological
argument \cite{'tHooft:1973jz} suggests that any large $N$ gauge
theory should be dual to a closed string theory. However, the four-dimensional
gauge theories for which an  independent definition of the dual string theory is presently available
are rather special.  Even among conformal field theories, which are the best
understood,  an explicit dual string description is known only for a  sparse subset of models.  In some sense
 all 
examples are close relatives of the original paradigm of ${\cal N}=4$ 
super Yang-Mills \cite{Maldacena:1997re,Gubser:1998bc,Witten:1998qj} and are found by considering stacks of 
 branes at local singularities in critical string theory, or variations of this setup,
{\it e.g.}  \cite{Kachru:1998ys,Lawrence:1998ja,Witten:1998xy, Klebanov:1998hh, Morrison:1998cs,  Benvenuti:2004dy, Lunin:2005jy}.\footnote{We should perhaps emphasize from the outset that our focus  is on {\it string} duals of gauge theories. There are
  strongly coupled field theories that admit {\it gravity} duals with no perturbative
string limit, see {\it e.g.}  \cite{Gauntlett:2004zh, Gaiotto:2009gz}.}  
Conformal field theories in this class can have lower or no supersymmetry, but are  far from
being ``generic''.
Some of their special features are:
 \begin{enumerate}
\item [(i)] The $a$ and $c$ conformal anomaly coefficients are equal
at large $N$ \cite{Henningson:1998gx}. 
\item [(ii)] The fields are in the adjoint or in bifundamental representations
of the gauge group. (Except possibly for a small number of fundamental flavors -- ``small'' in the large $N$ limit -- as in \cite{Aharony:1998xz}). 
\item [(iii)] The dual geometry is ten dimensional. 
\item [(iv)] The conformal field theory has an exactly marginal coupling $\lambda$, which corresponds to a geometric modulus on the dual string side. For large $\lambda$  the string sigma model is weakly coupled and the supergravity approximation is valid.\footnote{In some cases, as in ${\cal N} = 4$ SYM, the opposite limit of small $\lambda$ corresponds to a weakly coupled Lagrangian description on the field theory side. In other cases, like the Klebanov-Witten theory \cite{Klebanov:1998hh}, the Lagrangian description is never weakly coupled.}
\end{enumerate}
The situation certainly does not improve if one breaks conformal invariance --
   the field theories for which we can directly describe the string dual remain a very special set,
which does not include some of the most relevant cases, such as pure Yang-Mills theory. Many more field theories, including pure Yang-Mills,
can be described indirectly,  as low-energy limits of deformations of ${\cal N} = 4$ SYM (as \emph{e.g.} in \cite{Polchinski:2000uf} for ${\cal N}=1$ SYM)
 or of other UV fixed points, not necessarily four-dimensional (as  in \cite{Witten:1998zw} for ${\cal N}=0$ YM or  \cite{Klebanov:2000hb, Maldacena:2000yy} for  ${\cal N}=1$ SYM).
 These constructions count as physical ``existence proofs'' of the string duals, 
but if one wishes to focus just on the  low-energy dynamics, 
one invariably encounters  strong coupling  on the dual string side. 
 In the limit where  the unwanted UV degrees of freedom decouple, the dual appears to be described (in the most favorable duality frame)
by a closed-string sigma model with strongly curved target.
This may well be only a technical problem, which would be overcome by an analytic or even a numerical solution of the worldsheet CFT. The more fundamental
problem is that we lack a  precise recipe to write, let alone solve,
the limiting sigma model that describes only the low-energy degrees of freedom.

To break this impasse and enlarge the list of dual pairs outside the ${\cal N}=4$ SYM universality class,
we can try  to attack the ``next simplest case''. A natural candidate for
this role is ${\cal N}=2$ SYM with gauge group $SU(N_{c})$ and $N_{f}=2N_{c}$
flavor hypermultiplets in the fundamental representation of $SU(N_{c})$.
The number of flavors is tuned to obtain a vanishing beta function.
We  refer to this model as ${\cal N}=2$ superconformal QCD (SCQCD).
The theory violates properties (i) and
(ii)  but it still has a large amount of symmetry (half the maximal
superconformal symmetry) and it shares with ${\cal N}=4$ SYM the
crucial simplifying feature  of a tunable, exactly marginal gauge coupling
$g_{YM}$. (The theory also exhibits $S$-duality \cite{Seiberg:1994aj, Argyres:2007cn, Gaiotto:2009we}, though this will not be important for our considerations,
since we will work in the large $N$ limit, which does not commute with $S$-duality.)

The large $N$ expansion of ${\cal N}=2$ SCQCD is the one defined
by Veneziano  \cite{Veneziano:1976wm}: the number of colors $N_{c}$ and the number of fundamental
flavors $N_{f}$ are both sent to infinity, keeping fixed their ratio
($N_{f}/N_{c}\equiv2$ in our case) and the combination $\lambda\equiv g_{YM}^{2}N_{c}$.
Which, if any, is the dual string theory? And what happens to it  for large $\lambda$?

\section{The Veneziano Limit and Dual Strings}

\subsection{A general conjecture}

To understand in which sense we should expect a dual string description of a gauge theory in the Veneziano limit,
we start by reviewing general elementary facts about large $N$ counting,
Feynman-diagrams topology, and operator mixing.
At this stage we have in mind a generic field theory that contains 
both adjoint fields, which we collectively denote by $\phi_{\; b}^{a}$, with
$a,b=1,\dots,N_{c}$, and fundamental fields, denoted by $q_{\; i}^{a}$,
with $i=1,\dots,N_{f}$.  We can consider the theory both in the 't Hooft limit of large $N_c$ with $N_f$ fixed,
and in the Veneziano limit of large $N_c \sim N_f$.

\medskip

\noindent 
{\it  \bf $\mathbf{N_c \to \infty}$, $\mathbf{N_f}$ fixed}

\smallskip

 \noindent 
 Let us  first recall the familiar analysis in the
 't Hooft limit \cite{'tHooft:1973jz}, where
the number of colors $N_c$ is sent to infinity, with $\lambda=g_{YM}^2 N_c$ {\it and} the number of flavors $N_f$ kept fixed.  
 In this limit it is useful to represent
 propagators for adjoint  fields  with {\it double} lines, and propagators
 for fundamental fields  with {\it single} lines -- the lines keep track
 of the flow of the $a$ type (color) indices.
Vacuum Feynman diagrams admit a topological classification
as  Riemann surfaces with boundaries: each flavor loop is interpreted
as a boundary. The  $N$ dependence is
$N_{c}^{2-2h-b} N_f^b$, for $h$ the genus and
$b$ the number of boundaries.

 The natural dual interpretation  is then in terms of  
a string theory with coupling $g_s \sim 1/N_c$, 
 containing both a closed and an open sector --  the latter
arising from the presence of $N_f$ explicit ``flavor'' branes where open
strings can end.  Indeed this is the familiar way to introduce a small number of flavors
in the AdS/CFT correspondence \cite{Karch:2002sh}: by adding explicit flavor branes to the bulk geometry
(the simplest examples is adding D7 branes to the $AdS_5 \times S^5$ background).
Since $N_f \ll N_c$, the backreaction of the flavor branes can be neglected (probe approximation).

According to the standard AdS/CFT dictionary, single-trace ``glueball'' composite operators, of the schematic form $ \Tr\, \phi^\ell $
(where $\Tr$ is a color trace) are dual to closed string states, while ``mesonic'' composite operators,
 of the schematic form $\bar q^i \phi^\ell q_j$, are dual to  open string states. 
At large $N_c$, these two classes of operators play a special role since they can be regarded as 
``elementary'' building blocks: all other gauge-invariant composite
 operators of finite dimension  can be built by taking products  of the elementary (single-trace and mesonic)
 operators, and their correlation functions 
 factorize into the correlation functions of the elementary constituents.\footnote{Note  that in this discussion we are not
 considering baryonic operators, since they have infinite dimension in the strict large $N_c$ limit. Baryons are interpreted
 as solitons of the large $N_c$ theory; as familiar, in AdS/CFT
 they correspond  to non-perturbative (D-brane) states on the string theory side \cite{Witten:1998xy}.}
 This factorization is dual  to the fact for $g_s \to 0$ the string Hilbert space  becomes the free multiparticle Fock space of open and closed strings.
 
 Flavor-singlet mesons, of the form  $\sum_{i=1}^{N_f} \bar q^i \phi^\ell q_i$,
 mix with glueballs in perturbation theory,
 but the mixing is suppressed by a factor of $N_f/N_c \ll 1$, so  the distinction
 between the two classes of operators is meaningful in the 't Hooft limit.
 On the dual side, this translates into the statement that the mixing of open and closed strings
 in subleading since each boundary comes with a suppression factor of $g_s N_f \sim N_f/N_c$.

\medskip

\noindent 
{\ \bf   $\mathbf{N_c \sim N_f \to \infty}$}

\smallskip

\begin{figure}
\begin{centering}
\includegraphics[scale=0.6]{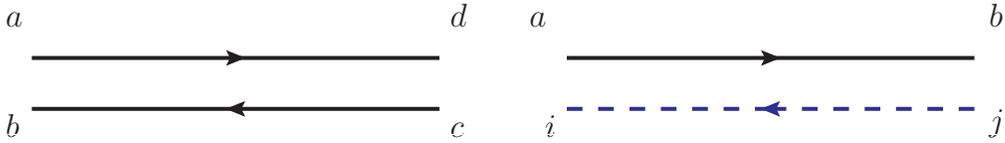}
\par\end{centering}
\caption{\label{doublelinepropagators}Double line propagators. The adjoint 
propagator $\langle \phi^{a}_{\; b} \;  \phi^{c}_{\; d} \rangle$
on the left, represented by two color lines,
and the fundamental propagator $\langle   q^{a}_{\; i} \;  \bar q ^{j}_{\; b} \rangle$ 
on the right, represented by  a color
and a flavor line.}
\end{figure}

 \noindent  We can now repeat the analysis in the Veneziano limit of large $N_c$ and large $N_f$ with $\lambda = g_{YM}^2 N_c$ and $N_f/N_c$ fixed.
In this limit it is appropriate to use
a double-line notation with two distinct types of lines \cite{Veneziano:1976wm}: color lines (joining $a$
indices) and flavor lines (joining $i$ indices). A $\phi$ propagator
decomposes as two color lines with opposite orientations, while a
$q$ propagator is made of a color and a flavor line (Figure \ref{doublelinepropagators}).
Since $N_{f}\sim N_{c}\equiv N$, color and flavor lines are on the
same footing in the counting of factors of $N$. It is natural to regard all
vacuum Feynman diagrams as  \textit{closed} Riemann surfaces, whose $N$ dependence
 is $N^{2-2h}$, for $h$ the genus.  At least at this  topological level,
 by  the same  logic of  \cite{'tHooft:1973jz}, we should expect
a gauge theory in the Veneziano limit to be described by the perturbative
expansion of a closed string theory, with coupling $g_{s} \sim1/N$. 
More precisely,  there should be a dual {\it purely closed}
string description of the \textit{flavor-singlet} sector of the gauge
theory.

This point can be sharpened looking at operator mixing. It is consistent to truncate
the theory to flavor-singlets, since they close under operator product expansion.
The new feature that arises in the Veneziano limit is the order-one mixing
of {}``glueballs'' and  flavor-singlet ``mesons''.  For large $N_{c}\sim N_{f}$,
the basic ``elementary'' operators are what we may call
 {} {\it generalized single-trace} operators,
of the form 
\begin{equation}
{\rm Tr}\left(\phi^{k_{1}}{\cal M}^{\ell_{1}}\phi^{k_{2}}\dots\phi^{k_{n}}{\cal M}^{\ell_{n}}\right)\,,\qquad{\cal M}_{\; b}^{a}\equiv\sum_{i=1}^{N_{f}}q_{\; i}^{a}\,\bar{q}_{\, b}^{i}\,.\label{generalizedsingletrace}\end{equation}
 Here we have introduced a flavor-contracted combination of a fundamental and an antifundamental
field, ${\cal M}_{\;\, b}^{a}$, which for the purpose of the large $N$ expansion
plays the role of just another adjoint field. The usual large $N$ factorization theorems apply: correlators
of generalized multi-traces factorize into correlators of 
generalized single-traces. In the conjectural duality with a closed
string theory, generalized single-trace operators are dual to single-string states.

We can imagine to start with a dual closed string description of the field theory
with {$N_f= 0$}, and first introduce a small number of flavors $N_f \ll N_c$ by
adding flavor branes in the probe approximation. As we increase $N_f$ to be $\sim N_c$, the probe
approximation breaks down: 
  boundaries are not suppressed and
for fixed genus we must sum over worldsheets with arbitrarily many
boundaries. The result of this resummation -- we are saying -- is
a new closed string background dual to the flavor-singlet sector of the field theory.
 The large mixing of closed strings
and flavor singlet open strings gives rise to {\it new effective closed-string degrees of freedom}, propagating in a backreacted geometry. This is the string
theory interpretation of the generalized single-trace operators (\ref{generalizedsingletrace}).

In stating the conjectured duality
we have been careful to restrict ourselves
 to the flavor-singlet sector of the field theory. One may entertain the idea  that ``generalized mesonic operators''
of the schematic form $\bar q^i \,   \phi^{k_{1}}{\cal M}^{\ell_{1}}\phi^{k_{2}}\dots\phi^{k_{n}}{\cal M}^{\ell_{n}} \, q_j$
(with open flavor indices $i$ and $j$) would map to elementary open string states in the bulk.
However this cannot be correct, because
 generalized mesons and generalized single-trace operators are not
independent -- already in free field theory they are constrained by algebraic relations -- so adding  an independent open string sector in the dual theory would amount to overcounting.

\subsection{Outline of the paper}

In this paper we focus on the concrete example of  ${\cal N} = 2$ SCQCD and
 look for  a closed string theory description of its flavor-singlet sector. 
 We work at the superconformal point (zero vev for all the scalars) and thus look
 for a string background with unbroken $AdS_5$ isometry.
We attack the problem from two fronts: from
the bottom-up, using the weakly-coupled Lagrangian description, and from the top-down,
studying  brane constructions in string theory. Correspondingly,   the paper is divided into two main parts.
The field theory analysis occupies sections 3-5, the string theory analysis sections  7-8. Section 6 provides a bridge, 
a first attempt to put together the clues of the field theory analysis and guess features of the  dual string theory.
 In the field theory sections
we pose and answer in rigorous detail a well-defined question: what is the protected spectrum
of ${\cal N} = 2$ SCQCD in the generalized single-trace sector? 
The string theory analysis is more qualitative and our program not yet complete.  We review brane constructions 
and argue that the decoupling limit leads to a {\it sub-critical} string background.  We carry the analysis
far enough to see that  the  string dual, which is largely constrained by symmetry,
matches several field theory expectations, but we leave the  determination  of the precise non-critical background for future work.

In both the bottom-up and top-down approaches it is very useful to view
  ${\cal N} = 2$ SCQCD as part of  an ``interpolating''   ${\cal N} =2$ superconformal field theory (SCFT) that
has product gauge group  $SU(N_c) \times SU(N_{\check c})$ and correspondingly two exactly marginal couplings $g$ and $\check g$.
For $\check g \to 0$ one finds  ${\cal N} = 2$ SCQCD {\it plus} a decoupled vector multiplet, while for $\check g= g$
one finds the  $\mathbb{Z}_2$ orbifold of ${\cal N} = 4$ SYM. The orbifold theory has a well-known closed string dual, type IIB on $AdS_5 \times S^5/\mathbb{Z}_2$,
and changing $\check g /g$ amounts to changing the period of the NSNS $B$-field  through the blow-down cycle of the orbifold.
As we are going to  discuss in detail,  the flavor-singlet operators of ${\cal N} =2$ SCQCD are a subsector of the operators of the interpolating SCFT.
So in a sense we are guaranteed success: we know a priori that the flavor-singlet sector of ${\cal N} =2$ SCQCD must be described
by the  closed string theory obtained by following the limit $\check g \to 0$ in the bulk. This is however a rather subtle limit, and making sense
of it will occupy us in the second part of the paper.

In a companion paper \cite{spinchain}  we have taken the next step of the bottom-up analysis. We have evaluated the planar one-loop dilation operator
in the scalar sector of ${\cal N} = 2$ SCQCD, as well as of the interpolating SCFT,
 and written it as the Hamiltonian of a  spin-chain system. The spin-chain for  ${\cal N} = 2$ SCQCD is  novel,
 since the chain is of the  ``generalized single-trace''  form (\ref{generalizedsingletrace}). The dynamics of magnon excitations
 is quite interesting. In particular it is amusing to see how the flavor-contracted fundamental/antifundamental
 pairs ${\cal M}_{\;\, b}^{a}$ arise as $\check g \to 0$ by a process of ``dimerization'' of the magnons of the interpolating SCFT.
    Some results of \cite{spinchain} will be an input in section 4 to the analysis  of the protected spectrum 
of ${\cal N} =2$ SCQCD.
 
 A more detailed outline of the rest of paper is as follows. We begin in section 3 with a review of the Lagrangian
 and symmetries of ${\cal N} = 2$ SCQCD and of the interpolating SCFT that connects it to the $\mathbb{Z}_2$ orbifold
 of ${\cal N} = 4$ SYM. In sections 4 and 5 we study the protected spectrum of short supermultiplets\footnote{We use the word ``short'' casually,
 to denote a multiplet that obeys any of type of shortening condition, unlike some authors who distinguish  between ``short'' and ``semi-short''.
 We use the precise notation for multiplets reviewed in appendix A when we need to make such distinctions.}
of  ${\cal N} =2$ SCQCD and its relation with the spectrum of the interpolating SCFT. This turns out to be a rather intricate
 exercise in superconformal representation theory. A part of the 
 protected spectrum of ${\cal N} = 2$ SCQCD
 is easy to determine, namely the supermultiplets built on primaries  made of scalar fields: 
  (\ref{list}) is the complete list of such primaries,  as shown in \cite{spinchain}  using the one-loop spin-chain. In section 4 we 
  follow in detail the evolution of the protected states of the interpolating SCFT, starting at the orbifold
  point $\check g = g$ where the complete protected spectrum is easily determined.    In the limit $\check g \to 0$ we recover  (\ref{list})   as the subsector of  protected primaries of the interpolating SCFT  that are flavor singlets.
  Now there are {\it many more} protected states in ${\cal N} =2$ SCQCD than there are for generic $\check g$ in the interpolating SCFT: the extra protected states arise from 
 long multiplets of the interpolating SCFT that split into short multiplets
   at $\check g = 0$. In section 5 we    use the superconformal index to
   demonstrate the existence of these extra protected states. We show that the number of extra states grows exponentially
   with the conformal dimension. We also  characterize the quantum numbers of the first    few of them using a ``sieve'' algorithm; 
   this characterization is up to a certain intrinsic ambiguity   of the superconformal index, which can only determine ``equivalence classes'' of  short
   multiplets, as we review in detail. Still, we have enough information to unambiguously demonstrate the existence of {\it higher-spin} protected states in the generalized
   single-trace sector, in sharp contrast with ${\cal N} = 4$ SYM.
   
    In section 6 we use the clues offered by the protected spectrum to argue  that the dual of ${\cal N} =2$ SCQCD should
    be a sub-critical string background, with seven ``geometric'' dimensions, containing both an $AdS_5$ and an $S^1$ factor.
    There must be a sector of light string states, with mass of the order of the AdS scale for all $\lambda$, 
    dual to the higher-spin protected states detected by the superconformal index --
so    even for large $\lambda$ the supergravity approximation cannot be entirely valid. 
    We suggest that there is also a separate sector of heavy string states, with $m^2 R_{AdS} \gg 1$  for $\lambda \to \infty$. We have in mind a scenario where in the interpolating
    SCFT there are two effective string lengths $l_s$ and $\check l_s$,  corresponding to the two `t Hooft couplings $\lambda$ and $\check \lambda$: for $\check \lambda \to 0$ and fixed $\lambda \gg 1$, the string length    $l_s \ll R_{AdS}$ is   associated with the massive sector, while $\check l_s \sim R_{AdS}$ is associated with the light sector.
     In section 7 we review brane constructions of the interpolating SCFT and of ${\cal N} =2$ SCQCD.
 The most useful construction is the Hanany-Witten setup with D4 branes suspended between NS5 branes.  
 We argue that the relevant dynamics is captured by a sub-critical brane setup, with color D3  and flavor D5 boundary states  in the exact IIB worldsheet CFT $\mathbb{R}^{5,1} \times SL(2)_2/U(1)/\mathbb{Z}_2$. We identify the   dual of  ${\cal N} = 2$ SCQCD with the backreacted background, where the D-branes are replaced by flux.
We do not yet know the precise background, but  it is largely constrained by symmetries. In section 8 we show
that just assuming a solution exists, the results of the top-down approach are in nice qualitative
agreement with the bottom-up expectations. A useful tool is the  spacetime ``effective action'' of the non-critical theory, which we identify  as  the seven-dimensional 
maximal supergravity with the (non-standard) $SO(4)$ gauging. 
We conclude in section 9 with a brief  discussion.

Several technical appendices supplement the text. 
In appendix A  we review the shortening conditions  of the ${\cal N} = 2$ superconformal algebra.
   In  appendix B we review the ${\cal N} =1$ chiral ring of ${\cal N} =2$ SCQCD and of the interpolating SCFT. In appendix C we evaluate
   the superconformal index for various combinations of short multiplets.
    In appendix D we review the Kaluza-Klein reduction on $AdS_5 \times S^1$ of the $(2,0)$ tensor multiplet,
   with a new detailed treatment of the zero modes. 
    In appendix E we review  the sub-critical IIB background  $\mathbb{R}^{5,1} \times SL(2)_2/U(1)/\mathbb{Z}_2$ and its spectrum.
    We make a new claim about the $7d$ ``effective action'' describing the lowest plane-wave states, which we identify with
 maximally supersymmetric $SO(4)$-gauged  supergravity.

 \subsection{Relation to previous work}
 
The idea that sub-critical string theories play a role in the gauge/gravity correspondence is of course not new.  
Polyakov's conjecture
that pure Yang-Mills theory should be dual to a $5d$ string theory, with the Liouville field playing the role of the fifth dimension,
predates  the AdS/CFT correspondence (see {\it e.g.} \cite{Polyakov:1997tj, Polyakov:1998ju, Polyakov:2000fk}). In fact one of the surprises
of AdS/CFT was that some supersymmetric gauge theories are dual to simple backgrounds of {\it critical} string theory.
General studies of AdS solutions of non-critical spacetime effective actions include \cite{Kuperstein:2004yk, Alishahiha:2004yv}. 
Non-critical holography has been mostly considered, starting with \cite{Polyakov:2004br, Klebanov:2004ya},
 in the ${\cal N} =1$ supersymmetric case, notably for  ${\cal N} =1$ super QCD  in the Seiberg conformal window,
 which is argued to be dual to $6d$ non-critical backgrounds of the form $AdS_5 \times S^1$ with string-size curvature.
 There is an interesting literature on the RNS worldsheet description of these $6d$ non-critical  backgrounds and
their gauge-theory interpretation, see {\it e.g.} \cite{Fotopoulos:2005cn, Ashok:2005py, Bigazzi:2005md, Murthy:2006xt}. Non-critical RNS superstrings were formulated 
in \cite{Kutasov:1990ua, Kutasov:1991pv} and shown in \cite{Ooguri:1995wj, Aharony:1998ub, Giveon:1999zm, Giveon:1999zm, Giveon:1999px, Giveon:1999tq}
 to describe subsectors of critical string theory -- the degrees of freedom localized near NS5 branes or (in the mirror description) Calabi-Yau singularities. 
 Non-critical superstrings have been also considered in the Green-Schwarz and pure-spinor formalisms, see {\it e.g.} \cite{Siegel:1995hq, Grassi:2005kc, Wyllard:2005fh, Adam:2006bt, 
 Adam:2007ws}.

Our analysis in sections 6 and 7 for ${\cal N} =2$ SCQCD will be in the same spirit as the analysis of \cite{Fotopoulos:2005cn, Murthy:2006xt} for ${\cal N}=1$ super QCD.
We will use the double-scaling limit defined  in  \cite{Giveon:1999px, Giveon:1999tq} and further studied in {\it e.g.}  \cite{Aharony:2003vk, Aharony:2004xn, Murthy:2003es}.  
One of our points is that the ${\cal N} =2$  supersymmetric case should be the simplest
for non-critical gauge/string duality. On the string side, more symmetry does  not hurt, but the real advantage
is on the field theory side. Little is known about the SCFTs in the Seiberg conformal window, since  generically they are strongly coupled, isolated
fixed points. By contrast ${\cal N} =2$ SCQCD has an exactly marginal coupling $\lambda$, which takes arbitrary non-negative values.
 There is  a weakly coupled Lagrangian description for $\lambda \to 0$, and we can bring to bear all the perturbative technology that has been so successful
 for ${\cal N} = 4$ SYM, for example in uncovering integrable structures.\footnote{${\cal N}=1$
 SQCD at the Seiberg self-dual point $N_f = 2 N_c$ admits an exactly marginal coupling (the coefficient of a quartic superpotential),
 which however is bounded from below -- the theory is never weakly coupled.} At the same time we may  hope, again in analogy with
 ${\cal N}=4$ SYM, that the string dual will simplify in the strong coupling limit   $\lambda \to \infty$.
 
There are also interesting approaches to holography for gauge theories with a large number of fundamental flavors in {\it critical} string theory/supergravity,
see {\it e.g.} \cite{Burrington:2004id, Casero:2006pt, Paredes:2006wb, Benini:2006hh,  Benini:2007gx,  Casero:2007jj,  Caceres:2007mu,  HoyosBadajoz:2008fw, Bigazzi:2008qq}.   The critical setup inevitably implies that the boundary gauge theory will have  
  UV completions with extra degrees of freedom   ({\it e.g.} higher supersymmetry and/or higher dimensions).

\section{Field Theory Lagrangian and Symmetries}

In this section we  briefly review the structure and symmetries
of ${\cal N} =2$ SCQCD, and its relation to the $\mathbb{Z}_2$ orbifold of  ${\cal N} = 4$ SYM.
 Much insight is gained by viewing ${\cal N} = 2$ SCQCD, which has {\it one}
exactly marginal parameter (the $SU(N_c)$ gauge coupling $g_{YM}$), as the limit  of a {\it two}-parameter 
family of ${\cal N} = 2$ superconformal field theories. This is the family of ${\cal N} = 2$  theories
with product gauge group\footnote{The ranks of the two groups coincide, $N_c \equiv N_{\check c}$,
but it will be useful to always distinguish graphically  with a ``check'' all quantities pertaining to the second group $SU(N_{\check c})$.}
 $SU(N_c) \times SU(N_{\check c})$ and two bifundamental hypermultiplets; its
 exactly marginal parameters are the
 two gauge-couplings $g_{YM}$ and $\check g_{YM}$. For $\check g_{YM} \to 0$ one recovers ${\cal N} = 2$ SCQCD {\it plus} a decoupled free vector multiplet
 in the adjoint of $SU(N_{\check c})$. At $\check g_{YM}= 0$, the second gauge group is interpreted as a subgroup of the global
 flavor symmetry, $ SU(N_{\check c}) \subset U(N_f = 2 N_c)$. For $\check g_{YM} = g_{YM}$, we have  instead the familiar
$\mathbb{Z}_2$ orbifold  of ${\cal N} = 4$ SYM. Thus by tuning $\check g_{YM}$ we interpolate continuously between ${\cal N} = 2$ SCQCD
and the ${\cal N} = 4$ universality class. 

The $a$ and $c$ anomalies are constant, and equal to each other, along this exactly marginal line: 
at the end point $\check g_{YM} = 0$, the $SU(N_{\check c})$ vector multiplets decouples, accounting for the ``missing'' $a-c$ in ${\cal N} = 2$ SCQCD.

\subsection{${\cal N} =2$ SCQCD} 

Our main interest is
${\cal N}=2$  SYM theory with gauge group $SU(N_{c})$ and $N_{f} = 2 N_c$ fundamental hypermultiplets. 
We refer to this theory as ${\cal N} = 2$ SCQCD. 
Its global symmetry group is $U(N_f) \times SU(2)_R \times U(1)_r$,
where  $SU(2)_R \times U(1)_r$  is the R-symmetry subgroup of the superconformal group. 
We use indices $\II, \JJ =\pm$  for $SU(2)_R$, $i, j=1, \dots N_f$ for the flavor group $U(N_f)$
and $a,b=1, \dots N_c$ for the color group $SU(N_c)$.

Table \ref{charges} summarizes the field content and quantum numbers  of the model:
The Poincar\'e supercharges  ${\cal Q}^{\II}_\alpha$, $\bar {\cal Q}_{\II \, \dot \alpha}$ and the conformal
supercharges ${\cal S}_{\II \, \alpha}$, $\bar {\cal S}^{\II}_ {\dot \alpha}$
 are $SU(2)_R$ doublets with charges $\pm 1/2$ under $U(1)_r$.
The $\NN=2$ vector multiplet consists of a gauge field $A_m$, two Weyl spinors
$\lambda_{\alpha}^{\II}$, $\II=\pm$, which form a doublet under $SU(2)_{R}$,
and one complex scalar $\phi$, all in the adjoint representation of $SU(N_{c})$. 
Each $\NN=2$ hypermultiplet consists of 
 an $SU(2)_{R}$ doublet $Q_{\II}$ of complex scalars
and of two Weyl spinors $\psi_{\alpha}$ and $\tilde{\psi}_{\alpha}$, $SU(2)_{R}$ singlets.
It is convenient to define the flavor contracted mesonic operators 
\[
\MM_{\JJ \, \, \, b}^{\, \, \II a} \equiv \frac{1}{\sqrt{2}} Q_{\JJ \mbox{ }i}^{\mbox{ }a}\,\bar{Q}_{\mbox{ }b}^{\II \mbox{ }i} \, ,
\]  
which may be  decomposed  into the $SU(2)_{R}$ singlet and triplet combinations
  \begin{equation} \label{M1M3}
\MM_{{\bf {1}}} \equiv \MM^{\, \,  \II}_{\II}\quad\mbox{and}\quad\MM_{ {\bf {3} } \JJ  }^{\quad \II}  \equiv
\MM^{\, \, \II}_{\JJ}-\frac{1}{2}\MM^{\, \, \KK}_{\KK}\, \delta^{\II}_{\JJ} \, .
\end{equation}
The operators ${\cal M}$ decompose  into adjoint plus singlet representations of the color group $SU(N_c)$;
the singlet piece is however subleading in the large $N_c$ limit.

\begin{table}
\begin{centering}
\begin{tabular}{|c||c|c|c|c|}
\hline 
 & $SU(N_{c})$  & $U(N_{f})$  & $SU(2)_{R}$  & $U(1)_{r}$\tabularnewline
\hline
\hline 
$\mathcal{Q}_{\alpha}^{\II}$  & \textbf{$\mathbf{1}$}  & \textbf{$\mathbf{1}$}  & \textbf{$\mathbf{{2}}$}  & $+1/2$\tabularnewline
\hline 
$\SS_{\II \,\alpha}$ & \textbf{$\mathbf{1}$}  & \textbf{$\mathbf{1}$}  & \textbf{$\mathbf{2}$}  & $-1/2$\tabularnewline
\hline
\hline 
$A_{m}$  & Adj  & \textbf{$\mathbf{1}$}  & \textbf{$\mathbf{1}$}  & $0$\tabularnewline
\hline 
$\f$  & Adj  & \textbf{$\mathbf{1}$}  & \textbf{$\mathbf{1}$}  & $-1$\tabularnewline
\hline 
$\la_{\alpha}^{\II}$  & Adj  & \textbf{$\mathbf{1}$}  & \textbf{$\mathbf{2}$}  & $-1/2$\tabularnewline
\hline 
$Q_{\II}$  & $\Box$  & $\Box$  & \textbf{$\mathbf{2}$}  & $0$\tabularnewline
\hline 
$\psi_{\alpha}$  & $\Box$  & $\Box$  & \textbf{$\mathbf{1}$}  & $+1/2$\tabularnewline
\hline 
$\tilde{\psi}_{\alpha}$  & $\overline{\Box}$  & $\overline{\Box}$  & \textbf{$\mathbf{1}$}  & $+1/2$\tabularnewline
\hline
\hline 
$\MM_{\bf 1}$ & Adj + {\bf1}  & \textbf{$\mathbf{1}$}  & \textbf{$\mathbf{1}$}  & $0$\tabularnewline
\hline 
$\MM_{\bf 3}$ & Adj  + {\bf 1}& \textbf{$\mathbf{1}$}  & \textbf{$\mathbf{3}$}  & $0$\tabularnewline
\hline
\end{tabular}
\par\end{centering}
\caption{\label{charges} Symmetries of  $\NN=2$ SCQCD.
We show the quantum numbers of  the supercharges ${\cal Q}^\II$, ${\cal S}_\II$,
of the elementary components fields and of the mesonic operators ${\cal M}$. 
Conjugate objects (such as $\bar {\cal Q}_{\II \dot{\alpha}}  $ and $\bar \phi$) are not written explicitly.}
\end{table}

\subsection{\label{sec:orbifold} $\mathbb{Z}_2$ orbifold of ${\cal N} = 4$ and interpolating family of SCFTs}

${\cal N} = 2$ SCQCD can be viewed as a limit of a family of superconformal theories;
 in the opposite limit the family reduces to a $\mathbb{Z}_2$ orbifold of ${\cal N} =4$ SYM.  In this subsection we first describe
the orbifold theory and then its connection to ${\cal N} = 2$ SCQCD.

As familiar, the field content of ${\cal N}=4$ SYM 
 comprises the gauge field $A_m$, four Weyl fermions
$\lambda^A_\alpha$ and six real scalars $X_{AB}$, where $A,B=1,\dots 4$ are indices of the $SU(4)_R$
R-symmetry group. Under $SU(4)_R$, the fermions are in the ${\bf 4}$ representation, while
the scalars are in ${\bf 6}$ (antisymmetric self-dual)  and  obey the reality condition\footnote{The $\dagger$ indicates
hermitian conjugation of the matrix in color space. We choose hermitian generators for the color group.}
\begin{equation}
X_{AB}^\dagger = \frac{1}{2} \epsilon^{ABCD} X_{CD}  \,. 
\end{equation}
We may parametrize $X_{AB}$ in terms of six real scalars $X_k$, $k=4,\dots 9$,
 \begin{equation}
X_{AB}=\frac{1}{\sqrt{2}}\left(\begin{array}{cc|cc}
0 & X_{4}+iX_{5} & X_{7}+iX_{6} & X_{8}+iX_{9}\\
-X_{4}-iX_{5} & 0 & X_{8}-iX_{9} & -X_{7}+iX_{6}\\
\hline -X_{7}-iX_{6} & -X_{8}+iX_{9} & 0 & X_{4}-iX_{5}\\
-X_{8}-iX_{9} & X_{7}-iX_{6} & -X_{4}+iX_{5} & 0\end{array}\right)\label{Xmatrix}\end{equation}
Next, we pick an $SU(2)_L \times SU(2)_R \times U(1)_r$ subgroup of $SU(4)_R$,
\begin{equation}
\begin{array}{cc}
1 & +\\
2 & -\\
3 &  \hat + \\
4 &  \hat -\end{array}
\left(\begin{array}{cc|cc}
SU(2)_{R}  \times U(1)_r&  & \,\\
 &  & \,\\
\hline  &  & \,\\
 &  & \, & SU(2)_{L} \times U(1)_r^* \end{array}\right) \, .\end{equation}
We use indices $\II, \JJ = \pm$ for $SU(2)_R$ (corresponding to $A,B =1,2$) and indices
$\hat \II, \hat \JJ = \hat \pm$  for $SU(2)_L$ (corresponding to $A,B =3,4$). To make more manifest
their transformation properties, the scalars are rewritten as the $SU(2)_L \times SU(2)_R$ singlet
$Z$ (with  charge $-1$ under $U(1)_r$) and as the bifundamental  $\mathcal{X}_{\II\hat{\II}}$ (neutral under $U(1)_r$),
\begin{equation}
\mathcal{Z} \equiv \frac{X_{4}+iX_{5}}{\sqrt{2}} \, ,
\qquad \mathcal{X}_{\II\hat{\II}}\equiv \frac{1}{\sqrt{2}}\left(\begin{array}{cc}
X_{7}+iX_{6} & X_{8}+iX_{9}\\
X_{8}-iX_{9} & -X_{7}+iX_{6}\end{array}\right) \,. \end{equation}
Note the reality condition $\mathcal{X}_{\II\hat{\II}}^\dagger = -\epsilon_{\II \JJ } \epsilon_{\hat \II \hat \JJ } \mathcal{X}_{\JJ \hat{\JJ}}$.
Geometrically,  $SU(2)_{L}\times SU(2)_{R}\cong SO(4)$
is the group of $6789$ rotations  and $U(1)_{R}\cong SO(2)$
the group of $45$ rotations.  Diagonal 
$SU(2)$ transformations $\mathcal{X} \rightarrow U\mathcal{X} U^{-1}$ ($U_R = U, U_L = U^*$)
preserve the trace, $\Tr[\mathcal{X}]=2iX_{6}$,  and thus
 correspond to $789$ rotations.

We  are now ready to discuss the orbifold projection. 
In  R-symmetry space, the orbifold group is  chosen to be $\mathbb{Z}_2 \subset SU(2)_L$ 
with elements  $\pm \mathbb{I}_{2 \times 2}$. This is the well-known
quiver theory \cite{Douglas:1996sw}
obtained by placing $N_c$ D3 branes at the $A_1$ singularity
$ \mathbb{R}^{2}\times\mathbb{R}^{4}/\mathbb{Z}_{2}$,  with $(X_6, X_7, X_8, X_9) \to \pm (X_6, X_7, X_8, X_9)$ and $X_4$ and $X_5$ invariant. Supersymmetry
is broken to $\NN =2$, since the supercharges with $SU(2)_L$ indices are projected out.
The $SU(4)_R$ symmetry is broken to $SU(2)_L \times SU(2)_R \times U(1)_r $, or more precisely
to $SO(3)_L \times SU(2)_R \times U(1)_r$ since only objects with integer $SU(2)_L$ spin survive.
The $SU(2)_R \times U(1)_r$ factors are the R-symmetry of the unbroken ${\cal N} = 2$ superconformal group,
while $SO(3)_L$ is an extra global symmetry under which the unbroken supercharges are neutral.

In color  space, we start with gauge group $SU(2N_c)$, and declare
 the non-trivial element of the orbifold to be 
 \begin{equation} \label{tau}
 \tau\equiv\left(\begin{array}{cc}
\mathbb{I}_{N_{c}\times N_{c}} & 0\\
0 & -\mathbb{I}_{N_{c}\times N_{c}}\end{array}\right) \, .
\end{equation} 
All in all the $\mathbb{Z}_2$ action on the ${\cal N} =4$ fields is
\begin{equation}
A_m  \rightarrow  \tau A_m\tau \, , \quad
Z_{\II\JJ}  \rightarrow  \tau Z_{\II\JJ}\tau \, ,\quad   \lambda_{\II} \to \tau \lambda_{\II} \tau \, , \quad 
\mathcal{X}_{\II\IIh}  \rightarrow  -\tau\mathcal{X}_{\II\IIh}\tau\, ,\quad\lambda_{\hat \II} \to -\tau \lambda_{\hat \II} \tau \,.
\end{equation}
The components that  survive the projection are
 \begin{eqnarray}  \label{survive}
A_m & = &   \left( \begin{array}{cc}
A_{\mu b}^{a} & 0 \\
0 &  \topp A_{\mu \topp b}^{\topp a} \end{array} \right) \quad
Z  =   \left( \begin{array}{cc}
 \f_{\mbox{ } \mbox{ }b}^{a} & 0 \\
0 &  \topp{\f}_{\mbox{ } \mbox{ } \topp b}^{\topp a} \end{array} \right) 
\\
 \lambda_{\II}& =&  \left( \begin{array}{cc}
 \lambda_{\II b}^{a} & 0 \\
0 &  \topp{\la}_{\II \topp b}^{\topp a} \end{array} \right)\quad
 \lambda_{\hat{\II}}  =   \left( \begin{array}{cc}
0 &  \psi_{\hat{\II} \topp a}^{a} \\
 \tilde{\psi}_{\hat{\II}b}^{\topp b} & 0 \end{array} \right)
 \label{fermident}
\\
 \mathcal{X}_{\II \IIh} & = &  \left( \begin{array}{cc}
0 & Q_{\II \IIh \topp a}^{\mbox{ }a} \\
- \epsilon_{\II \JJ} \epsilon_{\hat{\II} \hat{\JJ}} \bar{Q}_{\mbox{ } \mbox{ }b}^{\topp b \hat{\JJ} \JJ} & 0 \end{array} \right) 
  \,.
 \end{eqnarray}
The gauge group 
is broken to $SU(N_c) \times SU(N_{\check c}) \times U(1)$,
where the $U(1)$ factor is the {\it relative}\footnote{Had we started with $U(2 N_c)$ group,
we would also have an extra {\it diagonal} $U(1)$, which would completely decouple since no fields are charged under it.}
 $U(1)$ generated by $\tau$ (equ.(\ref{tau})): it
must be removed by hand, since its beta function is non-vanishing. The process of removing the relative $U(1)$ modifies the scalar potential
by double-trace terms, which arise from the fact that the auxiliary fields (in ${\cal N}=1$ superspace) are now missing the $U(1)$ component. Equivalently
we can evaluate the beta function for the double-trace couplings, and tune them to their fixed point \cite{Dymarsky:2005nc}.

Supersymmetry organizes the component fields  into the $\NN =2$ vector multiplets of each factor of the gauge group,
 $(\phi, \lambda_\II, A_m)$ and  $(\check \phi, \check \lambda_\II,\check A_m)$,
 and into two bifundamental hypermultiplets,  $(Q_{\II, \hat +}, \psi_{\hat +}, \tilde \psi_{\hat +})$
 and  $(Q_{\II, \hat -}, \psi_{\hat -}, \tilde \psi_{\hat -})$. Table 2 summarizes the field
 content and quantum numbers of the orbifold theory.

\begin{table}
\begin{centering}
\begin{tabular}{|c||c|c|c|c|c|}
\hline 
 & $SU(N_{c})_{1}$  & $SU(N_{c})_{2}$  & $SU(2)_{R}$  & $SU(2)_{L}$  & $U(1)_{R}$\tabularnewline
\hline
\hline 
$\QQ_{\alpha}^{\II}$ & \textbf{${\bf 1}$}  & \textbf{${\bf 1}$}  & \textbf{${\bf 2}$}  & \textbf{${\bf 1}$}  & +1/2\tabularnewline
\hline 
$\SS_{\II \, \alpha}$ & \textbf{${\bf 1}$}  & \textbf{${\bf 1}$}  & \textbf{${\bf 2}$}  & \textbf{${\bf 1}$}  & --1/2\tabularnewline
\hline
\hline 
$A_{m}$  & Adj  & \textbf{${\bf 1}$}  & \textbf{${\bf 1}$}  & \textbf{${\bf 1}$}  & 0\tabularnewline
\hline 
$\topp A_{m}$  & \textbf{${\bf 1}$}  & Adj  & \textbf{${\bf 1}$}  & \textbf{${\bf 1}$}  & 0\tabularnewline
\hline 
$\f$  & Adj  & \textbf{${\bf 1}$}  & \textbf{${\bf 1}$}  & \textbf{${\bf 1}$}  & --1\tabularnewline
\hline 
$\topp{\f}$  & \textbf{${\bf 1}$}  & Adj  & \textbf{${\bf 1}$}  & \textbf{${\bf 1}$}  & --1\tabularnewline
\hline 
$\la^{\II}$  & Adj  & \textbf{${\bf 1}$}  & \textbf{${\bf 2}$}  & \textbf{${\bf 1}$}  & --1/2\tabularnewline
\hline 
$\topp{\la}^{\II}$  & \textbf{${\bf 1}$}  & Adj  & \textbf{${\bf 2}$}  & \textbf{${\bf 1}$}  & --1/2\tabularnewline
\hline 
$Q_{\II\hat{\II}}$  & $\Box$  & $\overline{\Box}$  & \textbf{${\bf 2}$}  & \textbf{${\bf 2}$}  & 0\tabularnewline
\hline 
$\psi_{\hat{\II}}$  & $\Box$  & $\overline{\Box}$  & \textbf{${\bf 1}$}  & \textbf{${\bf 2}$}  & +1/2\tabularnewline
\hline 
$\tilde{\psi}_{\hat{\II}}$  & $\overline{\Box}$  & $\Box$  & \textbf{${\bf 1}$}  & \textbf{${\bf 2}$}  & +1/2\tabularnewline
\hline
\end{tabular}
\par\end{centering}
\caption{\label{orbifoldcharges}
Symmetries of  the $\mathbb{Z}_2$ orbifold of ${\cal N} = 4$ SYM and of the interpolating family of ${\cal N} = 2$ SCFTs.
}
\end{table}

The two gauge-couplings $g_{YM}$ and $\check g_{YM}$ can be independently varied
while preserving ${\cal N} =2$ superconformal invariance, thus defining a two-parameter family
of ${\cal N} = 2$ SCFTs. Some care is needed in adjusting the
 Yukawa and scalar potential terms so that ${\cal N} =2$ supersymmetry is preserved.
We find 
\begin{eqnarray} 
\label{yukawa}
\LL_{Yukawa}(g_{YM},\check{g}_{YM}) & = & i\sqrt{2}\mbox{Tr}\big[-g_{YM}\epsilon^{\II\JJ}\bar{\lambda}_{\II}\bar{\lambda}_{\JJ}\f-\check{g}_{YM}\epsilon^{\II\JJ}\bar{\topp{\la}}_{\II}\bar{\topp{\la}}_{\JJ}\topp{\f}\nonumber \\
 &  & +g_{YM}\epsilon^{\hat{\II}\hat{\JJ}}\tilde{\psi}_{\hat{\II}}\f\psi_{\hat{\JJ}}+\check{g}_{YM}\epsilon^{\hat{\II}\hat{\JJ}}\psi_{\hat{\JJ}}\topp{\f}\tilde{\psi}_{\hat{\II}}\nonumber \\
 &  & +g_{YM}\e^{\IIh\JJh}\tilde{\psi}_{\hat{\JJ}}\lambda^{\II}Q_{\II\IIh}+\check{g}_{YM}\e^{\IIh\JJh}Q_{\II\IIh}\topp{\la}^{\II}\tilde{\psi}_{\hat{\JJ}}\nonumber \\
 &  & -g_{YM}\e_{\II\JJ}\bar{Q}^{\hat{\JJ}\II}\lambda^{\JJ}\psi_{\hat{\JJ}}-\check{g}_{YM}\e_{\II\JJ}\psi_{\hat{\JJ}}\topp{\la}^{\II}\bar{Q}^{\hat{\JJ}\JJ}\big]+h.c.
 \end{eqnarray}
\begin{eqnarray}
 \label{potential}
\mathcal{V}(g_{YM},\check{g}_{YM}) & = & g_{YM}^{2}\mbox{Tr}\big[\frac{1}{2}[\fbar,\f]^{2}+\MM_{\II}^{\:\:\II}(\f\fbar+\fbar\f)+\MM_{\II}^{\:\:\JJ}\MM_{\JJ}^{\:\:\II}-\frac{1}{2}\MM_{\II}^{\:\:\II}\MM_{\JJ}^{\:\:\JJ}\big]\nonumber \\
 &  & +\topp g_{YM}^{2}\mbox{Tr}\big[\frac{1}{2}[\bar{\fh},\fh]^{2}+\topp{\MM}_{\:\:\II}^{\II}(\fh\bar{\fh}+\bar{\fh}\fh)+\topp{\MM}_{\:\:\JJ}^{\II}\topp{\MM}_{\:\:\II}^{\JJ}-\frac{1}{2}\topp{\MM}_{\:\:\II}^{\II}\topp{\MM}_{\:\:\JJ}^{\JJ}\big]\nonumber \\
 &  & +g_{YM}\topp g_{YM}\mbox{Tr}\big[-2Q_{\II\IIh}\fh\bar{Q}^{\IIh\II}\bar{\f}+h.c.\big]-\frac{1}{N_{c}}\mathcal{V}_{d.t.}  \, ,
 \end{eqnarray}
 where  the mesonic operators $\MM$ are defined as\footnote{Note that $\mbox{Tr}[\MM_{\II}^{\:\:\JJ}]=\mbox{Tr}[\topp{\MM}_{\:\:\II}^{\JJ}]$.}
\begin{equation} 
\MM_{\JJ\:\:\: b}^{\:\:\II a}  \equiv  \frac{1}{\sqrt{2}}Q_{\JJ\hat{\JJ}\topp a}^{a}\bar{Q}_{\quad\:\:\:\: b}^{\JJh\II\topp a}\, \, , \qquad
\topp{\MM}_{\:\:\JJ\topp b}^{\II\topp a}  \equiv  \frac{1}{\sqrt{2}}\bar{Q}_{\quad\:\:\:\: a}^{\JJh\II\topp a}Q_{\JJ\hat{\JJ}\topp b}^{a} \, ,
\end{equation}
 and the double-trace terms in the potential are
\begin{eqnarray} \label{doubletraces}
\mathcal{V}_{d.t.} & = & g_{YM}^{2}\big(\mbox{Tr}[\MM_{\II}^{\:\:\JJ}]\mbox{Tr}[\MM_{\JJ}^{\:\:\II}]-\frac{1}{2}\mbox{Tr}[\MM_{\II}^{\:\:\II}]\mbox{Tr}[\MM_{\JJ}^{\:\:\JJ}]\big)\\
 &  & +\topp g_{YM}^{2}\big(\mbox{Tr}[\topp{\MM}_{\:\:\JJ}^{\II}]\mbox{Tr}[\topp{\MM}_{\:\:\II}^{\JJ}]-\frac{1}{2}\mbox{Tr}[\topp{\MM}_{\:\:\II}^{\II}]\mbox{Tr}[\topp{\MM}_{\:\:\JJ}^{\JJ}]\big)  \nonumber \\
 & = & \big(g_{YM}^{2}+\topp g_{YM}^{2}\big)\big(\mbox{Tr}[\MM_{\II}^{\:\:\JJ}]\mbox{Tr}[\MM_{\JJ}^{\:\:\II}]-\frac{1}{2}\mbox{Tr}[\MM_{\II}^{\:\:\II}]\mbox{Tr}[\MM_{\JJ}^{\:\:\JJ}]\big) \,. \nonumber
 \end{eqnarray}

The $SU(2)_L$ symmetry is present for all values of the couplings (and so is the $SU(2)_R \times U(1)_r$ R-symmetry, of course).
At the orbifold point $g_{YM} = \check g_{YM}$
there is an extra $\mathbb{Z}_2$ symmetry (the quantum symmetry of the orbifold) acting as 
\begin{equation} \label{quantumZ2}
\phi \leftrightarrow \check \phi \,, \quad
\lambda_\II \leftrightarrow \check \lambda_\II \, , \quad  A_m \leftrightarrow \check A_m \, ,\quad \psi_{\hat \II} \leftrightarrow \tilde \psi_{\hat \II} \, ,\quad
Q_{\II \hat \II} \leftrightarrow  -\epsilon_{\II\JJ}\epsilon_{\hat{\II}\hat{\JJ}}\bar{Q}^{ \JJ \hat \JJ } \, .
\end{equation}

Setting $\check g_{YM} = 0$, the second vector multiplet $(\check \phi, \check \lambda_\II, \check A_m)$ becomes free and completely decouples 
from the rest of theory, which happens to coincide 
with ${\cal N} = 2$ SCQCD (indeed the field content is the same and ${\cal N} = 2$ susy does the rest). 
The $SU(N_{\check{c}})$
symmetry can now be interpreted as a global flavor symmetry. In fact there is a symmetry enhancement
$SU(N_{\check c}) \times SU(2)_L \to U(N_f = 2 N_c)$: one sees in (\ref{yukawa}, \ref{potential}) that for $\check g_{YM} = 0$
 the $SU(N_{\check c})$ index $\check a$
and the $SU(2)_L$ index $\hat \II$
 can  be combined  into a single flavor index $i \equiv (\check a, \hat I) =1, \dots 2 N_c$.

In the rest of the paper, unless otherwise stated, we will work in the large $N_c \equiv N_{\check c}$ limit, keeping fixed
 the `t Hooft  couplings
\be
\lambda \equiv g_{YM}^2 N_c  \equiv 8 \pi^2 g^2 \, , \qquad  \check \lambda \equiv \check g_{YM}^2 N_{\check c}  \equiv 8 \pi^2 \check g^2 \, .
\ee
The normalizations of $g$ and $\check g$ are convenient for the perturbative calculations of \cite{spinchain},
in this paper it is just important to keep in mind that they are (square roots of) the 't Hooft couplings.
We will refer to the theory with arbitrary $g$ and $\check g$ as the ``interpolating SCFT'', thinking of keeping $g$ fixed
as we vary $\check g$ from  $\check g = g$  (orbifold theory) to  $\check g = 0$ 
 (${\cal N} =2$ SCQCD  $\oplus$ extra  $N_{\check c}^2-1$ free vector multiplets).

\section{\label{orbifoldspectrum} Protected Spectrum of the Interpolating Theory}

In the present and in the following section we will study the protected spectrum
of ${\cal N} = 2$ SCQCD at large $N$, in the flavor singlet sector, and its relation with the protected spectrum of the interpolating SCFT.
We have argued that in the large $N$ Veneziano limit,  flavor singlets that diagonalize the dilation operator take the  ``generalized single-trace'' form (\ref{generalizedsingletrace}).
We will look for  the generalized single-trace operators belonging to short multiplets of the superconformal algebra.
These are the operators expected to map to the Kaluza-Klein tower of massless single
 closed string states, so they are the first place to look in  a ``bottom-up'' search for the string dual.

The determination of the complete list of short multiplets of ${\cal  N} = 2$ SCQCD in this ``generalized single-trace'' sector
 turns out to be more subtle than expected. A class of short multiplets is relatively easy to isolate,
  namely the multiplets based on the following  superconformal primaries:
\begin{equation} \label{list} 
\mbox{Tr} \, \MM_{\bf 3} = (Q^a_{i} \bar Q_a^i)_{\bf 3} \, , \qquad \mbox{Tr} \, \f^{\ell +2} \, ,\qquad\mbox{Tr}[\, T\f^{\ell}]\, , \qquad \ell \geq 0\,.
\quad 
\end{equation}
Here $T \equiv \phi \bar \phi - {\cal M}_{\bf 1}$. We hasten to add that this will turn out to be only
a small fraction of the complete set of protected operators. The set 
 (\ref{list})
is the complete list of one-loop protected primaries {\it in the scalar sector}, as we show in \cite{spinchain} by
a systematic evaluation of the one-loop anomalous dimension of all operators that are  made out of scalars and obey shortening conditions.  The operators ${\rm Tr}  \, \phi^\ell$ correspond to the vacuum of the spin-chain studied in \cite{spinchain}, while
the  ${\rm Tr} \, T  \phi^\ell$ correspond to the $p\to 0$ limit of a gapless magnon $T(p)$ of momentum $p$.

The operators $\mbox{Tr} \, \MM_{\bf 3}$ 
and  $\mbox{Tr} \, \f^{\ell +2}$ obey the familiar BPS condition  $\Delta = 2 R + |r|$, where $R$ is the $SU(2)_R$ spin and $r$ the $U(1)_r$ charge,
and they are generators of the chiral ring (with respect to an ${\cal N} = 1$ subalgebra), see appendix B.\footnote{\label{openfootnote}
Incidentally, the analysis of the chiral ring extends immediately to flavor non-singlets. The only chiral ring generator which is not a flavor singlet is 
 the
$SU(2)_R$ triplet bilinear
\be \label{protectedopen}
 {\cal O}^i_{{\bf 3} \, j} \equiv (\bar Q^i_{\,\,a} Q^a_{\,\,j})_{\bf 3} =  \bar Q^i_{ a \, \{ \II } Q^a_{ \JJ \} \,j}\, ,
\ee
in the adjoint of $SU(N_f)$.  The  conserved  currents for the $SU(N_f) \subset U(N_f)$ flavor symmetry belong to the short multiplet with bottom component ${\cal O}^i_{{\bf 3} \, j}$.
Similarly the current for the $U(1) \subset U(N_f)$ baryon number belongs to the $\Tr \,{\cal M}_{\bf 3}$ multiplet.}
By contrast $\mbox{Tr}[\, T\f^{\ell}]$ obey a ``semi-shortening'' condition and it may be missed in a naive analysis;
 in  these operators there is a large mixing of ``glueballs'' and ``mesons'' and the idea of considering ``generalized single-traces'' is essential.
The  $\mbox{Tr} \, T$ multiplet plays a distinguished role since it contains  the stress-energy tensor and $R$-symmetry currents. 

Protection of the operators  (\ref{list}) can be understood from the viewpoint of the interpolating SCFT connecting
${\cal N} = 2$ SCQCD with the $\mathbb{Z}_2$ orbifold of ${\cal N} = 4$ SYM, as follows. The complete spectrum
of short multiplets at the orbifold point $g  = \check g$ is well-known.
We  will argue, using superconformal representation
theory \cite{Dolan:2002zh}, 
 that the protected multiplets found at the orbifold point {\it cannot} recombine into long multiplets
as we vary $\check g$, so in particular  taking $\check g \to 0$ they must evolve into protected multiplets of
the theory
\be \label{decoupledtheory}
\{ \NN = 2 \; {\rm SCQCD} \; \oplus \; {\rm decoupled} \; SU(N_{\check{c}}) \; {\rm vector \; multiplet} \} \, .
\ee
The list (\ref{list}) is precisely recovered by restricting to $U(N_f)$ singlets. 
Remarkably however,  the superconformal index of  ${\cal N} = 2$ SCQCD, evaluated in the next section,
 will show the existence of  {\it many  more} protected states. The extra protected states  arise from the splitting
 {\it long} multiplets of the interpolating theory into short multiplets as $\check g \to 0$.

We will make extensive use of the  the list given by Dolan and Osborn\cite{Dolan:2002zh} of all possible shortening conditions
of the ${\cal N} = 2$ superconformal algebra. We summarize their results and establish notations in appendix \ref{rep-theory}.

\subsection{Protected Spectrum at the Orbifold Point}

At the orbifold point ($g = \check g$) the state space of the field theory
 is the direct sum of an untwisted and a twisted sector, respectively 
 even and odd  under the ``quantum'' $\mathbb{Z}_2$ symmetry (\ref{quantumZ2}).

 \subsubsection{\label{sec:untwisted}Untwisted sector}

 Operators in the untwisted sector of the orbifold descend from operators of ${\cal N} = 4$ SYM
 by projection onto the $\mathbb{Z}_2$ invariant subspace.
 Their correlators coincide at large $N_c$ with ${\cal N} = 4$ correlators  \cite{Bershadsky:1998cb,Bershadsky:1998mb}.
 In particular   the complete list of   untwisted { protected} states is obtained by projection
 of the protected states of ${\cal N} = 4$. We will be interested in single-trace
 operators;  as is well-known,
 the only protected single-trace operators of ${\cal N} = 4$ belong to the $\frac{1}{2}$ BPS multiplets  $\BB_{[0,p,0]}^{\frac{1}{2},\frac{1}{2}}$,
 built on the chiral primaries ${\rm Tr} X^{\{ i_1} \dots X^{i_p \} }$, with $p \geq 2$,  in the $[0, p, 0]$ representation of $SU(4)_R$ (symmetric traceless of $SO(6)$)
 The decomposition  of each $\frac{1}{2}$ BPS multiplet $\NN=4$ into $\NN=2$
multiplets reads \cite{Dolan:2002zh}
\begin{eqnarray}
\BB_{[0,p,0]}^{\frac{1}{2},\frac{1}{2}} & \simeq & (p+1)\hat{\BB}_{\frac{1}{2}p}\oplus\EE_{p(0,0)}\oplus\bar{\EE}_{-p(0,0)}\nonumber \\
 &  & \oplus(p-1)\hat{\CC}_{\frac{1}{2}p-1(0,0)}\oplus p(\DD_{\frac{1}{2}(p-1)(0,0)}\oplus\bar{\DD}_{\frac{1}{2}(p-1)(0,0)}\nonumber \\
 &  & \oplus\bigoplus_{k=1}^{p-2}(k+1)(\BB_{\frac{1}{2}k,p-k(0,0)}\oplus\bar{\BB}_{\frac{1}{2}k,k-p(0,0)})\nonumber \\
 &  & \oplus\bigoplus_{k=0}^{p-3}(k+1)(\CC_{\frac{1}{2}k,p-k-2(0,0)}\oplus\bar{\CC}_{\frac{1}{2}k,k-p+2(0,0)})\nonumber \\
 &  & \oplus\bigoplus_{k=0}^{p-4}\bigoplus_{l=0}^{p-k-4}(k+1)\AA_{\frac{1}{2}k,p-k-4-2l(0,0)}^{p}\label{decomposition} \, ,\end{eqnarray}
 which can be  understood by considering
 all possible ways to substitute $X^i \to {\cal Z}, \bar {\cal Z}, {\cal X}_{\II \hat \II}$, 
 {\it i.e.} $\mathbf{6} \to (0,0)_1 \oplus (0,0)_{-1} \oplus (\frac{1}{2}, \frac{1}{2})_0$ in the  branching $SU(4)_R \to
 SU(2)_L \times SU(2)_R \times U(1)_r$.
   The $\mathbb{Z}_2$
orbifold projection is then accomplished by the substitution (\ref{survive}); 
states with an even (odd) number of ${\cal X}$s are kept (projected out), or equivalently,
 states with integer  (half-odd) $SU(2)_R$ spin are kept (projected out). 
Table \ref{untwistedtable} lists all the superconformal primaries  of the orbifold
theory obtained by  this procedure.

Let us explain the notation. The explicit expressions in terms of fields are schematic. The symbol $\sum$ indicates  summation over all ``symmetric traceless''
permutations of the component fields allowed by the index structure. 
The symbol $\Torb$ stands for the appropriate combination of two scalar fields,  neutral under the R symmetry. In the case of the multiplet $\hat \CC_{0(0,0)}$, $\Tr \, \Torb = \Tr \, [T +  \check \phi \bar {\check \phi} ]$, the bottom component of the stress tensor multiplet of the orbifold theory. 
 The $SU(2)_R \times U(1)_r$ quantum numbers are manifest as labels
of the ${\cal N} = 2$ multiplets, while the  $SU(2)_{L}$ quantum numbers
can be  seen from  the multiplicity of each multiplet on the right hand side of (\ref{decomposition}) -- 
the $SU(2)_L$ spin always equals the $SU(2)_R$ spin of the multiplet,
because  $SU(2)_R$ and $SU(2)_L$ indices  always come in
pairs $(\II \hat \II)$ and are separately symmetrized. 
\begin{table}
\begin{centering}
\begin{tabular}{|l|l|}
\hline 
Multiplet  & Orbifold operator ($R,\ell\geq0,\,n\geq2$) \tabularnewline
\hline
\hline 
$\hat{\BB}_{R+1}$  & $\mbox{Tr}[(Q^{+\hat{+}}\bar{Q}^{+\hat{+}})^{R+1}]$ \tabularnewline
\hline 
$\bar \EE_{-(\ell+2)(0,0)}$  & $\mbox{Tr}[\f^{\ell+2}+\fh^{\ell+2}]$  \tabularnewline
\hline 
$\hat{\CC}_{R(0,0)}$  & $\mbox{Tr}[\sum \Torb (Q^{+\hat{+}}\bar{Q}^{+\hat{+}})^{R}]$  \tabularnewline
\hline 
$\bar \DD_{R+1(0,0)}$  & $\mbox{Tr}[\sum(Q^{+\hat{+}}\bar{Q}^{+\hat{+}})^{R+1}(\f +\fh)]$  \tabularnewline
\hline 
$\bar \BB_{R+1,-(\ell+2)(0,0)}$  & $\mbox{Tr}[\sum_{i}(Q^{+\hat{+}}\bar{Q}^{+\hat{+}})^{R+1}\f^{i}\fh^{\ell+2-i}]$ \tabularnewline
\hline 
$\bar \CC_{R,-(\ell+1)(0,0)}$  & $\mbox{Tr}[\sum_{i} \Torb(Q^{+\hat{+}}\bar{Q}^{+\hat{+}})^{R}\f^{i}\fh^{\ell+1-i}]$  \tabularnewline
\hline
\hline $\AA_{R,-\ell(0,0)}^{\Delta=2R+\ell+2n}$ & $\mbox{Tr}[\sum_{i} \Torb^n(Q^{+\hat{+}}\bar{Q}^{+\hat{+}})^{R}\f^{i}\fh^{\ell-i}]$ \tabularnewline
\hline
\end{tabular}
\par\end{centering}
\caption{\label{untwistedtable} Superconformal primary
operators in the untwisted sector of the orbifold theory. They descend from the $\frac{1}{2}$ BPS primaries of ${\cal N} =4$ SYM.
 The symbol $\sum$ indicates  summation over all ``symmetric traceless''
permutations of the component fields allowed by the index structure.}
\end{table}

\begin{table}
\begin{centering}
\begin{tabular}{|l|l|}
\hline 
Multiplet  & Orbifold operator $(\ell\geq0)$ \tabularnewline
\hline
\hline 
$\hat{\BB}_{1}$  & $\mbox{Tr}[(Q^{+\hat{+}}\bar{Q}^{+\hat{-}}-Q^{+\hat{-}}\bar{Q}^{+\hat{+}})]=\Tr\,\MM_{\bf 3}$ \tabularnewline
\hline 
$\bar \EE_{-\ell-2(0,0)}$  & $\mbox{Tr}[\f^{\ell+2}-\fh^{\ell+2}]$  \tabularnewline
\hline 
\end{tabular}
\par\end{centering}
\caption{\label{twistedtable} Superconformal primary
operators in the twisted sector of the orbifold theory.}
\end{table}

\bigskip

\subsubsection{Twisted sector}
\label{twistedsubsection}
\medskip

\noindent
 In the twisted sector, we claim that
the complete  list of  single-trace superconformal primary operators obeying shortening conditions is
\begin{equation} \label{protectedtwisted}
{\rm Tr} [ \tau  Z^\ell ]  = \mbox{Tr}[\f^{\ell}-\fh^{\ell}] \, \; {\rm for} \; \ell \geq 2 \,  \quad\mbox{and}\quad\mbox{Tr}[\tau {\cal X}_{\II \hat \II} {\cal X}_{\JJ \hat \JJ} 
\epsilon^{\II \JJ}] =- \Tr  [Q_{   \IIh \{ \II  }\bar{Q}_{\JJ  \} }^{ \hat{\II}}] = - \Tr {\cal M}_{\bf 3} \, .
\end{equation}
That these operators are protected can be seen by the fact that they are the generators of the ${\cal N} =1$ chiral ring in the twisted sector, 
as we show in appendix \ref{chiralring}.  A priori there could be extra twisted states that do not belong to the chiral ring, as is the case for the untwisted sector. In the next section
we will evaluate the superconformal index of the orbifold theory and find  that  it matches  perfectly with the contribution of our claimed list of short multiplets.

The primary  $\mbox{Tr}[\f^{\ell}- \fh^{\ell}]$ corresponds  for each $\ell \geq 2$ to
 a second copy of the chiral multiplet $\bar \EE_{-\ell (0,0)}$ --  the first copy being the one in the untwisted sector built on  $\mbox{Tr}[\f^{\ell} + \fh^{\ell}]$.
The operator $\Tr  [Q_{   \IIh \{ \II  }\bar{Q}_{\JJ  \} }^{ \hat{\II}}]$ is an $SU(2)_R$ triplet with vanishing $U(1)_r$ charge and $\Delta = 2$,
and must be identified with the primary of a  $\hat {\cal B}_1$ multiplet. 
This protected multiplet has  been overlooked in previous discussions of the orbifold field theory. It is protected only in the theory
where the relative $U(1)$ has been correctly subtracted (see section 3.2), as seen both in the chiral ring analysis of appendix B and 
in an explicit one-loop calculation.

\subsection{\label{orbifoldaway}From the orbifold point to  ${\cal N} = 2$ SCQCD}

As we move away from the orbifold point by changing $\check g$,
the short multiplets that we have just enumerated may {\it a priori} recombine into
long multiplets and acquire a non-zero anomalous dimension. The  possible recombinations of  short  multiplets of the  $\NN=2$ superconformal
algebra were classified in \cite{Dolan:2002zh}. For short multiplets with a Lorentz-{\it{scalar}} bottom component, the relevant rule is
\[
\AA_{R,-\ell(0,0)}^{2R+\ell+2} \simeq
\bar\CC_{R,-\ell(0,0)}\oplus\bar\BB_{R+1,-(\ell+1)(0,0)} \,.
\]
 In the special case $\ell=0$, the short multiplets on the right hand side
 further decompose into even shorter multiplets as
 \[
\AA_{R,0(0,0)}^{2R+2}\simeq\hat{\CC}_{R(0,0)}\oplus\DD_{R+1(0,0)}\oplus\bar{\DD}_{R+1(0,0)}\oplus\hat{\BB}_{R+2(0,0)}\]\,.
It follows that the short multiplets of the orbifold theory that
that could  in principle recombine are
\begin{eqnarray}
&&  
\mbox{Tr}[\sum_{i}T(Q^{+\hat{+}}\bar{Q}^{+\hat{+}})^{R}{\f}^{i}{\fh}^{\ell-i}]\oplus\mbox{Tr}[\sum_{i}(Q^{+\hat{+}}\bar{Q}^{+\hat{+}})^{R+1}{\f}^{i}{\fh}^{\ell-i}] 
\longrightarrow \AA_{R,-\ell(0,0)}^{2R+\ell+2} 
 \\
&&
\mbox{Tr}[\sum T(Q^{+\hat{+}}\bar{Q}^{+\hat{+}})^{R}]\oplus\mbox{Tr}[\sum_{i}(Q^{+\hat{+}}\bar{Q}^{+\hat{+}})^{R+1}\bar{\f}^{i}\bar{\fh}^{1-i}]  \oplus\nonumber \\
 &&
\mbox{Tr}[\sum_{i}(Q^{+\hat{+}}\bar{Q}^{+\hat{+}})^{R+1}\f^{i}\topp{\f}^{1-i}]\oplus\mbox{Tr}[\sum(Q^{+\hat{+}}\bar{Q}^{+\hat{+}})^{R+2}]
\longrightarrow  \AA_{R,0(0,0)}^{2R+2} \,.
 \label{recomb}
 \end{eqnarray}
However we see that  the proposed recombinations entail short multiplets with {\it different} $SU(2)_L$ quantum
numbers, which is impossible
since the supercharges are neutral under $SU(2)_L$. Thus $SU(2)_L$ selection
rules forbid the recombination, and the protected multiplets of the orbifold theory
remain short for all values of $g$ and $\check g$. 
This conclusion was reached using  superconformal
representation theory,
 and it is a rigorous result valid at the full quantum level.\footnote{
We will rephrase the same result in the next section by computing
 a refined superconformal index that also keeps track of the $SU(2)_L$ quantum number.}

 In the limit $\check g \to 0$, we must be able to match the protected states of the interpolating SCFT
with  protected states of $\{ {\cal N} = 2$ SCQCD $\oplus$ decoupled vector multiplet$\}$.
In  \cite{spinchain} we follow this evolution in detail using the one-loop spin chain Hamiltonian.
The basic features of this evolution can be understood just from group theory.
The protected states naturally splits into two sets:
$SU(2)_L$ singlets and $SU(2)_L$ non-singlets. It is clear that all the
(generalized) single-trace operators of ${\cal N} = 2$ SCQCD must arise from the
$SU(2)_L$ singlets.

The $SU(2)_L$ singlets are:
\begin{enumerate}
\item[(i)] One  $\hat {\cal B}_1$ multiplet, corresponding to the primary $ \Tr  [Q_{   \IIh \{ \II  }\bar{Q}_{\JJ  \} }^{ \hat{\II}}] = \Tr \,{\cal M}_{\bf 3}$. Since this is the only
operator with these quantum numbers, it cannot mix with anything and its form is independent of $\check g$.
\item[(ii)] Two $\bar \EE_{-\ell (0, 0)}$ multiplets for each $\ell \geq 2$,  corresponding
to  the primaries $\Tr\, [ \phi^\ell \pm {\check \phi}^\ell ]$.
  For each $\ell$, there is a two-dimensional space
of protected operators, and we may choose whichever basis is more convenient.
For $g = \check g$, the natural basis vectors are the untwisted and twisted  combinations
(respectively even and odd under $\phi \leftrightarrow \check \phi$), while for $\check g = 0$ the natural basis vectors
are $\Tr\,  \phi^\ell$ (which is an operator of ${\cal N} =2$ SCQCD) and $\Tr\,  {\check \phi}^\ell$
(which belongs to the decoupled sector).
\item[(iii)]  One $\hat {\cal C}_{0 (0,0)}$ multiplet (the stress-tensor multiplet), corresponding to the primary $\Tr \, \Torb = \Tr \, [T +  \check \phi \bar {\check \phi} ]$.
We have checked that this combination is an eigenstate with zero eigenvalue for all $\check g$.
For $\check g = 0$,  we may trivially subtract out the decoupled piece $\Tr  \, \check \phi \bar {\check \phi}$ and recover
 $\Tr \, T$, the stress-tensor multiplet of ${\cal N} = 2$ SCQCD.
 \item[(iv)] One $\bar \CC_{0, -\ell (0,0)}$ multiplet for each $\ell \geq 1$. In the limit $\check g \to 0$, we expect this multiplet to evolve to the  $\Tr\, T \phi^\ell$
multiplet of ${\cal N} = 2$ SCQCD. We have checked this in detail in \cite{spinchain}.

\end{enumerate}
All in all, we see that this list reproduces the list (\ref{list}) of one-loop protected scalar operators of ${\cal N} = 2$ SCQCD, {\it plus} the extra states $\Tr \check \phi^\ell$ that decouple for $\check g = 0$.

 The basic protected primary of ${\cal N} = 2$ {SCQCD}
 which
   is charged under
 $SU(2)_L$    is the $SU(2)_L$ triplet contained in the mesonic operator ${\cal O}^i_{{\bf 3_R } \, j} = (\bar Q^i_a Q^a_j)_{\bf3_R}$ (see footnote \ref{openfootnote}).
Indeed writing the $U(N_f = 2 N_c)$ flavor 
indices $i$ as $i = (\check a, \hat \II)$, with $\check a =1, \dots N_f/2 = N_c$ ``half'' flavor indices and $\II = \hat \pm$ $SU(2)_L$ indices,
we can decompose
\be
{\cal O}^i_{{\bf 3_R } \, j}  \rightarrow {\cal O}^{\check a}_{{\bf 3_R  3_L} \, \check b}  \, , \quad {\cal O}^{\check a}_{{\bf 3_R  1_L} \, \check b}  \, .
\ee
In particular we may consider the highest weight combination for both $SU(2)_L$ and $SU(2)_R$,
\be \label{QQmeson}
(\bar{Q}^{+\hat{+}}Q^{+\hat{+}})^{\check a}_{\; \check b} \, .
\ee
States with higher $SU(2)_L$ spin can be built by taking products of   ${\cal O}_{{\bf 3_R  3_L} }$ with $SU(2)_L$ and $SU(2)_R$
indices separately symmetrized  -- and  this is the only way
to obtain protected states of ${\cal N} =2$ SCQCD charged under $SU(2)_L$ which have finite conformal dimension in the Veneziano limit.
 It  is  then a priori clear that a protected primary of the interpolating theory with $SU(2)_L$ spin $L$
  must evolve as $\check g \to 0$
  into a   {product} of $L$ copies of $(\bar{Q}^{+\hat{+}}Q^{+\hat{+}})$ and of as many additional decoupled scalars $\check \phi$ and $\bar {\check \phi}$
as needed to make up for the correct $U(1)_r$ charge and conformal dimension.  Examples of this evolution are given in \cite{spinchain}.

\subsection{Summary}

In summary all the short multiplets of the interpolating theory remain short as $\check g \to 0$, and 
 have a natural interpretation in this limit.  The $SU(2)_L$-singlet protected states
 evolve into the list (\ref{list}) of protected states of SCQCD, plus some extra states made purely from the decoupled
vector multiplet.  The interpolating theory has also many single-trace protected states with non-trivial $SU(2)_L$ spin,
which are flavor non-singlets from the point of view of ${\cal N} = 2$ SCQCD:
we have seen that  in the limit $\check g \to 0$, a  state with $SU(2)_L$ spin $L$ can be  interpreted as a ``multiparticle state'',
 obtained by linking together  $L$  short ``open''  spin-chains with of SCQCD and decoupled fields $\check \phi$.
This is also suggestive of
a dual string theory interpretation: as $\check g \to 0$,  single closed string states carrying $SU(2)_L$ quantum numbers disintegrate into multiple open strings.

Thus by embedding ${\cal N} = 2$ SCQCD into the interpolating SCFT
we have confirmed that the operators (\ref{list}) are protected at the full quantum level, 
since they arise as the limit of operators whose protection can be shown at the orbifold point and is preserved  by the exactly marginal deformation.
However this argument does {\it not} guarantee that (\ref{list}) is the {\it complete} set of  protected 
generalized single-trace primaries of ${\cal N} = 2$ SCQCD. 
 Indeed we will exhibit many more such states in the next section: they arise from {\it long} multiplets
 of the interpolating theory splitting into short multiplets at $\check g = 0$.

\section{Extra Protected Operators of ${\cal N} = 2$ SCQCD from the Index}

The superconformal index \cite{Kinney:2005ej} (see also \cite{Romelsberger:2005eg})
computes ``cohomological'' information about the protected
spectrum of a superconformal field theory. It counts (with signs) the multiplets obeying shortening conditions,
up to equivalence relations that set to zero all sequences of short multiplets that may in principle
recombine into long multiplets.  The index is invariant under exactly marginal deformations and can thus
be evaluated in the free field limit (if the theory admits a Lagrangian description). 
It should be kept in mind that the index does not  completely fix the protected spectrum. A first issue is a certain ambiguity in the
quantum numbers of the protected multiplets detected by the index.
Short multiplets can be organized into ``equivalence classes'', such that each short multiplet in a class
gives the same contribution to the index.
 For ${\cal N} = 2$ 4d superconformal
theories these equivalence classes contain a finite number of short multiplets.
This finite ambiguity could in principle be resolved by an explicit one-loop calculation, but  in practice
this is difficult since the diagonalization of the one-loop dilation operator becomes rapidly complicated
as the conformal dimension increases.
A second issue is  that some sequences of short multiplets that are kinematically allowed to recombine into long multiplets  may  in fact remain protected for dynamical reasons. This dynamical protection is known
to occur at large $N_c$ in ${\cal N} = 4$ SYM for certain multi-trace operators, but not for single-trace operators.

Despite these caveats, the index is a very valuable tool. 
In this section, after reviewing the definition of the index \cite{Kinney:2005ej}, we  explain
exactly what kind of information can be extracted from it, by characterizing the ``equivalence classes'' of short multiplets
that give the same contribution to the index.  We then proceed to concrete calculations, evaluating
the index for  the interpolating  SCFT and for ${\cal N} = 2$ SCQCD.
The free field contents of the two theories, and thus their indices, are different:  recall that the interpolating SCFT has an extra vector multiplet
in the adjoint of $SU(N_{\check c})$.
The index for the interpolating theory confirms the protected spectrum of single-trace operators discussed in the previous section. 
By contrast, the index for ${\cal N} = 2$  SCQCD reveals the existence of many more 
generalized single-trace operators obeying shortening conditions: their
  degeneracy grows exponentially with the conformal dimension. Interestingly, we find protected operators with arbitrarily high spin,
  though none of them is a higher-spin conserved  current.  We account for  the origin of these extra protected states by identifying long multiplets of the interpolating
 theory that at $\check g = 0$  split into short multiplets:
  some of the resulting short multiplets belong purely to ${\cal N} = 2$ SCQCD ({\it i.e}. do not contain fields in the decoupled
  vector multiplet) and comprise the extra states.

\subsection{Review of the Superconformal Index}

The superconformal index  \cite{Kinney:2005ej} is just the Witten index with respect to one of the Poincar\'e supercharges, call it  ${\cal Q}$,
of the superconformal algebra. Let ${\cal S} = {\cal Q}^\dagger$ be the conformal supercharge conjugate to ${\cal Q}$,
and $\delta\equiv2\{\mathcal{S},\mathcal{Q}\}$. Every state in a unitary representation of the superconformal algebra has $\delta\geq0$.
The index is defined as
\be
\II = \Tr\, (-1)^F e^{-\alpha \delta+ M}\, ,
\ee
where the trace is over the Hilbert space of the theory on $S^3$, in the usual radial quantization,
and $M$ is any operator that commutes with $\QQ$ and $\SS$. 
The index receives contributions only from states with $\delta=0$, which are in one-to-one correspondence
with the cohomology classes of ${\cal Q}$. It is thus independent of $\alpha$.

There are in fact two inequivalent possibilities for the choice of ${\cal Q}$,
leading to a ``left'' index $\II^{\ind L}$ and a ``right'' index $\II^{\ind R}$. The choice $\QQ = \QQ^{1}_-$ leads to the ``left'' index $\II^{\ind L}$. In this case 
\begin{equation}
\delta^{\ind L}=\Delta-2j-2R-r \, .
\end{equation}
Introducing chemical potentials for all the operators that commute with $\QQ$ and $\SS$, 
one defines
\be
\II^{\ind L}(t,y,v) \equiv   \Tr\, (-1)^{F}\, t^{2(\Delta+j)}\, y^{2\bar j}v^{r -R} \,.
\label{indexformulaL}
\ee
The choice $\QQ = \bar \QQ_{2+}$ gives instead the ``right'' index  $\II^{\ind R}$. In this case
\bea
 \delta^{\ind R} & \equiv & \Delta-2\bar{j}-2R+r \\
\II^{\ind R}(t,y,v)&=&\Tr\, (-1)^{F}\, t^{2(\Delta+\bar{j})}\, y^{2j}v^{-r-R}\label{indexformulaR}.
\eea
The relation between the left and right index is simply $j \leftrightarrow \bar j$ and $r \leftrightarrow -r$.
For an ${\cal N} = 2$ theory, which is necessarily non-chiral, the left and right indices are in fact equal as functions
of the chemical potentials, $\II^{\ind L}(t,y,v) = \II^{\ind R}(t,y,v)$, but it will be useful to have introduced the definitions of both.

\subsection{Equivalence Classes of Short Multiplets}

We have mentioned that there is a certain finite ambiguity in extracting from the index
which are the actual multiplets that remain short. Schematically, the issue is the following.
Suppose that two short multiplets, $S_1$ and $S_2$,
can recombine to form a long multiplet $L_1$, 
\be
S_1 \oplus S_2 = L_1 \, ,
\ee
and similarly that $S_2$ can recombine with a third short multiplet $S_3$ to give another long multiplet $L_2$,
\be
S_2 \oplus S_3 = L_2\,.
\ee
By construction, the index evaluates to zero on long multiplets, so
\be
\II(S_1) = - \II(S_2) = \II(S_3) \,.
\ee
We say that the two multiplets $S_1$ and $S_3$ belong to the same equivalence class, since their indices are the same.
Note that $S_2$ {\it can} be distinguished from $S_1 \sim S_3$ by  the overall sign of its index.

The recombination rules for ${\cal N}=2$ superconformal algebra are \cite{Dolan:2002zh}
\begin{eqnarray}
\AA_{R,r(j,\bar{j})}^{2R+r+2j+2} & \simeq & \CC_{R,r(j,\bar{j})}\oplus\CC_{R+\frac{1}{2},r+\frac{1}{2}(j-\frac{1}{2},\bar{j})}
\label{recomb2}\\
\AA_{R,r(j,\bar{j})}^{2R-r+2\bar j+2} & \simeq & \bar\CC_{R,r(j,\bar{j})}\oplus\bar\CC_{R+\frac{1}{2},r-\frac{1}{2}(j,\bar{j}-\frac{1}{2})}
\label{eq:3rd recomb}\\
\AA_{R,j-\bar{j}(j,\bar{j})}^{2R+j+\bar{j}+2} & \simeq & \hat{\CC}_{R(j,\bar{j})}\oplus\hat{\CC}_{R+\frac{1}{2}(j-\frac{1}{2},\bar{j})}\oplus\hat{\CC}_{R+\frac{1}{2}(j,\bar{j}-\frac{1}{2})}\oplus\hat{\CC}_{R+1(j-\frac{1}{2},\bar{j}-\frac{1}{2})} \,.
\label{recomb1}
\end{eqnarray}
Notations are reviewed in appendix \ref{rep-theory}. The ${\cal C}$, $\bar {\cal C}$ and $\hat {\cal C}$ multiplets
 obey certain ``semi-shortening'' conditions, see Table \ref{shortening}, while $\AA$ multiplets are generic long multiplets. 
 A long multiplet whose conformal dimension is exactly at the unitarity threshold can be decomposed  into shorter multiplets according to (\ref{recomb2},\ref{eq:3rd recomb},\ref{recomb1}).
  We can formally regard any  multiplet obeying some shortening condition (with the exception of the $\EE$ and $\bar \EE$ types)
 as a multiplet of  type ${\cal C}$, $\bar {\cal C}$ or $\hat \CC$  by allowing the spins $j$ and $\bar j$, whose natural range is over the non-negative
 half-integers,  to  take the value $-1/2$ as well.   The translation is as follows:
\begin{equation}
\CC_{R,r(-\frac{1}{2},\bar{j})}\simeq\BB_{R+\frac{1}{2},r+\frac{1}{2}(0,\bar{j})}.
\label{translation}
\end{equation}
\[
\hat{\CC}_{R(-\frac{1}{2},\bar{j})}\simeq\DD_{R+\frac{1}{2}(0,\bar{j})},\qquad\hat{\CC}_{R(j,-\frac{1}{2})}\simeq\bar{\DD}_{R+\frac{1}{2}(j,0)}\,. 
\]
\be
\hat{\CC}_{R(-\frac{1}{2},-\frac{1}{2})}\simeq \DD_{R+\frac{1}{2}(0,-\frac{1}{2})} \simeq
\bar{\DD}_{R+\frac{1}{2}(-\frac{1}{2},0)} \simeq \hat{\BB}_{R+1}.
\ee
Note how these rules flip statistics: a multiplet with bosonic primary ($j + \bar j$ integer) is turned
into a multiplet with fermionic primary ($j + \bar j$ half-odd), and viceversa.  
With these conventions, the rules (\ref{recomb2}, \ref{eq:3rd recomb}, \ref{recomb1}) are the most general recombination rules. 
The  $\EE$ and $\bar \EE$ multiplets never recombine.

\begin{figure}[htbp]
\begin{centering}
\includegraphics[scale=0.55]{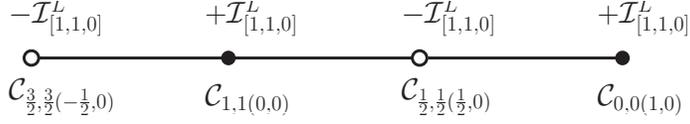}
\par\end{centering}
\caption{The equivalence classes $[1,1,0]^{\ind L}_{\pm}$. The multiplets belonging to $[1,1,0]^{\ind L}_{\pm}$ have index $ \pm \II^{\ind L}_{[1,1,0]}$. The sum of the indices  of adjacent multiplets is zero, as 
required by the recombination rule. 
\label{line1}
}
\end{figure}

Let us start by characterizing the equivalent classes for ${\cal C}$-type multiplets.
The right index vanishes identically on ${\cal C}$ multiplets.  From (\ref{recomb2}), we have
 \be
\II^{\ind L}[\CC_{R,r(j,\bar{j})}]+\II^{\ind L}[\CC_{R+\frac{1}{2},r+\frac{1}{2}(j-\frac{1}{2},\bar{j})}]  =  0\,.
\label{eq:index2}
\ee
Clearly $\tilde R \equiv R+j$, $\tilde r \equiv r+j$ and $\bar j$ and the overall sign are the invariant quantum numbers that label  an equivalence class.  We denote by
$[\tilde R,\tilde r,\bar j]^{\ind L}_{+}$ the equivalence class of ${\cal C}$ multiplets with 
 $\II^{\ind L}= \II^{\ind L}[{\cal C}_{\tilde R,\tilde r (0, \bar{j})} ]$, and by $[\tilde R,\tilde r,\bar j]^{\ind L}_{-}$ the class 
 with  $\II^{\ind L}= -\II^{\ind L}[{\cal C}_{\tilde R,\tilde r (0, \bar{j})} ]$,
\begin{eqnarray}
\,[\tilde R,\tilde r,\bar j]^{\ind L}_{+}  & = &\{   {\cal C}_{\tilde R-m,\tilde r -m \, (m, \bar j )} \, | \, m = 0, 1, 2 \dots\, , m \leq   \tilde R \} \, 
\\ 
\,[\tilde R,\tilde r,\bar j]  ^{\ind L}_{-}  &= & \{   {\cal C}_{\tilde R-m,\tilde r -m \, (m, \bar{j})} \, | \, m = -\frac{1}{2}, \frac{1}{2}, \frac{3}{2} \dots\, ,   m \leq \tilde R  \} \, .
\end{eqnarray}
Explicitly, the left index of the class $[\tilde R,\tilde r,\bar j]_\pm^{\ind L}$ is:
\be
\II^{\ind L}_{[\tilde R,\tilde r,\bar j]_{\pm}^{\ind L}}=  \pm(-1)^{2\bar j+1}t^{6+4\tilde R+2\tilde r}v^{-2+\tilde r-\tilde R}\frac{(1-t^2 v)(t-\frac{v}{y})(t-vy)}{(1-t^3 y)(1-\frac{t^3}{y})}(y^{2\bar j}+\ldots +y^{-2\bar j})
\label{indexC}
\ee
We have illustrated the equivalence classes $[1,1,0]^{\ind L}_{\pm}$ in Figure \ref{line1} by listing multiplets  on the $j$ axis. 
 The allowed values of $\tilde R$ and $\bar j$ are  $-\frac{1}{2}$, $0$, $\frac{1}{2}$, $1$, $\dots$, with the proviso
 that $j = -\frac{1}{2}$ or $\bar j = -\frac{1}{2}$ must be interpreted according to (\ref{translation}).
 For the lowest value of $\tilde R$, $\tilde R=-\frac{1}{2}$, the class $[-\frac{1}{2},\tilde r ,\bar j]^{\ind L}_{+}$ is empty while the class $[-\frac{1}{2},\tilde r ,\bar j]^{\ind L}_{-}=\BB_{\frac{1}{2},\tilde r+1(0,\bar j)}$ consists
 of a single multiplet, which can then  be determined without any ambiguity. 
  For $\tilde R=0$,  $[0 ,\tilde r ,\bar j]^{\ind L}_{+}=\CC_{0,\tilde r(0,\bar j)}$ and
  $[0,\tilde r ,\bar j]^{\ind L}_{-}=\BB_{1,\tilde r+1(0,\bar j)}$ both contain a single multiplet and again there is no ambiguity. Finally for $\tilde R=\frac{1}{2}$,  $[\frac{1}{2}, \tilde r, \bar j ]_{+}=\CC_{\frac{1}{2},\tilde r(0,\bar j)}$ contains
  a single multiplet, but  $[\frac{1}{2}, \tilde r, \bar j ]_{-}$ already has two and from  the index alone cannot decide which of the two actually remains protected.
Clearly the ambiguity grows linearly with  $\tilde R$.
 
The analysis for the  $\bar \CC$ multiplets is entirely analogous, and follows 
from the previous discussion by the substitutions $j \leftrightarrow \bar j$, $r \leftrightarrow -r$.
 One needs to consider $\II^{\ind R}$, since now it is $\II^{\ind L}$ that evaluates to zero. 
The equivalence classes are defined to be the set of all the $\bar \CC$ multiplets with same $\II^{\ind R}$ up to sign, and are denoted as $[\bar {\tilde R} , \bar{\tilde r} ,j]^{\ind R}_{\pm}$, where $ \bar {\tilde R} \equiv  R+\bar j$, $ \bar{\tilde r} \equiv -r+\bar j$.
 
\begin{figure}[htbp]
\begin{centering}
\subfloat[$\hat{\CC}_{0(\frac{1}{2},\frac{1}{2})}$ and $\hat{\CC}_{2(-\frac{1}{2},-\frac{1}{2})}\equiv\hat{\BB}_{3(0,0)}$]
{
\begin{centering}
\includegraphics[scale=0.7]{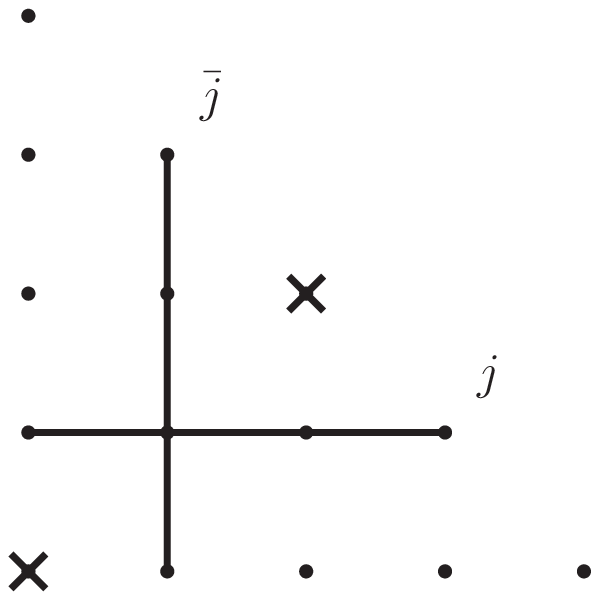}
\par\end{centering}
}
$\qquad\qquad\qquad$
\subfloat[$\hat{\CC}_{1(-\frac{1}{2},\frac{1}{2})}\equiv\DD_{\frac{3}{2}(0,\frac{1}{2})}$
and $\hat{\CC}_{1(\frac{1}{2},-\frac{1}{2})}\equiv\bar{\DD}_{\frac{3}{2}(0,\frac{1}{2})}$]
{
\begin{centering}
\includegraphics[scale=0.7]{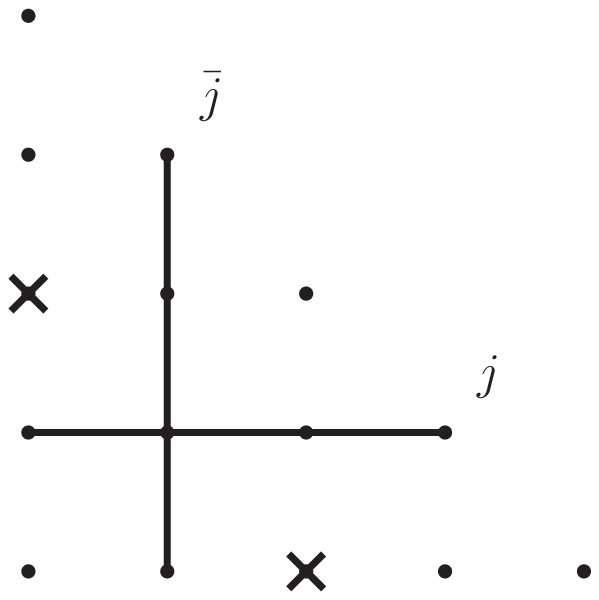}
\par\end{centering}
}
\par\end{centering}
\caption{\label{indexChat}Example of two configurations of the $\hat{\CC}$
multiplets with $R+j+\bar{j}=1$ contributing the same to both
$\II^{\ind L}$ and $\II^{\ind R}$. The multiplets are denoted by crosses on the $(j , \bar j)$ grid.  The indices are the same
for (a) and (b) because the
{\it projections} on the $j$ and $\bar j$ ({\it i.e.} the {\it sets} of $j$ and ${\bar j}$ values) are the same.
}
\end{figure}
 
The analysis for the $\hat{\CC}$ multiplets is slightly more involved. 
Unlike $\CC$ and $\bar{\CC}$ multiplets, $\hat{\CC}$ multiplets contribute to both $\II^{\ind L}$ and $\II^{\ind R}$. Moreover the quantum number $r$ is
fixed by the additional shortening condition $r = \bar j - j$. The left and right equivalence classes of 
$\hat \CC_{R(j,\bar j)}$ are  $[R+j,\bar j,\bar j]^{\ind L}_\pm$ and $[R+\bar j,j,j]^{\ind R}_\pm$ respectively.
The left index determines $\tilde R = R + j$ and the right index $\bar {\tilde R} = R + \bar j$,
so all in all  no two  different $\hat \CC$ multiplets give the same contribution to both  $\II^{\ind L}$ and $\II^{\ind R}$.  Nevertheless
different {\it direct sums} of $\hat \CC$ multiplets can have the same $\II^{\ind L}$ and $\II^{\ind R}$. 
It is convenient to introduce the quantum number $\hat R  \equiv R +  j + \bar j$, which is an invariant for both the left and the right  equivalence classes,
and to label the equivalence classes for $\hat \CC$ multiplets as $[\hat R, \bar j]^{\ind L}_\pm$ and $[\hat R,  j]^{\ind R}_\pm$. This new way to label  the classes
does not entail any loss of information, and makes it more convenient to analyze both the indices simultaneously. 
Explicitly, the left and right indices for these equivalence classes are:
\bea
\II^{\ind L}_{[\hat R, \bar j]^{\ind L}_\pm}&=&\pm(-1)^{2\bar j}\frac{t^{6-2\bar j+4\hat R}v^{-1+2\bar j-\hat R}(1-t^2 v)}{(1-t^3 y)(1-t^3 /y)}\nonumber
\\
&& (t(y^{2\bar j+1}+\ldots+y^{-(2\bar j+1)})-v(y^{2\bar j}+\ldots+y^{-2\bar j}))\\
\II^{\ind R}_{[\hat R,  j]^{\ind R}_\pm}&=&\pm(-1)^{2 j}\frac{t^{6-2 j+4\hat R}v^{-1+2 j-\hat R}(1-t^2 v)}{(1-t^3 y)(1-t^3 /y)}\nonumber
\\
&& (t(y^{2 j+1}+\ldots+y^{-(2 j+1)})-v(y^{2 j}+\ldots+y^{-2 j})) \,.
\eea
Now the point is that given a collection of $\hat \CC$ multiplets with the same value of $\hat R$, 
the left index determines the {\it set} of  $\bar j$ values  while
the right index determines the set of $j$ values, but in general there is not enough information to fix uniquely all quantum
numbers.  Figure \ref{indexChat} illustrates the ambiguity in a simple example: two different configurations, each consisting of two $\hat \CC$ multiplets, 
    give the same  contribution to both $\II^{\ind L}$ and $\II^{\ind R}$. 
 
\begin{table}
\begin{centering}
\begin{tabular}{|l|l|}
\hline 
Multiplet  & Equivalence class   \\
\hline
\hline 
$\CC$  & $[\tilde R,\tilde r,\bar j]^{\ind L}_{\pm}  \equiv [R+j,r+j, \bar j]^{\ind L}_{\pm}$\\
\hline
$\bar \CC$  & $[\bar {\tilde R},\bar {\tilde r}, j]^{\ind R}_{\pm}  \equiv [R+\bar j,-r+\bar j,  j]^{\ind R}_{\pm}$\\
\hline
$\hat \CC$  & $[\hat R,\bar j]^{\ind L}_{\pm}  \equiv [R+j+\bar j, \bar j]^{\ind L}_{\pm}$\\
  & $[\hat R,j]^{\ind R}_{\pm}  \equiv [R+j+\bar j,  j]^{\ind R}_{\pm}$\\
\hline
\end{tabular}
\par\end{centering}
\caption{Summary of notation for equivalence classes of short multiplets.}
\label{notation}
\end{table}

\subsection{The Index of the Interpolating Theory}

We now review the calculation of  the index for the orbifold theory  \cite{Kinney:2005ej, Nakayama:2005mf}.\footnote{While we agree
 with the general procedure followed in \cite{Nakayama:2005mf}, we disagree with the final result, equ.(3.5) of \cite{Nakayama:2005mf}.
 The discrepancy can be traced  to an incorrect subtraction of the  $U(1)$ factors in \cite{Nakayama:2005mf},
they are apparently taken to be ${\cal N} =1$ rather than ${\cal N} =2$ vector multiplets (equ.(2.12) of \cite{Nakayama:2005mf}). For the same reason we disagree with the expression 
((3.7) of \cite{Nakayama:2005mf}) for the  contribution to the index of the  $6d$ $(2,0)$ massless tensor multiplet, which we evaluate in appendix \ref{indexappendix}.}
 The index is invariant under exactly
marginal deformation and is thus  the same   for the whole family of 
 interpolating SCFTs.  The procedure is well-established. One enumerates
 the ``letters'' of the theory with $\delta = 0$ and then counts all possible
 gauge-invariants words. This is done efficiently by a matrix model, which for  large $N$ can be evaluated by saddle point. 
 Tables \ref{vectorindex} and \ref{hyperindex} list  the  $\delta^{\ind R}=0$ letters
 from the ${\cal N}=2$ vector and hyper multiplets.\footnote{For definiteness we 
 evaluate $\II^{\ind R}$, but recall that $\II^{\ind L} (t, y, v)  = \II^{\ind R}(t, y, v)$. The concrete letters with $\delta^{\ind L} = 0$ are different but 
 the left and right single-letter indices coincide.}
  Equations of motion are accounted for by introducing words with ``wrong'' statistics.
\begin{table}
\begin{centering}
\begin{tabular}{|r||r|r|r|r|r||r|}
\hline 
Letters  & $\Delta$  & $j$  & $\bar{j}$  & $R$  & $r$    & $\II^{\ind R}$ \tabularnewline
\hline
\hline 
$\phi$  & 1  & 0  & 0  & 0  & -1  &  $t^{2}v$ \tabularnewline
$\lambda_{+}^{1}$  & 3/2  & 1/2  & 0  & 1/2  & -1/2  &  $-t^{3}y$\tabularnewline
$\lambda_{-}^{1}$  & 3/2  & -1/2  & 0    & 1/2  & -1/2  & $-t^{3}y^{-1}$\tabularnewline
\hline 
$\bar{\lambda}_{2+}$  & 3/2  & 0  & 1/2    & 1/2  & 1/2  &  $-t^{4}v^{-1}$ \tabularnewline
$\bar{F}_{++}$  & 2  & 0  & 1   & 0  & 0  & $t^{6}$ \tabularnewline
\hline 
$\p_{++}$  & 1  & 1/2  & 1/2   & 0  & 0  & $t^{3}y$ \tabularnewline
$\p_{-+}$  & 1  & -1/2  & 1/2    & 0  & 0  &  $t^{3}y^{-1}$\tabularnewline
\hline 
$\p_{-+}\la_{+}^{1}+\p_{++}\la_{-}^{1}=0$  & 5/2  & 0  & 1/2   & 1/2  & 1/2  &  $t^{6}$ \tabularnewline
\hline
\end{tabular}
\par\end{centering}
\caption{Letters with $\delta^{\ind R}=0$ from the ${\cal N} = 2$ vector multiplet}
\label{vectorindex}
\end{table}
\begin{table}
\begin{centering}
\begin{tabular}{|r||r|r|r|r|r||r|}
\hline 
Letters  & $\Delta$  & $j$  & $\bar{j}$   & $R$  & $r$  &  $\II^{\ind R}$ \tabularnewline
\hline
\hline 
$q$  & 1  & 0  & 0   & 1/2  & 0   & $t^{2}v^{-1/2}$ \tabularnewline
$\bar{\psi}_{+}$  & 3/2  & 0  & 1/2    & 0  & -1/2    & $-t^{4}v^{1/2}$\tabularnewline
\hline 
$\tilde{q}$  & 1  & 0  & 0   & 1/2  & 0    & $t^{2}v^{-1/2}$ \tabularnewline
$\bar{\tilde{\psi}}_{+}$  & 3/2  & 0  & 1/2    & 0  & -1/2   & $-t^{4}v^{1/2}$ \tabularnewline
\hline
\end{tabular}
\par\end{centering}
\caption{Letters with $\delta^{\ind R}=0$ from the hyper multiplet}
\label{hyperindex}
\end{table}
One finds the single-letter indices for the vector multiplet and the ``half'' hyper multiplet
\begin{eqnarray}
f_{V}(t,y,v)&=&\frac{t^{2}v-t^{3}\left(y+y^{-1}\right)-t^{4}v^{-1}+2t^{6}}{\left(1-t^{3}y\right)\left(1-t^{3}y^{-1}\right)}\\
f_{H}(t,y,v)&=&\frac{t^{2}}{v^{1/2}}\frac{(1-t^{2}v)}{\left(1-t^{3}y\right)\left(1-t^{3}y^{-1}\right)} \, .
\end{eqnarray}
The
 single-letter index then reads 
\begin{eqnarray}
i_{orb}(t,y,v; U,\check U) & = & f_V(t,y,v)(\mbox{Tr}U\, \mbox{Tr}U^{\dagger} -1)+ f_V(t,y,v)(\mbox{Tr}{\check U}\, \mbox{Tr}{\check U}^{\dagger}-1)
\nonumber
\\
& & +\left(w+\frac{1}{w}\right) f_H(t,y,v)(\mbox{Tr}U \,\mbox{Tr}{\check U}^{\dagger}+  \mbox{Tr}U^\dagger \,\mbox{Tr}{\check U})\,.
\end{eqnarray}
Here $U$ and $\check U$ are   is an $N_c \times N_c$ unitary matrices out of which we construct the relevant characters of $SU(N_{c})$ and $SU(N_{\check{c}})$.
We have also introduced a potential $w$ that keeps track  of $SU(2)_L$ quantum numbers: $w+\frac{1}{w}$  is the character of the fundamental representation of $SU(2)_L$.
The index is obtained by enumerating all gauge-invariant operators in terms of the matrix integral
\bea
\II_{orb}&=&\int [dU] [d \check U] \exp\left(\sum_{n} \frac{1}{n} i_{orb}(t^n,y^n,v^n ;  U^n \check U^n) \right) \, ,
\label{matrixindex}
\eea
which for large $N_c$ can be carried out explicitly,
\be
\II_{orb} \cong \prod_{n=1}^{\infty}\frac{e^{-\frac{2}{n} f_V(t^n, y^n, v^n)}}{(1-f_V(t^{n},y^{n},v^{n}))^{2}-(w^{2n}+w^{-2n}+2)f_H^{2}(t^{n},y^{n},v^{n})} \equiv \II_{orb}^{m.t.}\,.
\label{multitrace}
\ee
This expression contains the contribution from all the gauge-invariant operators of the theory, which at large $N_c$ are multi-traces, hence the superscript in  $\II_{orb}^{m.t.}$.
To extract the contribution from single-traces  we evaluate the 
plethystic logarithm (see {\it e.g.} \cite{Feng:2007ur})
\begin{eqnarray}
\II_{orb}^{s.t.} & = & \sum_{n=1}^{\infty}\frac{\mu(n)}{n}\log[\II_{orb}^{m.t}(t^n,y^n,v^n)]
\\
& = & -\sum_{n=1}^{\infty}\frac{\varphi(n)}{n}\log[(1-f_V(t^{n},y^{n},v^{n}))^{2}-(w^{2n}+w^{-2n}+2)f_H^{2}(t^{n},y^{n},v^{n})]
\\
&&- 2 f_V(t,y,v) \nonumber \\
 & = & 2\left[\frac{t^{2}v}{1-t^{2}v}-\frac{t^{3}y}{1-t^{3}y}-\frac{t^{3}y^{-1}}{1-t^{3}y^{-1}}\right]+\frac{\frac{t^{4}w^2}{v}}{1-\frac{t^{4}w^2}{v}}+\frac{\frac{t^{4}}{vw^2}}{1-\frac{t^{4}}{vw^2}}- 2 f_V(t,y,v)\,.
 \label{factored}
\end{eqnarray}
Here $\mu(n)$ is the Moebius function ($\mu(1)\equiv 1$, $\mu(n) \equiv 0$ if $n$ has repeated prime factors,  and $\mu(n) = (-1)^k$ if $n$ is the product of $k$ distinct primes), and
$\varphi (r)$ is the Euler Phi function, defined as the number of positive integers less than or equal to  $r$ that are coprime  with respect to $r$. 
We have used the properties
\[
\sum_{d | n}  d \, \mu\left( \frac{n}{d}  \right) = \varphi(n) \, , \qquad  \sum_r \frac{\varphi(r)}{r} \log (1-x^r)=\frac{-x}{1-x} \,.
\]
The index is of course independent of $g$ and $\check g$. 
At the orbifold point $g = \check g$ it makes sense organize the spectrum into a twisted and an untwisted sector. Protected operators in the untwisted sectors are known
from inheritance from ${\cal N} = 4$ SYM. To evaluate the contribution to the index from the untwisted sector
we start with the single-trace  index for $SU(N_c)$ $\NN=4$ SYM and project onto the ${\mathbb Z}_2$ invariant subspace. 
The single-trace index for ${\cal N} = 4$ is found by regarding ${\cal N} = 4$ as an ${\cal N} = 2$ theory with one adjoint
vector and one adjoint hyper. A short calculation gives \cite{Kinney:2005ej}\footnote{Our notations for the chemical potentials are slightly different from \cite{Kinney:2005ej}.}
\bea
\II_{{\cal N} = 4} &=&\frac{t^2v}{1-t^2v}+\frac{\frac{t^2w}{\sqrt v}}{1-\frac{t^2w}{\sqrt v}}+\frac{\frac{t^2}{w\sqrt v}}{1-\frac{t^2}{w\sqrt v}}-\frac{t^{3}y}{1-t^{3}y}-\frac{t^{3}y^{-1}}{1-t^{3}y^{-1}}\nonumber\\
&&-f_V(t,y,v)-(w+\frac{1}{w})f_H(t,y,v)\,.
\eea
The $\mathbb{Z}_2$ acts as $w\to -w$ leaving invariant the under potentials, so
 the index  of the untwisted sector of the ${\mathbb Z}_2$ orbifold theory is
\begin{eqnarray}
\II_{untwist} &  =& \frac{1}{2} ( \II_{{\cal N} = 4} (t,y,v,w)  +  \II_{{\cal N} = 4}(t,y,v,-w)) \\
 & = & \frac{t^{2}v}{1-t^{2}v}-\frac{t^{3}y}{1-t^{3}y}-\frac{t^{3}y^{-1}}{1-t^{3}y^{-1}}+\frac{\frac{t^{4}w^2}{v}}{1-\frac{t^{4}w^2}{v}}+\frac{\frac{t^{4}}{vw^2}}{1-\frac{t^{4}}{vw^2}} - f_V(t,y,v)\,. \nonumber
\label{untwisted}
\end{eqnarray}
Subtracting the contribution of the untwisted sector from the total index (\ref{factored}),  we finally find 
\begin{equation}
\II_{twist}  =  \frac{t^{2}v}{1-t^{2}v}-\frac{t^{3}y}{1-t^{3}y}-\frac{t^{3}y^{-1}}{1-t^{3}y^{-1}} - f_V(t,y,v) \,.
\label{twisted} 
\end{equation}
In appendix \ref{indexappendix} we confirm that this precisely matches with the contribution from the twisted multiplets $\{ {\cal M}_{\bf 3}, \Tr (\phi^{2+\ell} - \check \phi^{2+\ell}) \,, \ell \geq0 \}$,
which are
the generators of the ${\cal N} = 1$ chiral ring in the twisted sector.

\subsection{\label{SCQCDindex} The  Index of ${\cal N}=2$ SCQCD and the Extra States}

The single-letter index for ${\cal N}=2$ SCQCD is
\be
i_{QCD}(t,y,v; U, V)  =  f_V(t,y,v)(\mbox{Tr}U\, \mbox{Tr}U^{\dagger} -1)
 + f_H(t,y,v)(\mbox{Tr}U \,\mbox{Tr}{ V}^{\dagger}+  \mbox{Tr}U^\dagger \,\mbox{Tr}{ V})\, ,
\ee
where $U$ an $N_c \times N_c$ matrix and $V$ an $N_f \times N_f$ matrix, with $N_f = 2N_c$.
 We are interested in gauge {\it and} flavor-singlets, so we integrate over both $U$ and $V$,
 \be
 \II_{QCD} = \int [dU] [dV] 
\exp\left(\sum_{n} \frac{1}{n} i_{QCD}(t^n,y^n,v^n ;  U^n  V^n) \right) \, .
\label{SCQCDmatrixindex}
 \ee
For large $N_c$ and $N_f$ with $N_f/N_c$ fixed we can again use saddle point,
\be
 \II_{QCD}  \cong 
 \prod_{n=1}^{\infty}\frac{e^{-\frac{1}{n} f_V(t^n, y^n, v^n)}}{(1-f_V(t^{n},y^{n},v^{n})) -f_H^{2}(t^{n},y^{n},v^{n})} 
 \equiv  \II_{QCD}^{m.t.} \,.
\ee
The index that enumerates (generalized) single-trace operators is then
\be \label{correctindex}
\II_{QCD}^{s.t.} =  -\sum_{n=1}^{\infty}\frac{\varphi(n)}{n}\log[(1-f_V(t^{n},y^{n},v^{n}))\,
-f_H^{2}(t^{n},y^{n},v^{n})]  - f_V(t,y,v)\,.
\ee
Unlike the orbifold theory, there is no nice factorization of the single-letter index and we cannot extract
the plethystic log explicitly. This is already an indication of a more complicated structure than expected.
The naive expectation is that all protected generalized single-trace multiplets of ${\cal N} = 2$ SCQCD
are exhausted by the list $\{ {\cal M}_{\bf 3}\,, \Tr \, \phi^{2+\ell} \, , \Tr\, T \phi^\ell \, , \ell \geq 0 \}$,
obtained  by projecting the protected single-trace spectrum of the interpolating theory onto $U(N_f)$ singlets.
We evaluate the corresponding index in appendix \ref{indexappendix},
\be
\II_{naive} = \frac{1}{(1-t^3y)(1-\frac{t^3}{y})}[-t^6(1-\frac{t}{v}(y+\frac{1}{y}))-\frac{t^{10}}{v}+\frac{t^4v^2(1-\frac{t}{vy})(1-\frac{ty}{v})}{1-t^2v}+\frac{t^4}{v}(1-t^2v)]\, ,
\ee
which is different from the correct index (\ref{correctindex}). Expanding in powers of $t$,  the first discrepancy  appears at $O(t^{13})$.

To get some insight, let us rewrite the single-trace index of the 
 orbifold theory as
\bea
\II^{s.t.}(h,k) & = & -\sum_{n=1}^{\infty}\Big[ \frac{\varphi(n)}{n}\log[(1-f_V(t^{n},y^{n},v^{n}))(1-h f_V(t^{n},y^{n},v^{n}))  \nonumber\\
&& -(k(w^{2n}+1+w^{-2n})+1)f_H^{2}(t^{n},y^{n},v^{n})
\Big] - f_V(t,y,v)\,.
\eea
We have introduced a variable $h$ that keeps track of  the number of $SU(N_{\check c})$ vector multiplets, and a variable $k$ associated
with the {\it triplet}  combination of two neighboring $SU(2)_L$ indices. The index  (\ref{correctindex}) for ${\cal N} = 2$  SCQCD is recovered in the limit  $(h,k)\to (0,0)$.
Indeed setting $(h,k) = (0,0)$:
 this amounts to omitting the ``second'' vector multiplet and to project onto $U(N_f)$ singlets, which is equivalent
to first projecting onto $SU(N_{\check c})$ singlets (automatically done in the interpolating theory) and then contracting all neighboring $SU(2)_L$ indices
into the singlet combination. The grading of gauge-invariant words by powers of $h$ (number of letters in the $SU(N_{\check c})$ vector multiplet)
makes sense only for $\check g = 0$. Similarly, for $\check g \neq 0$ only the overall $SU(2)_L$ spin
of a state is a meaningful quantum number, not the specific way neighboring $SU(2)_L$ indices are contracted. (For example
it is clearly possible to construct $SU(2)_L$ singlets which are not $U(N_f)$ singlets.)
 At $\check g \neq 0$ words with different $h$ or $k$ grading will generically mix. 

The origin of the extra protected states is then clear. As $\check g \to 0$, a long multiplets of the interpolating
theory, which obviously does not contribute to $\II_{orb}$,  may hit the unitarity bound and decompose
into a sum of short multiplets, some of which are $U(N_f)$ singlets and thus belong to ${\cal N} = 2$ SCQCD,
but some of which have instead non-trivial $h$ or $k$ grading. Schematically
\be
\lim_{\check g \to 0}  L  = \oplus S_{(h,k)=(0,0)} \; \oplus S_{(h, k) \neq (0,0)}\,.
\ee
The operators $\{ S_{(h,k)=(0,0)}   \}$ are the extra  states. They are protected in ${\cal N} = 2$ SCQCD
because they have no partners to recombine with.

Remarkably the extra protected states are vastly more numerous than the naive list. The asymptotic growth
of states in the naive list is clearly linear in the conformal dimension -- the number of states with $\Delta < N$ grows as $\sim 2N$,
in other terms the density of states $\rho (\Delta)$ is constant.
This modest growth  is consistent with the fact that the naive single-trace index does not ``deconfine'', {\it i.e.} it does not
diverge as a function of $t = e^{-1/T}$ for any finite temperature $T$. The same behavior  holds for  the orbifold theory or for ${\cal N} = 4$ SYM.
By contrast, the single-trace
index of ${\cal N} = 2$ SCQCD exhibits 
  Hagedorn behavior. Setting for simplicity all other potentials to 1, we encounter a divergence at $t = t_H$  such that
 \be
 1-f_V(t_H, 1, 1) - f_H^2(t_H, 1, 1) = 0 \longrightarrow t_H \cong 0.897769 \,.
 \ee
This implies an exponential   growth in the density of states contributing to the index,
\be
\rho(E')\sim e^{\beta_H  E'} \, ,\quad E' \equiv \Delta + j \, ,\quad \beta_H=-\ln t_H\cong 0.107842 \,.
\ee
 It is interesting to compare this behavior with the density of {\it generic} generalized single-trace operators of ${\cal N} = 2$ SCQCD. The density
of generic states, unlike the density of protected states, is  of course a function of the coupling.
For $g = 0$, it is obtained by calculating the phase transition temperature of the complete generalized single-trace  partition function (with no $(-1)^F$). We find
$\sim e^{\beta^\prime_H (\Delta+j)}$ with $\beta_H^\prime=1.34254$. Not surprisingly, $\beta_H < \beta'_H$.
 The density of  protected states, while exponential,  grows at a much slower rate than the density of the generic states, {or at least this is the behavior for small $g$.}

\subsection{Sieve Algorithm}

We would like to  list the quantum numbers of the extra protected states, up to the finite
equivalence class ambiguity intrinsic to the index.  There is no closed-form expression for 
$\II_{QCD}^{s.t.}$ but we can identity the equivalence classes contributing to it
in a systematic expansion in powers of $t$, by implementing a ``sieve'' algorithm similar in spirit to the one of \cite{Beisert:2003te}.

The first discrepancy between $\II_{QCD}^{s.t.}$ is the $O(t^{13})$ term 
\be
 \II_{QCD} - \II_{naive} =  -\frac{t^{13}}{v} (y+\frac{1}{y}) + \dots
\label{discr}
\ee
On the other hand, expanding  (\ref{indexC}) in powers of $t$, the lowest term is
\be
-t^{6+4 \tilde R+2 \tilde r }v^{\tilde r -\tilde R}(y^{2\bar j}+\ldots +y^{-2\bar j}) \,.
\ee 
Matching with (\ref{discr}) 
 we determine the equivalence class of the first new protected multiplet to be $[\tilde R, \tilde r, \bar j]^{\ind L}_+ = [\frac{3}{2},\frac{1}{2},\frac{1}{2}]^{\ind L}_+$. 
 Since $\tilde r=\bar j$, this is actually a $\hat \CC$ multiplet so we rewrite its equivalence class as $[ \hat R, \bar j ]^{\ind L}=
 [2,\frac{1}{2}]^{\ind L}_+$. Subtracting the whole index of the class
 from the discrepancy we  proceed to the next mismatch in the $t$ expansion, and so on. In this way, we can systematically construct the equivalence classes of all the extra protected multiplets of the SCQCD. The results from $\II^{\ind L}$ for first few multiplets are:
\begin{itemize}
\item $\CC$ multiplets: $[2,2,0]^{\ind L}_+,\,[2,3,0]^{\ind L}_+,\,[2,4,0]^{\ind L}_+,\,[3,2,0]^{\ind L}_-,\,[3,2,1]^{\ind L}_-,\, \ldots$
\item $\hat \CC$ multiplets: $[2,\frac{1}{2}]^{\ind L}_+,\,[4,1]^{\ind L}_+,\,[4,\frac{3}{2}]^{\ind L}_+,\, \ldots$
\end{itemize}
From the analysis of $\II^{\ind R}$ we can write down the {\it right} equivalence classes of the protected multiplets. Since $\II^{\ind R} = \II^{\ind L}$, 
the list of  {\it  right} equivalence classes is  obtained immediately from the list  of {\it left} equivalence classes  by the substitutions $\CC\rightarrow \bar \CC$ and $\ind L\rightarrow \ind R$. 

Protected $\bar \CC$ multiplets are just conjugates of protected $\CC$ multiplets. The $\hat \CC$ multiplets, however, appear in both left and right classes,
and as we discussed this gives  more information. 
For example the $\hat \CC$ multiplet in $[2,\frac{1}{2}]^{\ind L}_+ $ also belongs to $[2,\frac{1}{2}]^{\ind R}_+$ and furthermore it is the only multiplet with $\hat R=R+j+\bar j=2$. The left equivalence class determines $\bar j=\frac{1}{2}$, the right equivalence class  $j=\frac{1}{2}$ and both  also imply $R=\hat R - j-\bar j=1$.  This determines
the lowest-lying extra protected $\hat \CC$  multiplet to be $\hat \CC_{1(\frac{1}{2},\frac{1}{2})}$. 
For $\hat R=4$, there are two multiplets with 
$\bar j=1,\,\frac{3}{2}$ and with same values of $j$. Two possible $(j,\bar j)$ Lorentz spins are $(1,1),\,(\frac{3}{2},\frac{3}{2})$ or $(1,\frac{3}{2}),\,(\frac{3}{2},1)$ but we also know that it is a bosonic multiplet from the subcript $+$. This picks out the pair $(1,1),\,(\frac{3}{2},\frac{3}{2})$ with $R=4-1-1=2$ and $R=4-\frac{3}{2}-\frac{3}{2}=1$ respectively. This determines the next protected $\hat \CC$ multiplets to be $\hat \CC_{1(\frac{3}{2},\frac{3}{2})}$ and $\hat \CC_{2(1,1)}$. To summarize, the first three protected $\hat \CC$ multiplets are:
\begin{itemize}
\item $\hat \CC$ multiplets: $\hat \CC_{1(\frac{1}{2},\frac{1}{2})}$, $\hat \CC_{1(\frac{3}{2},\frac{3}{2})}$, $\hat \CC_{2(1,1)}$, $\ldots$
\end{itemize}
A striking feature of the extra protected multiplets is that they contain
states with higher spin, in fact we believe that the sieve will produce arbitrarily high spin. To the best
of our knowledge this is the first time that higher-spin protected multiplets are found in an {\it interacting} 4d superconformal
field theory. Note that {\it none} of the protected states we find are higher spin {\it conserved currents}, which correspond to the multiplets $\hat \CC_{0 (j, \bar j)}$.
This is not surprising: higher spin conserved currents are the hallmark of a free theory, but ${\cal N} = 2$ SCQCD is most definitely
an interacting quantum field theory. As in ${\cal N} = 4$ SYM \cite{Beisert:2004di}, higher spin conserved currents exist at strictly zero coupling, but they are
 anomalous and recombine into long multiplets at non-zero coupling.

\section{Dual Interpretation of the Protected Spectrum}

As we have repeatedly emphasized,  ${\cal N} = 2$ SCQCD can be obtained as the $\check g_{YM} \to 0$ limit
of a family of ${\cal N} = 2$ superconformal field theories, which reduces for  $g_{YM} = \check g_{YM}$ to 
  the ${\cal N}=2$  $\mathbb{Z}_2$ orbifold of ${\cal N} = 4$ SYM. 
  This latter  theory has a familiar dual description has IIB string theory 
  on $AdS_5 \times S^5/\mathbb{Z}_2$ \cite{Kachru:1998ys}, so it would seem that to find the dual of $\NN = 2$ SCQCD we simply
  need to follow the fate of the bulk string theory under the exactly marginal deformation. 
   Recall that at the orbifold point 
the NSNS $B$-field  has half-unit period through the  blown-down $S^2$ of the orbifold singularity, $\int_{S^2} B_{NS} = 1/2$ \cite{Aspinwall:1995zi}. 
  Taking $\check{g}_{YM} \neq g_{YM}$ is dual to changing the  period of  $B$-field, according to the dictionary
   \cite{Lawrence:1998ja, Klebanov:1999rd} 
 \begin{eqnarray} \label{Bdictionary}
&& \frac{1}{g_{YM}^2} +   \frac{1}{\check g_{YM}^2}   =  \frac{1}{2 \pi g_s} \, \\
  && \frac{\check g_{YM}^2}{g_{YM}^2}  = \frac{\beta}{1-\beta} \, , \qquad  \beta \equiv \int_{S^2} B_{NS}\,.
 \end{eqnarray}
 The catch is that
the limit  $\check g_{YM}  \to 0$ translates on the dual side to the singular limit of vanishing $B_{NS}$  and vanishing  string coupling $g_s$,
and the IIB background $AdS_5 \times S^5/\mathbb{Z}_2$ becomes ill-defined. 
 We will study in the next section how to handle this subtle limit.  In this section we will try to learn about the string dual of ${\cal N} = 2$ SCQCD from the ``bottom-up'',
 collecting the clues offered by the spectrum of protected operators.
 We start by reviewing the well-known bulk-boundary dictionary  for the protected states of the
 orbifold theory.

\subsection{KK interpretation of the orbifold protected specrum}

The untwisted spectrum of the orbifold field theory (summarized in Table \ref{untwistedtable}), has a transparent dual interpretation as the Kaluza-Klein spectrum of
IIB supergravity on $AdS_5 \times S^5/\mathbb{Z}_2$.  It is appropriate to write the metric of $S^5/\mathbb{Z}_2$
as \cite{Aharony:1998xz}
\be \label{alphametric}
ds^2_{S^5/\mathbb{Z}_2} =  d\alpha^2 + \sin^2 \alpha \, d \varphi^2 + \cos^2 \alpha \, ds^2_{S^3/\mathbb{Z}_2} \, , \qquad 0 \leq \varphi \leq 2 \pi \,, \quad 0 \leq \alpha \leq \frac{\pi}{2} \,.
\ee
Momentum on  $S^1$ corresponds to the $U(1)_r$ charge $r$. The $SO(4) \cong SU(2)_L \otimes SU(2)_R$
isometry of the 3-sphere is broken to  $SO(3)_L \otimes SU(2)_R$ by the $\mathbb{Z}_2$ orbifold, which projects out
harmonics with $j_L$   half-odd.  Needless to say, $SU(2)_R$ and $SO(3)_L$ are interpreted as the field theory symmetry groups
 of the same name, so in particular the right spin $j_R$ is identified with the quantum number $R$.
 Finally the harmonics on the $\alpha$ interval 
are parametrized by an integer $n$,  dual  to the power of neutral scalar ${\cal T}$ (with $\Delta = 2$)
in the schematic expressions of the operators in Table \ref{untwistedtable}.  It is not difficult to carry an explicit KK expansion and confirm
that $\Delta = |r| + 2 R + 2n$. A nice shortcut is to consider the KK expansion of the ten dimensional dilaton-axion \cite{Aharony:1998xz}, 
since only {\it scalar} harmonics on $S^5/\mathbb{Z}_2$ are required.  Scalar harmonics on $S^3/\mathbb{Z}_2$ have $(j_L , j_R) = (2 R, 2 R)$ with $2 R$ a non-negative integer.  
One finds $\Delta =  |r| + 2 R + 2n + 4$ \cite{Aharony:1998xz}, as expected 
from the fact that the KK modes of the dilaton-axion are dual to the descendants obtained by acting with ${\cal Q}^4 \bar {\cal Q}^4$ on the superconformal primaries of Table \ref{untwistedtable}.

The twisted states of the orbifold field theory (shown in Table \ref{twistedtable}),  
must map on the dual side to twisted closed string states localized at the fixed locus of the orbifold,
which is $AdS_5 \times S^1$, corresponding to $\alpha = \pi/2$ in the parametrization (\ref{alphametric}). 
The massless twisted states of IIB on the $A_1$ singularity comprise one massless six-dimensional tensor multiplet, so the KK reduction of the tensor multiplet 
 on $AdS_5 \times S^1$ must reproduce the protected twisted states of the orbifold field theory. It does, as we review in  appendix \ref{tensorKK} following the analysis of
 \cite{Gukov:1998kk}, to which we add a detailed treatment of the zero modes. We find that the zero modes of the tensor multiplet correspond to the multiplet build on the ``exceptional state''
${\rm Tr}\, {\cal M}_{\bf 3}$.

\subsection{Interpretation for $\NN = 2$ SCQCD?}

 The protected spectrum of  ${\cal N} = 2$ SCQCD  (restricting as usual to flavor singlets, and in the large $N$ Veneziano limit)
 consists of two sectors: the ``naive''  list of protected primaries (\ref{list}) easily found by a one-loop calculation
  in the scalar sector \cite{spinchain}; and the many more extra ``exotic'' states found in the analysis of the superconformal
 index. 
 
 The ``naive'' spectrum arises from a truncation of the protected spectrum of the interpolating theory (as $\check g \to 0$)
 to $U(N_f)$ singlets. We have discussed in section 2 the reason to  focus on the flavor-singlet sector:
  flavor-singlet operators, which necessarily are of ``generalized single-trace type'' in the Veneziano limit, 
  are   expected to map to single closed string states.   The restriction to $U(N_f)$ singlets
  has an interesting geometric interpretation: flavor singlets are in particular $SU(2)_L$ singlets, and thus they are dual
  to supergravity states with {\it no} angular momentum  on $S^3/\mathbb{Z}_2$ in the parametrization (\ref{alphametric}).
  So in performing this restriction we are ``losing'' three spatial dimensions.   
  As explained around (\ref{QQmeson}), the protected primaries of the interpolating theory that are {\it not} flavor-singlets can be decomposed
  in the limit $\check g \to 0$ as products of ``mesonic'' operators $(\bar Q^{+ \hat +} Q^{+ \hat +})^{\check a}_{\, \check b}$ and decoupled
  scalars of the ``second'' vector multiplet. The dual interpretation in the bulk is that  as $\check g \to 0$  KK modes on $S^3/\mathbb{Z}_2$
   become  multi-particle states of open strings. The flavor singlet sector of $\NN = 2$ SCQCD
   does not ``see'' the $S^3/\mathbb{Z}_2$ portion of the geometry. 
      We regard the ``loss'' of    $S^3/\mathbb{Z}_2$ as  a first hint that the string dual to the singlet sector of $\NN = 2$ SCQCD should be a {\it sub-critical} string background.
The $S^1$ factor on the other hand is preserved.
  
 We may also ignore the relation of $\NN = 2$ SCQCD with  the orbifold theory, and consider
   the protected states (\ref{list}) at face value: they are immediately suggestive of Kaluza-Klein reduction on a circle. 
  The dual geometry must contain an $AdS_5$ factor to implement the conformal symmetry,
  and an $S^1$ factor to generate the two  KK towers dual to $\{{\rm Tr}\, T\, \phi^\ell \}$ and  $\{{\rm Tr}\, \phi^{\ell+2} \}$.
  Moreover the radii of the $AdS_5$ and $S^1$ factor must be equal. 
  Indeed Kaluza-Klein reduction on $S^1$ gives a mass spectrum $m^2 \sim \ell^2/R_{S^1}^2$ (for $\ell$ large),
  and correspondingly a conformal dimension $\Delta \cong m R_{AdS} \cong \ell \frac{R_{AdS}}{R_{S^1}}$.
  Inspection of (\ref{list}) gives $R_{AdS} = R_{S^1}$. The isometry of $S^1$ 
  is interpreted as the $U(1)_r$ R-symmetry. On the other hand, there is no hint in the protected spectrum (\ref{list})
  of a ``geometrically'' realized $SU(2)_R$. The relation with the interpolating theory makes
  it clear that indeed the geometric factor $S^3/\mathbb{Z}_2$, with isometry $SU(2)_R \otimes SO(3)_L$,
   is lost in the limit $\check g \to 0$.

  We can  further split the ``naive'' spectrum (\ref{list}) into the primaries $\{ \Tr \, {\cal M}_{\bf 3}$,  $\Tr \phi^\ell \}$
  and the primaries $\{ \Tr T \phi^\ell \}$. The first set,
 of course, is isomorphic to the twisted states of the orbifold, and 
can be precisely matched with the KK reduction on $AdS_5 \times S^1$ of one tensor multiplet of $(2,0)$ 
chiral supergravity. 
A first guess is that the primaries   $\{ \Tr \,T \phi^\ell \}$ correspond to the KK reduction
of the $6d$ $(2,0)$ {\it gravity} multiplet on $AdS_5 \times S^1$, but this is incorrect. The zero
modes of the $6d$ gravity multiplet correctly match the stress-energy tensor multiplet (whose bottom component is the primary $\Tr\, T$),
but there are not enough states in the higher KK modes to match the states in the $\Tr\, T \phi^\ell$ for $\ell >0$.
This could have been anticipated by tracing the origin of the states  $\{ \Tr \, T \phi^\ell \}$ in the orbifold theory:
the dual supergravity states have no angular momentum  on $S^3/\mathbb{Z}_2$ in the parametrization (\ref{alphametric}),
but they are extended in the remaining {\it seven} dimensions. So a better guess is that the states   $\{ \Tr \, T \phi^\ell \}$ should
have an interpretation in seven-dimensional supergravity. 

In summary, with some hindsight, the ``naive'' spectrum appears to indicate a sub-critical string background, with seven
``geometric'' dimensions, and containing both  an $AdS_5$ and an $S^1$ factor, with  $R_{AdS} = R_{S^1}$.

The extra exotic protected states teach another important lesson.  They arise in the limit $\check g \to 0$ from long multiplets on the interpolating
theory that hit the unitarity bound and split into short multiplets. In the dual string theory, this means that a fraction of the massive
closed string states become massless in the limit   $\check g \to 0$. It is a substantial enough fraction to give rise to a Hagedorn degeneracy,
as we saw in section \ref{SCQCDindex}. This has the crucial implication that  {\it the dual description of $\NN = 2$ SCQCD  is never
in terms of supergravity}, 
since even in the limit $ \lambda \equiv g_{YM}^2 N_c \to \infty$  there is an infinite tower of ``light'' closed string states,
 with a mass of the order of the AdS scale. However  it seems plausible to conjecture that there is also a  second sector of ``heavy'' string states that decouple for $\lambda \to \infty$.

The picture that we have in mind is the following. There are really two 't Hooft couplings in the interpolating theory,
$ \lambda \equiv g_{YM}^2 N_c$ and $ \check \lambda \equiv \check g_{YM}^2 N_c$, and correspondingly {\it two}
effective string tensions $T_s \sim 1/l_s^2$ and $\check T_s \sim 1/\check l_s^{\;2}$.  The idea of two effective string tensions is intuitive from
the spin chain viewpoint, since the bifundamental fields separate different regions of the chain, occupied by adjoint fields
of the two different groups $SU(N_c)$ and $SU(N_{\check{c}})$ and thus
 governed by the two different gauge couplings. At the orbifold point, of course, $ \lambda =\check  \lambda$. In the limit in which the unique
't Hooft coupling of the orbifold theory is sent to infinity the string length goes to zero in AdS units according to the usual AdS/CFT dictionary $R_{AdS_5}/l_s\sim \lambda^{1/4}$,
leading to the decoupling of all massive string states. To approach $\NN =2$ SCQCD
we are interested in what happens as $\lambda$ is kept large, but $\check \lambda$ is sent to zero.
At present we do not know how to modify the AdS/CFT dictionary in this limit. 
The most naive extrapolation would suggest a hierarchy between two different scales:
there should be one sector of closed string states governed by $l_s \sim \lambda^{-1/4} R_{AdS}$ and thus very massive, and another governed by $\check l_s \sim R_{AdS}$
and thus light. The latter  would correspond to the exotic protected states revealed by the index.

\section{Brane Constructions  and Non-Critical Strings}

The interpolating SCFT has a dual description as IIB on $AdS_5 \times S^5 /\mathbb{Z}_2$,
but this description breaks down in the $\check g \to 0$ limit that we wish to study. We must describe the theory in  a different duality frame.
We will argue that the correct description is in terms of a {\it  non-critical} superstring background.
In this section we reconsider the IIB brane setup leading to the interpolating SCFT, and review 
how it can be T-dualized to a IIA Hanany-Witten setup (see {\it e.g.} \cite{Giveon:1998sr} for a review). 
The T-dual frame allows for a more transparent understanding
of the limit $\check g \to 0$, as a double-scaling  limit in which two brane NS5 collide while the string coupling
is sent to zero. In this limit the near-horizon dynamics  is described a non-critical string background, which
(before the backreaction of the D-branes) admits  an exact worldsheet description as $\mathbb{R}^{5,1}$
times   $SL(2)_2/U(1)$, the supersymmetric cigar CFT. We are led to identify the near-horizon backreacted background,
where D-branes are replaced by flux, with the dual of ${\cal N} = 2$ SCQCD.  

\subsection{Brane Constructions}
\label{subEmbedding}

 The interpolating
 SCFT arises at the low-energy limit on $N_c$ D3 branes sitting at the orbifold singularity   $\mathbb{R}^2 \times \mathbb{R}^4 /\mathbb{Z}_2$.
 The blow-up modes of the orbifold are set to zero, since they correspond to massive deformations of the $4d$ field theory. The NSNS period $\beta$ 
is related to $g_{YM}$ and $\check g_{YM}$ by the dictionary (\ref{Bdictionary}).
 As $\beta \to 0$ the $D$-strings obtained by wrapping $D3$ branes on the blow-down cycle of the orbifold 
become tensionless and string perturbation theory
breaks down.  It is useful to T-dualize to a IIA Hanany-Witten description, where the  deformation $\beta$ can be pictured more easily.
To perform the T-duality we should first replace the  $A_1$ singularity  $\mathbb{R}^4/\mathbb{Z}_2$ 
with its $S^1$ compactification, a two-center Taub-NUT space of radius $\tilde R$.
The local singularity is recovered for $\tilde R \to \infty$.  

Recall, more generally, that the $S^1$ compactification of the resolved $A_{k-1}$ singularity is
 a  $k$-center Taub-NUT, a hyperk\"aler manifold which 
can be concretely described as an   $S^1$ fibration of $\mathbb{R}^3$. Let  $\tilde \tau$  be the coordinate of the $S^1$ fiber and $\vec y$ the coordinates of the $\mathbb{R}^3$ base.
The $S^1$ fiber degenerates to zero size at $k$ points on the base,  $\vec y = \vec y^{(a)}$, $a=1,\dots k$, 
and goes to a finite radius $\tilde R$ at  the infinity of $\mathbb{R}^3$. 
(Topologically the $S^1$ is non-trivially fibered over the $S^2$ boundary of $\mathbb{R}^3$, with monopole charge $k$.)
Rotations of the $\vec y$ coordinates are interpreted as the $SU(2)$ symmetry that rotates the complex structures. 
From the viewpoint
of the worldvolume theory of  D3 branes probing the singularity, this is the $SU(2)_R$ R-symmetry. The geometry has also an extra $U(1)_L$ symmetry
acting as angular rotation in the $S^1$ fiber.\footnote{The $A_1$ singularity ($k=2$, $\vec y_a =0$, $\tilde R = \infty$) has
a symmetry enhancement $U(1)_L \to SO(3)_L$, whose  field theory manifestation is the  $SO(3)_L$ global symmetry of the
$\mathbb{Z}_2$ orbifold of ${\cal N} = 4$ SYM, discussed in section \ref{sec:orbifold}. The symmetry is broken to $U(1)_L$ for finite $\tilde R$; the full $SO(3)_L$ is recovered
in the infrared.} (Finally the $U(1)_r$ of the $4d$ gauge theory corresponds to an isometry outside the Taub-NUT, namely rotations  in the $\mathbb{R}^2$ factor of  $\mathbb{R}^2 \times \mathbb{R}^4/\mathbb{Z}_2$.)

The metric of a  $k$-center Taub-NUT space  has  $3(k-1)$ non-trivial  hyperk\"ahler moduli (after setting say $\vec y^{(1)} \equiv 0$ by an overall translation), which correspond to the blow-up modes of the $(k-1)$ cycles -- one  $SU(2)_R$ triplet for each cycle.
In the string sigma model one  needs to further specify the periods
of  $B_{NSNS}$ and $B_{RR}$ on each cycle,which gives two extra real  moduli for each cycle, singlets under $SU(2)_R$. 
Altogether the $5 = 3 + 1 + 1 $  moduli for each cycle are the scalar components of a
 tensor multiplet living in the six transverse directions to the Taub-NUT (or ALE) space. 
  T-duality along the $\tilde \tau$ direction yields a string background 
with non-zero NSNS $H$ flux and  non-trivial dilaton, which is interpreted as the background produced by $k$ NS5 branes \cite{Ooguri:1995wj, Gregory:1997te}.
The NS5 branes sit at  $\vec y^a$ in the $\mathbb{R}^3$ directions,
and are localized on the dual circle.\footnote{Naive application of the T-duality rules
gives NS5 branes smeared on the dual circle. The localized solution arises after taking
into account worldsheet instanton corrections \cite{Tong:2002rq}.}
The NSNS periods map to the relative angles of the NS5 branes on the dual circle.

 Let us apply these rules to our case. We start
  on the IIB side with  the configuration
 \begin{center}
\begin{tabular}{c|cccccccccc}
 {\bf IIB }& $x_{0}$ & $x_{1}$ & $x_{2}$ & $x_{3}$ & $x_{4}$ & $x_{5}$ &
$\tilde{\tau}$ & $y_{1}$ & $y_{2}$ & $y_{3}$\tabularnewline
\hline
${\rm TN}_{2}$ &  &  &  &  &  &  & $\times$ & $\times$ & $\times$ &
$\times$\tabularnewline
${\rm D3}$ & $\times$ & $\times$ & $\times$ & $\times$ &  &  &  &  &
& \tabularnewline
\end{tabular}
\par\end{center}
 The two-center Taub-NUT $TN_2$ has radius $\tilde R$, 
vanishing blow-up modes $(\vec y^{(1)} = \vec y^{(2)} = 0)$ and $\int_{S^2} B_{NSNS} = \beta$. 
T-duality gives the IIA configuration
\begin{center}
\begin{tabular}{c|cccccccccc}
{\bf IIA} & $x_{0}$ & $x_{1}$ & $x_{2}$ & $x_{3}$ & $x_{4}$ & $x_{5}$ & $\tau$
& $y_{1}$ & $y_{2}$ & $y_{3}$\tabularnewline
\hline
$2\,{\rm NS5}$ & $\times$ & $\times$ & $\times$ & $\times$ & $\times$
& $\times$ &  &  &  & \tabularnewline
${\rm D4}$ & $\times$ & $\times$ & $\times$ & $\times$ &  &  &
$\times$ &  &  & \tabularnewline
\end{tabular}
\par\end{center}
\begin{figure}[h]
\begin{center}
\includegraphics[height=4cm,angle=0]{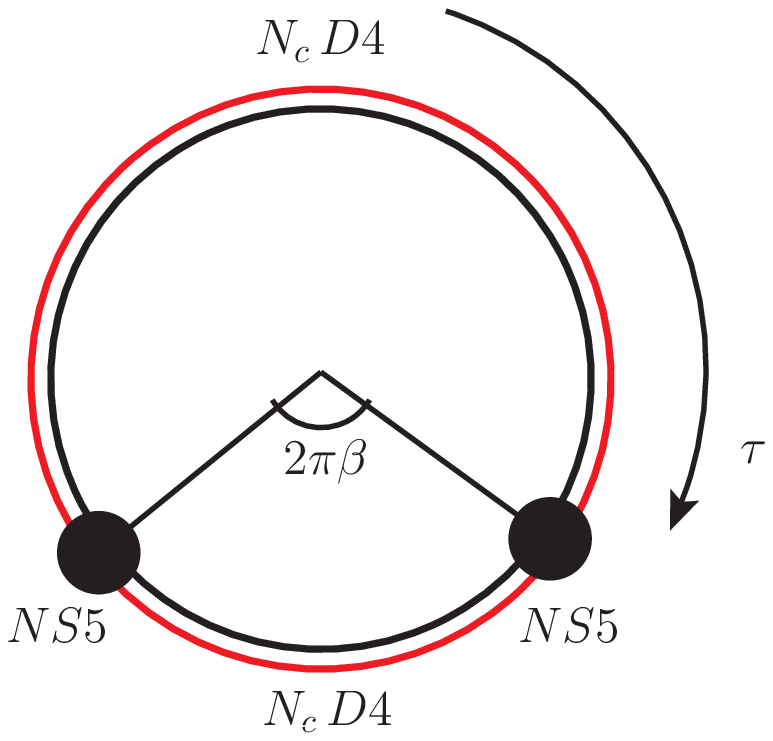}   \includegraphics[height=5cm,angle=0]{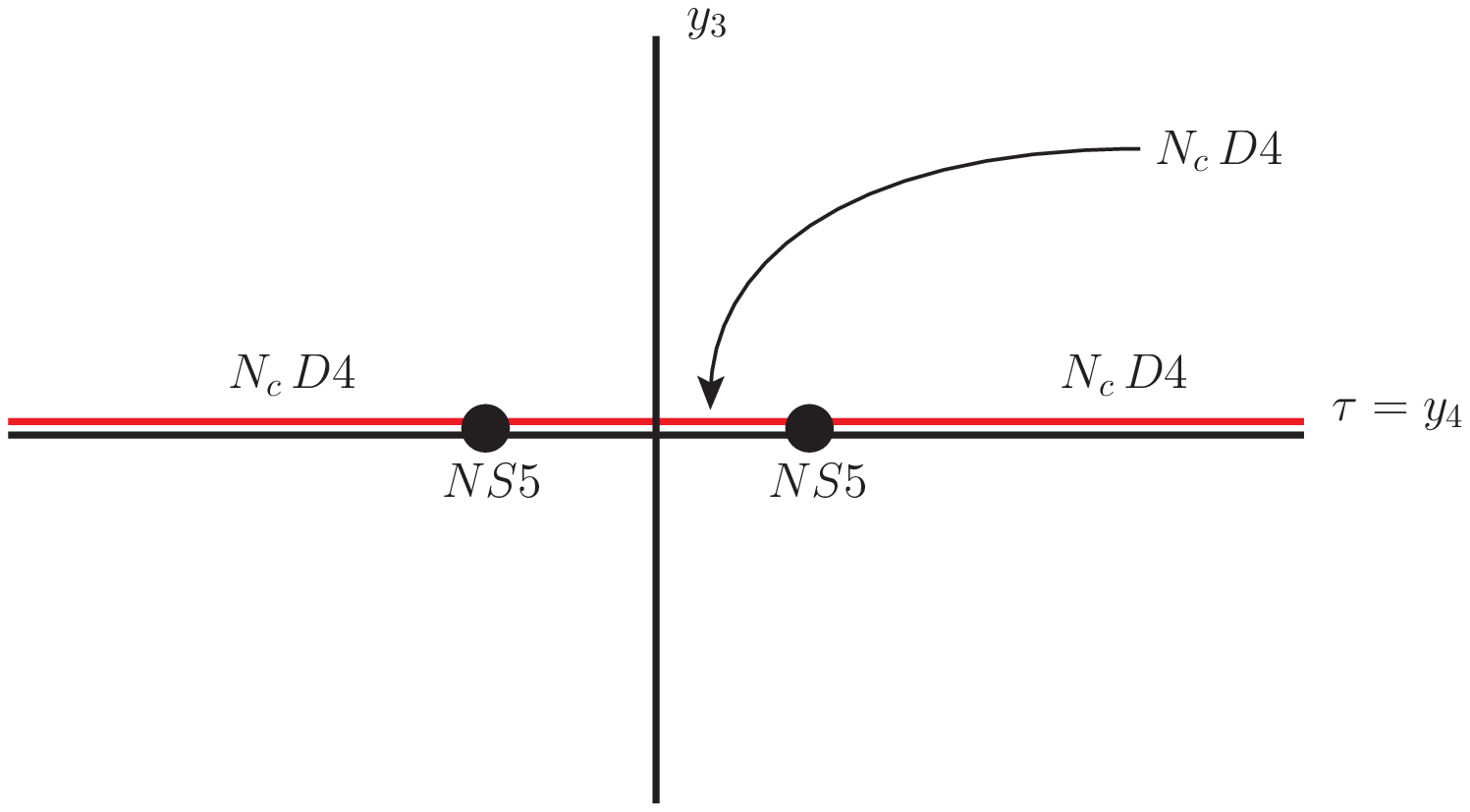}    
\\
\caption{\label{HWfig} Hanany-Witten setup for the interpolating SCFT (on the left) and for $\NN =2$ SCQCD (on the right).}
\end{center}
\end{figure} 
The two NS5 branes, at the origin of $\mathbb{R}^3$ are localized
on the dual circle of radius $R = \alpha'/\tilde R$ and at an angle $2 \pi \beta$ from each other.
The string couplings are related as
\be
g_s^{A} = \frac{R}{l_s} g_s^{B} =  \frac{l_s}{\tilde R} g_s^{B}  \,.
\ee
T-duality maps the $N_c$  $D3$ branes  on the IIB side (which can also be thought as {\it two stacks} of fractional branes \cite{Karch:1998yv}) 
  to two stacks of $N_c$ D4 branes on the IIA side, each stack ending on the two NS5 branes and
extended along either arc segment of the $\tau$ circle (see Figure \ref{HWfig}). 
This is the familiar Hanany-Witten setup
for the $\mathbb{Z}_2$ orbifold field theory. 
The four-dimensional field theory  living on the non-compact directions
0123 decouples  from the higher dimensional and stringy degrees of freedom in the limit
\begin{eqnarray}
&& g^{A}_s \to 0 \, \quad l_s \to 0 \, , \quad R \to 0 \, , \label{HWlimit}\\
&& {\rm with} \; \;   \frac{ \beta R}{2 \pi g_s^{A} l_s} \equiv \frac{1}{g_{YM}^2}  \; \;  {\rm and} \;  \; \frac{ (1-\beta) R}{2 \pi g_s^{A} l_s} \equiv \frac{1}{\check g_{YM}^2}  \; \; {\rm fixed} \,.\nonumber
\end{eqnarray}
At this stage we are still keeping both gauge couplings $g_{YM}$ and $\check g_{YM}$ finite.
If $L$ is the 4d length scale above which the field theory is a good description,
we have the hierarchy of scales 
\be
L \gg l_s \gg R \cong g_s^{A} l_s  \,.
\ee
Again, rotations in the $y_i$ directions  correspond to the $SU(2)_R$ R-symmetry of the ${\cal N} = 2$ $4d$ field theory,
while rotations in the $45$ plane correspond to the $4d$ $U(1)_r$ symmetry.
Finally the $U(1)_L$ symmetry, which was related to momentum conservation along the $S^1$ fiber in the IIB setup, is T-dualized to  {\it winding}  symmetry in the Hanany-Witten IIA setup.
It gets  enhanced in the infrared to the $SO(3)_L$ symmetry of the $4d$ field theory.

\subsection{From Hanany-Witten to a Non-Critical Background}
\label{FromTo}

The limit $\check g_{YM} \to 0$ (with $g_{YM}$ fixed) can now be understood more geometrically: 
it corresponds to $\beta \to 0$, the limit of coincident NS5 branes. In this limit we can ignore the periodicity of the $\tau$ direction
and think of two NS5 branes located in $\mathbb{R}^4$ at a distance $\tau_0 \equiv 2 \pi \beta R$ from each other, with $ \tau_0 \to 0$. 
There is a stack of $N_c$ D4 branes suspended between the two  NS5s and two stacks of $N_c$ semi-infinite D4s,  ending on either NS5 brane.
As is well-known, $k \geq 2$ {\it coincident} NS5 branes generate a string frame background with a strongly coupled near horizon region -- the string coupling blows up 
 down the infinite throat towards the location of the branes. The throat region
 is  the CHS background \cite{Callan:1991at}
 \be\label{CHS}
 \mathbb{R}^{5,1} \times SU(2)_k \times \mathbb{R}_\rho  \, , \quad {\rm with \; dilaton} \; \Phi = -\frac{\rho}{\sqrt{2k}} \, ,
 \ee
where $\rho$ is the radial direction (the NS5  branes are located at $\rho = -\infty$). 
The supersymmetric  $SU(2)_k$ WZW model describes the angular $S^3$;
 it arises by combining the bosonic $SU(2)_{k-2}$ and  three free fermions $\psi_i$, $i=1,2,3$, which make up an $SU(2)_2$.
This description breaks down for large negative $\rho$ where the string coupling $e^\Phi$ is large.
  In Type IIA (our case),  we must uplift  to M-theory to obtain the correct description of the near horizon region strictly coincident NS5 branes.
 However, what we are really interested in is bringing the branes together in a controlled fashion, simultaneously turning off the string  coupling $g_s^{A}$.
  We can break the limit (\ref{HWlimit}) into two steps:
 \begin{itemize}
 \item[(i)]
  We  first take the double scaling limit \cite{Giveon:1999px, Giveon:1999tq}
  \be \label{doublescaling}
  \tau_o \to 0 \, , \quad g_s^{A} \to 0 \, , \quad \frac{\tau_0}{l_s g_s^{A}} \equiv \frac{1}{g_{eff}} \sim  \frac{1}{g_{YM}^2} \; {\rm fixed}\, , \quad l_s   \; {\rm fixed}.
  \ee\item[(ii)] 
  We then send $l_s \to 0$. 
\end{itemize}
Let us first consider the purely closed background  without the D4 branes. 
The  double-scaling limit (i)  has been studied in detail  in  \cite{Giveon:1999px, Giveon:1999tq}, precisely with the motivation of avoiding  strong coupling.
In this limit the region near the location of the NS5  branes decouples from the rest of the geometry and is described
by a perfectly  regular background of {\it non-critical} superstring theory \cite{Giveon:1999px, Giveon:1999tq}.   To describe the background as a worldsheet CFT  it is useful to perform
a further T-duality, in an angular direction around the branes. If $\tau \equiv y_4$ is the direction
along which the branes are separated, we pick say the $y_3 y_4$ plane and perform a T-duality
around $\chi = \arctan y_3/y_4$. The result is the exact IIB background 
\be \label{noncritical}
\mathbb{R}^{5, 1} \times SL(2)_2/U(1)/\mathbb{Z}_2 \,.
\ee
The $\mathbb{Z}_2$ orbifold implements the GSO projection. 
The Kazama-Susuki coset $SL(2)_2/U(1)$ is  the  supersymmetric Euclidean 2d black hole, 
or supersymmetric cigar, at level $k=2$. The corresponding sigma-model background is 
\begin{eqnarray}
ds^{2} & = & d\rho^{2}+\tanh^{2}(\frac{Q\rho}{2})d\theta^{2}+dX^{\mu}dX_{\mu}\qquad\theta\sim\theta+\frac{4\pi}{Q}\\
\Phi & = & -\ln\cosh(\frac{Q\rho}{2}),\qquad B_{ab}=0 \,.
\end{eqnarray}
In appendix E we review several properties of this background.
An equivalent (mirror) description of $SL(2)/U(1)$ is as the ${\cal N} = 2$ superLiouville theory \cite{Hori:2001ax}. 
The two descriptions are manifestly equal in the asymptotic region $\rho \to \infty$,
where they reduce to $(S^1 \times$ linear dilaton).
At large $\rho$, the leading perturbation away from the
linear dilaton  takes a different form in the semiclassical cigar and Liouville descriptions, but in the complete quantum description
 both the cigar and Liouville perturbations are present.
The cigar description is more appropriate for $k \to \infty$, since in this limit  the cigar perturbation dominates
at large $\rho$ over the Liouville perturbation,
while the Liouville description is more appropriate for $k \to 0$, where the opposite is true. For $k=2$
both descriptions are precisely on the same footing --  the cigar and Liouville
perturbations are present with equal strength and are in fact rotated into each another
by the $SU(2)_R$ symmetry \cite{Murthy:2003es}. For $k=2$  the asymptotic radius of the cigar is $\sqrt{2 \alpha'}$, which
is the free fermion radius, implying that for large $\rho$ the angular coordinate $\theta$ and its superpartner $\psi_\theta$ can then be replaced
by three free fermions $\psi_i$, or equivalently by $SU(2)_2$.
The cigar background   is thus a smoothed out version of the CHS background (\ref{CHS}) -- the negative $\rho$ region of CHS
 has been cut-off and  the string coupling  is now  bounded from above
by its value $g_{eff}$ at the tip of the cigar.\footnote{
As an aside, it is worth recalling  the generalization of this discussion to $k$ NS5 branes, equally spaced on a contractible circle 
in the $y_3 y_4$ plane. T-duality around the angular coordinate $\chi$ produces the background \cite{Giveon:1999px}
\be\label{kCFT}
\mathbb{R}^{5, 1} \times (SL(2)_k/U(1) \times SU(2)_k/U(1) )/\mathbb{Z}_k \, .
\ee
The central charges are of the Kazama-Susuki cosets are
\be
c(SL(2)_k/U(1)) = 3 + \frac{6}{k} \, , \quad c(SU(2)_k/U(1)) = 3- \frac{6}{k} \,  .
\ee
The CFT (\ref{kCFT})
In the semiclassical  limit $k  \to \infty$ we have a weakly curved ``geometric'' $10d$ background, while in the opposite limit $k=2$ the curvature
is string scale, the $SU(2)/U(1)$ piece disappears and we have the ``non-critical'' string background (\ref{noncritical}).
}
 
To summarize, we started from a IIA configuration of  two separated NS5 branes in  flat space,
and took the double-scaling limit (\ref{doublescaling}). In this limit the near-horizon region  decouples from the asymptotic
flat space region, and is described by the exact non-critical IIB background (\ref{noncritical}). (The switch from IIA and IIB is due
to the angular T-duality along $\chi$.) The reduction of degrees of freedom from critical to non-critical strings happens because
we are focusing on a subsector of the full theory, namely the degrees of freedom near the singularity
produced by the colliding NS5 branes. 
The transverse direction $\rho$ can be  thought of as a worldsheet RG scale, with the asymptotically flat region at large $\rho$ playing the role
of the UV and the cigar geometry playing the role of the IR --  in focusing to the near horizon region we lose the asymptotic flat space degrees of freedom.
In particular, what remains of the transverse $S^3$  
 is just the ``stringy''  $SU(2)_2$ associated  with the free fermions $\psi_i$, $i=1,2,3$.

We can easily follow the fate of the D-branes through the double scaling limit and  T$_\chi$-duality:
the D4 branes suspended between the two NS5s become D3 branes localized at the tip of the cigar,
while the semi-infinite D4 branes  become D5 branes extended on the cigar.  This at least is the intuitive geometric picture.
Since the cigar background has string-size curvature near the tip, a more appropriate description of the D-branes
is in terms of the exact boundary states. Boundary states for  the Kazama-Susuki coset $SL(2)/U(1)$ (equivalently, for the superLiouville CFT)
have been studied in several papers \cite{Israel:2004jt, Fotopoulos:2004ut, Hosomichi:2004ph, Eguchi:2004kx,  Israel:2005fn}, following the construction of boundary states
in bosonic Liouville theory, and used in ${\cal N}=1$ non-critical holography in \cite{Fotopoulos:2005cn, Ashok:2005py, Murthy:2006xt}.
 There are indeed natural candidates for the two types of cigar D-branes
that we need. The branes localized near the tip of the cigar are the analog of Liouville ZZ  \cite{Zamolodchikov:2001ah} branes, while the branes extended
along the cigar are the analog of the Liouville FZZT \cite{Fateev:2000ik, Teschner:2000md} branes.  
The non-critical string setup can be summarized by the following diagram:
\begin{center}
\begin{tabular}{c|cccccccccc}
{\bf IIB} & $x_{0}$ & $x_{1}$ & $x_{2}$ & $x_{3}$ & $x_{4}$ & $x_{5}$ & $\rho$
& $\theta$ 
\tabularnewline
\hline
${\rm D3}$ & $\times$ & $\times$ & $\times$ & $\times$ &  &  &
 &  
\tabularnewline
${\rm D5}$ & $\times$ & $\times$ & $\times$ & $\times$ & 
&  & $\times$ & $\times$  
 \tabularnewline
\end{tabular}
\par\end{center}
We could have taken this as our starting point. The theory on the worldvolume of the $N_c$  D3 branes (the ``color'' branes)
reduces for energies much smaller than the string scale to ${\cal N} = 2$ $SU(N_c)$ SYM, coupled to $N_f = 2 N_c$ hypermultiplets arising
from the open strings stretched between the D3s and the ``flavor'' D5s.  This is true by construction, since
we obtained this non-critical setup as a limit of a well-known brane realization  of the same field theory, and it could
also be checked directly, by examining the open string spectrum and preserved supersymmetries. 

To decouple the field theory we need to take
 $l_s \to 0$  (step (ii) in our previous discussion of the field theory limit).   
This amounts on the gravity side to 
the near-horizon limit of the geometry produced by the D-branes. By the usual arguments \cite{Maldacena:1997re},
we are led to conjecture that the resulting non-critical string background is  dual  to ${\cal N} = 2$ SCQCD.

\section{Towards the String Dual of ${\cal N} = 2$ SCQCD}

The explicit construction of the background after the backreaction of the D-branes is left for future work.
 In this  section we outline a  line of attack, based on a $7d$ ``effective action'' which we identify as  maximal supergravity with $SO(4)$ gauging.
In fact several features of the background  can be determined from symmetry considerations alone, and just assuming that a solution
 exists we will find a nice qualitative agreement with the bottom-up field theory analysis, notably in the protected spectrum of operators.

\subsection{Symmetries}

Let us start  by recapitulating the symmetries. The obvious bosonic symmetries of the closed string background (\ref{noncritical}) (the background before introducing  D-branes,
henceforth the ``cigar background'')
are the Poincar\'e group in $\mathbb{R}^{5,1}$ and the $U(1)$ isometry of the $\theta$ circle. In fact since as $\rho \to \infty$ the $\theta$ circle is
at the free fermion radius, there is an asymptotic  ``stringy'' enhancement of the $U(1)$ symmetry to $SU(2)_{\psi_i} \times SU(2)_{\tilde \psi_i} \cong SO(4)$.
At finite $\rho$ the cigar and super-Liouville interactions break this symmetry to the {\it diagonal} $SU(2)$. This has a clear geometric
interpretation in the HW picture   (before the angular T$_\chi$-duality)
of the two colliding NS5 branes: the $SO(4)$  symmetry is the isometry of the transverse four directions
to  two {\it coincident} NS5 branes; separating the branes along one direction ($\tau  = y_4$ in the picture on the right of Figure 4)  breaks the symmetry to $SO(3) \cong SU(2)$
(rotations of $y_i$, $i=1,2,3$). This surviving diagonal $SU(2)$ is interpreted as the $SU(2)_R$ R-symmetry of the ${\cal N} = 2$ $4d$ gauge theory.
Adding the color D3 branes and the flavor D5 branes 
breaks the $6d$ Poincar\'e symmetry to $4d$ Poincar\'e symmetry in the directions $x_m$, $m=0,1,2,3$,  times the rotational symmetry in the $45$ plane.
The latter is interpreted as the $U(1)_r$ R-symmetry of the gauge theory. Note that the branes preserve the same (diagonal) $SU(2)$
as the cigar and super-Liouliville interactions. This is again transparent in the picture 
of colliding NS5 branes, since both the ``compact'' D4 branes and the ``non-compact'' D4 branes, which become respectively
the color D3s and the flavor D5s after T$_\chi$-duality, are oriented along the same $\tau  = y_4$ direction
in which the two NS5s are  separated. Finally we should mention the fermionic  symmetries. As we review in appendix E, the background
(\ref{noncritical}) has 16 real supercharges, corresponding to the $(2,0)$ Poincar\'e superalgebra in
 $\mathbb{R}^{5,1}$. Adding the D-branes breaks the supersymmetry in half, so that 8 Poincar\'e supercharges survive (that D3s and D5s break the {\it same} half is again
 obvious in the T-dual frame where they are both (parallel) D4 branes). Taking the near-horizon geometry is expected to give the usual supersymmetry enhancement,
 restoring a total of 16 supercharges that form the ${\cal N} = 4$ $AdS_5$ superalgebra (isomorphic to the ${\cal N} =2$ $4d$ superconformal algebra).

\subsection{The cigar background and $7d$ maximal $SO(4)$-gauged supergravity}

The cigar background (\ref{noncritical}) 
 is analyzed in some detail in appendix E, which the reader is invited to read at this point. Let us summarize some of the relevant points.
 The physical  spectrum of the cigar background consists of: (i) normalizable states localized at the tip of the cigar $\rho \sim 0$, living in $\mathbb{R}^{5,1}$: they fill a tensor multiplet
of $(2,0)$ $6d$ supersymmetry; (ii) delta-function normalizable states, corresponding to plane waves in the radial $\rho$ direction; (iii) non-normalizable
vertex operators, supported in the large $\rho$ region. 

We are only interested in the cigar background as
 an intermediate step towards the background dual to ${\cal N} = 2$ SCQCD, obtained in the near-horizon
 limit of the D3/D5 brane configuration.   A possible strategy is to use the cigar background,
 which admits an exact CFT description, to derive a spacetime ``effective action''. The spacetime action is expected to be background independent
 and should admit as classical solutions {\it both} the cigar background {\it and} the background dual to ${\cal N} = 2$ SCQCD.
 (In this respect, the cigar background is analogous to the $10d$ flat background of IIB string theory, which is described
 at low energies by $10d$  IIB supergravity; {\it another} solution of IIB supergravity is the $AdS_5 \times S^5$ background dual to ${\cal N} =4$ SYM.)
For the purpose of deriving an ``effective action'' the relevant part of the spectrum is (ii),  the continuum of plane-wave states. Performing
a KK reduction on the $\theta$ circle, the plane-wave states are naturally organized in a tower of increasing $7d$ mass (which gets contribution
both from the $\theta$ momentum and from string oscillators). There is is no real separation
of scales between the lowest mass level and the higher ones, because the linear dilaton has string-size gradient.
Nevertheless the states belonging to lowest level are special: although they obey
``massive'' $7d$ wave-equations, this is  an artifact of the linear dilaton; the counting of degrees of freedom
is that of massless $7d$ states because of  gauge invariances.

Remarkably, we find  that  for large $\rho$ the lowest-mass level of the continuum spectrum 
is described by seven dimensional {\it maximally supersymmetric}  supergravity (32 supercharges), 
but with a non-standard gauging: only an $SO(4)$ of the full $SO(5)$ R-symmetry is gauged. This supergravity has been constructed only quite recently \cite{Samtleben:2005bp, Weidner:2006rp}. 
The maximal supersymmetry (which, as we shall see momentarily, is spontaneously broken to half-maximal, consistently with our previous counting)
 can be understood as follows. After fermionizing the angular coordinate $\theta$, we have
a total of ten left-moving fermions, $\psi_\mu$, $\mu = 0 \dots 5$ along $\mathbb{R}^{5,1}$, $\psi_\rho$ and $\psi_i$, $i=1,2,3$ (the last three corresponding to $\partial \theta$, $\psi_\theta$),
and similarly ten right-moving fermions.  So the construction of the lowest-level physical states  of our sub-critical theory is entirely isomorphic to the construction
of the massless states of the standard critical  IIB string theory, except of course that the momenta are now seven dimensional. The $SO(4)$ 
that is being gauged is the asymptotic $SU(2)_{\psi_i} \times SU(2)_{\tilde \psi_i} \cong SO(4)$ that we have mentioned. It turns out that 
unlike the standard $SO(5)$-gauged $7d$ sugra, which admits the maximally supersymmetric $AdS_7$ vacuum,
 the  $SO(4)$-gauged theory breaks half of the supersymmetry spontaneously. The scalar potential of the $SO(4)$-gauged theory
 does not admit a stationary solution but only a domain wall solution \cite{Samtleben:2005bp, Weidner:2006rp}, which is nothing but the linear dilaton background, with 16 unbroken supercharges --
the  $6d$ $(2,0)$ super-Poincar\'e invariance discussed earlier.

 Incidentally, we believe   that this is a  general phenomenon:  
  non-critical superstrings in various dimensions must  admit (non-standard) gauged supergravities   as their spacetime ``effective actions'', in the sense that we  have discussed. 
It may be worth to explore this connection  systematically.

\subsection{An Ansatz}

We expect  the $SO(4)$-gauged $7d$ sugra that describes the ``massless'' fields to be a useful tool, though not a perfect one because
we know that the higher levels are not truly decoupled. The next step is to look for a solution of this supergravity with all the expected symmetries.  In the seven dimensional theory the $SU(2)_R$ symmetry is not realized
geometrically -- its last remnant was the (string-size) $\theta$ circle, over which we have KK reduced to get down to $7d$.
On the other hand, the $U(1)_r$ symmetry {\it is} geometric, and conformal
symmetry is expected to arise in the near-horizon geometry, which must then contain both an $S^1$ and an $AdS_5$ factor. The most general
ansatz for the $7d$ metric with the expected isometries is 
\be \label{7ansatz}
ds^2 =  f(y) ds^2_{AdS_5} + g(y) d\varphi^2 +  C(y) dy^2 \, .
\ee
Here $\varphi$ is the angular coordinate of the $S^1$ associated to  $U(1)_r$ isometry, while the $y$ has range in a finite interval, say $y\in [0,1]$.
Restoring the $\theta$ coordinate, the non-critical background would have the form
\be
ds^2 =  f(y) ds^2_{AdS_5} + g(y) d\varphi^2 + h(y) d\theta^2 + C(y) dy^2 \, .
\ee
Comparing  with the brane setup, which is again
\begin{center}
\begin{tabular}{c|cccccccccc}
{\bf IIB} & $x_{0}$ & $x_{1}$ & $x_{2}$ & $x_{3}$ & $x_{4}$ & $x_{5}$ & $\rho$
& $\theta$ 
\tabularnewline
\hline
${\rm D3}$ & $\times$ & $\times$ & $\times$ & $\times$ &  &  &
 &  
\tabularnewline
${\rm D5}$ & $\times$ & $\times$ & $\times$ & $\times$ & 
&  & $\times$ & $\times$  
 \tabularnewline
\end{tabular}
\par\end{center}
we identify $\varphi$  is angular coordinate in the $45$ plane, while
 $y$ could be taken to be a relative angle between the radial distance  in the 45 plane 
 and the radial distance $\rho$ along the cigar, $y = \frac{2}{\pi} \arctan(\rho/ \sqrt{x_4^2 + x_5^2})$.
 The D5 branes sit at $y=1$. 
 
 The program is then to look for a solution (\ref{7ansatz}) of the $SO(4)$-gauged $7d$ supergravity, possibly allowing for singular
 behavior at the original location $y=1$ of the flavor branes.  For fixed $N_c$ and $N_f (= 2 N_c)$, we expect a one-parameter family of solutions, 
 because  the 't Hooft coupling $\lambda$ is exactly marginal -- the AdS scale should be a modulus, as in the familiar $AdS_5 \times S^5$ case.
  The color (D3) branes are magnetically charged under
 the RR one-form $C_{\hat \mu}^{(2,2)}$ (see Table 18) and the flavor branes (which are actually D4 branes from the viewpoint in the $7d$ theory)
 are magnetically charged under the RR zero-form $C^{(2,2)}$. The corresponding fluxes will be turned on in the solution.
 As usual the color branes will be completely replaced by flux. Our analysis of the large $N$ Veneziano limit suggests
 that new effective closed string degrees of freedom,  dual to ``generalized single-trace'' operators,  arise from the resummation of open string perturbation theory.
 This favors the scenario in which also the flavor branes are completely replaced by flux. 
 This  fundamental issue would be illuminated by
 an explicit solution. 
 
The program of  finding a supergravity background for ${\cal N} =2$ SCQCD was also discussed
in  critical IIB supergravity \cite{Grana:2001xn} and in $11d$ supergravity  \cite{Gaiotto:2009gz}, but no explicit solutions are yet known. It would be interesting
to understand the relation of these approaches with our sub-critical setup. In particular a somewhat singular limit of solutions found in \cite{Gaiotto:2009gz} should correspond to ${\cal N} =2$ SCQCD,  
and it would be nice to understand this in detail.

 \subsection{Spectrum}

Already at this stage we can recognize that the top-down  (string theory)
 and bottom-up (field theory) analyses are in qualitative agreement. Both suggest that the string dual of ${\cal N} = 2$ SCQCD is  a sub-critical background with an $AdS_5$ and an $S^1$ factor.  
 In   the field theory protected spectrum we found a sharp difference between the  $U(1)_r$ and $SU(2)_R$ factors
 of the R-symmetry group: there are towers of states with  increasing $U(1)_r$,  but no analogous
 towers for $SU(2)_R$. The brane construction confirms 
the natural interpretation of this fact: while the $U(1)_r$ is realized geometrically as the isometry of a
  ``large''  $S_{\varphi}^1$, with its towers of KK modes,
  the $SU(2)_R$ is associated to the string-sized $S_{\theta}^1$ of the cigar (and in fact the very enhancement from the $\theta$ isometry $U(1) \subset SU(2)_R$ to the full $SU(2)_R$ is a stringy phenomenon).   The ``naive''
 part of the protected spectrum nicely matches:
 \begin{itemize}
 \item[(i)]
  The multiplets built on the primaries  $\{\Tr \, {\cal M}_{\bf 3}\, , \Tr \,  \phi^{2+\ell} \}$
 correspond to the KK modes on $S_{\varphi}^1$ of the $6d$ tensor multiplet  (see appendix D): these
 are the truly normalizable states of the cigar background, localized at the tip of the cigar ($y=0$ in the parametrization (\ref{7ansatz})). 
   \item[(ii)]
   The multiplets built on $\{ \Tr \, T \phi^\ell \}$ correspond to
the  KK modes on $S_{\varphi}^1$ of the bulk $7d$ $SO(4)$-gauged supergravity: this is the lowest level of  the plane-wave spectrum of the cigar background.
While we have not performed a detailed KK reduction, for which the precise geometry is required, it is clear
 that the bulk graviton maps to the stress tensor, which is part of the $\Tr \, T$ multiplet, and that the $\ell$-th KK mode
 of the graviton maps to the unique spin 2 state in the $\Tr \, T \phi^\ell$ multiplet. Supersymmetry should do the rest.
\end{itemize}
The ``extra'' protected states of the field theory must correspond
to light string states in the bulk, with mass of order of the AdS scale, but we do not know how to establish 
a more precise dictionary at this point.
We have suggested in section 6 that the string theory dual to ${\cal N} = 2$ SCQCD may contain two
sectors of string states, in correspondence with the two effective string scales $l_s$ and $\check l_s$ of the interpolating
theory: a light sector, controlled by $\check l_s \sim R_{AdS}$ for all $\lambda$,
and a heavy sector,  controlled by $l_s \ll R_{AdS}$ for $\lambda \gg 1$.
The string length of the cigar background should  be identified with $l_s$, so  the massive string states
of the cigar background would correspond to the heavy sector and decouple for large $\lambda$. 
The light sector is more mysterious. A tantalizing speculation is that the light states correspond to cohomology
classes with  non-normalizable   ${\cal N} = 2$ Liouville dressing, {\it i.e.}
supported at large $\rho$ (operators of type (iii) in the list of section E.4). It is clearly possible to tune
the $\rho$-momentum to achieve ``massless'' six-dimensional states, at the expense
of making them non-normalizable in the $\rho$ direction. Perhaps the extra protected states of ${\cal N} = 2$ SCQCD
are somewhat analogous to the discrete states of the $c=1$ matrix model, which are indeed dual to vertex
operators with non-normalizable Liouville dressing.\footnote{
Alternatively, our idea of two effective string scales may be wrong, and the unique scale $l_s$ may be of the order of $R_{AdS}$ for all $\lambda$.
In this case all anomalous dimensions would remain small for  large $\lambda$. The extra protected states would be special only
in that their anomalous dimension is exactly zero for all $\lambda$. This is certainly a  logical possibility.}

If indeed $l_s \ll R_{AdS}$ for large $\lambda$, the $7d$ supergravity, while not capturing
the whole theory even in this limit (as we know from the existence of the extra protected states),
may still offer a useful description of  a subsector. 

\section{Discussion}

We may now look back to section 1,  at the list of special features shared by all
 $4d$ CFTs for which an explicit string dual is presently known. We have studied in some detail
  perhaps the most symmetric theory that violates property (i) (since $a \neq c$ at large $N$)
and property (ii) (since it has a large number of fields in the fundamental representation), while still satisfying the nice simplifying feature (iv) of
 an exactly marginal coupling $\lambda$. We have argued that the 
dual string theory is not ten dimensional, thus violating (iii), and proposed  a sub-critical string dual in eight dimensions (including
the string-size $\theta$).
The  theory emerges as a  limit of a family of superconformal field theories that have $a=c$ and
admit ten dimensional string duals. In this singular limit some fields decouple on the field theory side, leading to $a \neq c$,
while on the string side two dimensions  are lost (counting $\theta$ as a dimension).
 It is tempting to  link the two phenomena.
The natural speculation is that the $4d$ gauge theories in the ``${\cal N} =4$ universality class''  (which among
other things  are characterized by $a=c$) have $10d$ string dual, while theories with ``genuinely'' fewer supersymmetries have sub-critical
duals. A plausible pattern for (susy |dimension) is (${\cal N}$  |$d$) =  (4|10), (2|8), (1|6), (0|5).
We have given evidence for the ${\cal N} =2 \leftrightarrow d=8$ connection,
while  \cite{Klebanov:2004ya, Fotopoulos:2005cn, Murthy:2006xt} focused on  ${\cal N}=1 \leftrightarrow d=6$.

Our example is in harmony with
the  no-go theorem 
that $a=c$ for all field theories with an $AdS_5$ gravity dual,
since we argued that even for large $\lambda$
 the supergravity approximation to the dual  of ${\cal N} =2$ SCQCD
 cannot be entirely valid. 
 The imbalance between $a$ and $c$ must arise from higher-curvature terms
in the $AdS_5$ gravity theory  \cite{Nojiri:1999mh}.  We believe that the stringy origin of these higher curvature terms is  the Wess-Zumino action of the flavor branes, as in the example studied
in \cite{Aharony:1999rz, Blau:1999vz}: the flavor Wess-Zumino terms  were shown to generate ${\cal R}^2$ corrections  to  the $5d$ Einstein-Hillbert action,
contributing  at order $O(N_f/N_c)$ to $a-c$. In the example of \cite{Aharony:1999rz, Blau:1999vz} $N_f \ll N_c$, while
in our case $N_f \sim N_c$ and $a-c = O(1)$, but the mechanism must be the same.
It is important to keep in mind that the  higher-curvature terms from the WZ action are topological in nature and are  on a  different footing from the higher-curvature
corrections due to  the closed string sigma-model loops, which are instead suppressed by powers of $l_s/R_{AdS}$. So
there is no contradiction in principle between  our suggestion that  for large $\lambda$ the non-critical background has a string length $l_s \ll R_{AdS}$, and
the  fact that $a-c = O(1)$, since $a-c$ arises from the higher-curvature terms coming from the WZ action, 
since they  are  not suppressed.

It is worth pointing out a simple relation between our ${\cal N}=2$  story and  the ${\cal N} =1$ story
of \cite{Klebanov:2004ya, Fotopoulos:2005cn, Murthy:2006xt},  if we specialize their setup to ${\cal N}=1$  super QCD with $N_f = 2 N_c$,
 the Seiberg self-dual theory. This theory can be viewed as the $\check g \to 0$ limit of a family of  ${\cal N} =1$
 SCFTs with product gauge-group $SU(N_c) \times SU(N_{\check c})$; when the couplings
are equal the family reduces to the Klebanov-Witten theory \cite{Klebanov:1998hh}, which is dual to  $AdS_5 \times T^{1,1}$. This is entirely analogous to the
relation between  ${\cal N} =2$ SCQCD  and the $\mathbb{Z}_2$ orbifold of ${\cal N}=4$ SYM,
and of course this is not a coincidence: the two-parameter family of ${\cal N} =1$ theories is obtained  from the two-parameter family of ${\cal N} =2$ theories
flowing in the IR by a {\it relevant} deformation. For $g = \check g$,  this is the well-known RG flow
from the $\mathbb{Z}_2$ orbifold to the KW theory triggered by $\Tr (\phi^2 - \check \phi^2)$ \cite{Klebanov:1998hh}.
Unlike the ${\cal N} =2$ family, for ${\cal N} =1$ the couplings are bounded from below and the family of ${\cal N} =1$ SCFTs
is never weakly coupled. The exactly marginal coupling of the self-dual ${\cal N}=1$ super QCD is the coefficient
 of a quartic superpotential -- it cannot be taken arbitrarily small but it can be taken arbitrarily large.
Our analysis of appendix E should easily generalize
to this case, to find  the gauged supergravity  describing  the lightest modes
of the continuum spectrum. 
Only an isolated supergravity solution exists \cite{Klebanov:2004ya} (for arbitrary
$N_f \sim N_c$), but in the special case $N_f =2 N_c$ a one-parameter family of solutions is expected.
This is also confirmed by the vanishing of the dilaton tadpole when $N_f = 2 N_c$ \cite{Murthy:2006xt}. It would be nice to understand this point better.

Clearly there are  many open questions. The bottom-up analysis would be 
greatly  enhanced if we could determine the large $\lambda$ behavior
 of generic non-protected operators. This may  eventually be possible if ${\cal N} =2 $ SCQCD exhibits an all-loop integrable
 structure. In our companion spin-chain paper \cite{spinchain} we find a preliminary hint of one-loop integrability.
In the top-down approach, work is in progress to verify
whether the ansatz (\ref{7ansatz}) 
is indeed a solution of the $SO(4)$-gauged supergravity. It will be interesting to  understand its physical
implications, especially  the role of the warping factors and their possible singularity at $y=1$.

  Ultimately an accurate description of the string dual will require  the full non-critical
 sigma-model in RR background. It would be very interesting to
  start with the sigma-model for $AdS_5 \times S^5 /\mathbb{Z}_2$, which can be quantized either in
 the generalized light-cone gauge or in the pure-spinor formalism, and understand the
 transition to  a non-critical sigma-model in the $\check g \to 0$ limit.   This may
 well be the simplest instance of such a transition -- we should learn  the rules of the game in this highly symmetric example.

\section*{Acknowledgements}

It is pleasure to thank Massimo Bianchi, Davide Gaiotto, Ami Hanany, Juan Maldacena, Luca Mazzucato, Andrei Parnachev, Joe Polchinski, Shlomo Razamat, Martin Rocek, 
Matthias Staudacher, Peter van Nieuwenhuizen  and Edward Witten for  stimulating discussions. LR  is grateful to  the University of Rome Tor Vergata,
to the KITP,  Santa Barbara (``Fundamental Aspects of Superstring Theory'', January '09), to the Galileo Galilei Institute, Florence (``New Perspectives in String Theory'', June `09), 
 and to the  IST, Lisbon (``IST Fest'', June `09),  for the  opportunity to present preliminary versions of this work, and for their warm hospitality.
 This work supported in part by the DOE grant DEFG-0292-ER40697 and by the NSF grant PHY-0653351-001. 
Any opinions, findings, and conclusions or recommendations expressed in this material are those of the authors
and do not necessarily reflect the views of the National Science Foundation.

\appendix


\section{\label{rep-theory}Shortening Conditions of the {\cal N} =2 Superconformal Algebra}

\begin{table}
\begin{centering}
\begin{tabular}{|c|l|l|l|l|}
\hline 
\multicolumn{4}{|c|}{Shortening Conditions} & Multiplet\tabularnewline
\hline
\hline 
$\BB_{1}$  & $\QQ_{\alpha}^{1}|R,r\rangle^{h.w.}=0$  & $j=0$ & $\Delta=2R+r$  & $\BB_{R,r(0,\bar{j})}$\tabularnewline
\hline 
$\bar{\BB}_{2}$  & $\bar{\QQ}_{2 \dot{\alpha}}|R,r\rangle^{h.w.}=0$  & $\bar j=0$ & $\Delta=2R-r$  & $\bar{\BB}_{R,r(j,0)}$\tabularnewline

\hline 
$\EE$  & $\BB_{1}\cap\BB_{2}$  & $R=0$  & $\Delta=r$  & $\EE_{r(0,\bar{j})}$\tabularnewline
\hline 
$\bar \EE$  & $\bar \BB_{1}\cap \bar \BB_{2}$  & $R=0$  & $\Delta=-r$  & $\bar \EE_{r(j,0)}$\tabularnewline
\hline 
$\hat{\BB}$  & $\BB_{1}\cap\bar{B}_{2}$  & $r=0$, $j,\bar{j}=0$  & $\Delta=2R$  & $\hat{\BB}_{R}$\tabularnewline
\hline
\hline 
$\CC_{1}$  & $\e^{\alpha\beta}\QQ_{\beta}^{1}|R,r\rangle_{\alpha}^{h.w.}=0$  &  & $\Delta=2+2j+2R+r$  & $\CC_{R,r(j,\bar{j})}$\tabularnewline
 & $(\QQ^{1})^{2}|R,r\rangle^{h.w.}=0$ for $j=0$  &  & $\Delta=2+2R+r$  & $\CC_{R,r(0,\bar{j})}$\tabularnewline
\hline 
$\bar \CC_{2}$  & $\e^{\dot\alpha\dot\beta}\bar\QQ_{2\dot\beta}|R,r\rangle_{\dot\alpha}^{h.w.}=0$  &  & $\Delta=2+2\bar j+2R-r$  & $\bar\CC_{R,r(j,\bar{j})}$\tabularnewline
 & $(\bar\QQ_{2})^{2}|R,r\rangle^{h.w.}=0$ for $\bar j=0$  &  & $\Delta=2+2R-r$  & $\bar\CC_{R,r(j,0)}$\tabularnewline
\hline 
$\mathcal{F}$  & $\CC_{1}\cap\CC_{2}$  & $R=0$  & $\Delta=2+2j+r$  & $\CC_{0,r(j,\bar{j})}$\tabularnewline
\hline 
$\bar{\mathcal{F}}$  & $\bar\CC_{1}\cap\bar\CC_{2}$  & $R=0$  & $\Delta=2+2\bar j-r$  & $\bar\CC_{0,r(j,\bar{j})}$\tabularnewline
\hline 
$\hat{\CC}$  & $\CC_{1}\cap\bar{\CC}_{2}$  & $r=\bar{j}-j$  & $\Delta=2+2R+j+\bar{j}$  & $\hat{\CC}_{R(j,\bar{j})}$\tabularnewline
\hline 
$\hat{\mathcal{F}}$  & $\CC_{1}\cap\CC_{2}\cap\bar{\CC}_{1}\cap\bar{\CC}_{2}$  & $R=0, r=\bar{j}-j$ & $\Delta=2+j+\bar{j}$  & $\hat{\CC}_{0(j,\bar{j})}$\tabularnewline
\hline
\hline 
$\DD$  & $\BB_{1}\cap\bar{\CC_{2}}$  & $r=\bar{j}+1$  & $\Delta=1+2R+\bar{j}$  & $\DD_{R(0,\bar{j})}$\tabularnewline
\hline 
$\bar\DD$  & $\bar\BB_{2}\cap{\CC_{1}}$  & $-r=j+1$  & $\Delta=1+2R+j$  & $\bar\DD_{R(j,0)}$\tabularnewline
\hline 
$\mathcal{G}$  & $\EE\cap\bar{\CC_{2}}$  & $r=\bar{j}+1,R=0$  & $\Delta=r=1+\bar{j}$  & $\DD_{0(0,\bar{j})}$\tabularnewline
\hline
$\bar{\mathcal{G}}$  & $\bar\EE\cap{\CC_{1}}$  & $-r=j+1,R=0$  & $\Delta=-r=1+j$  & $\bar\DD_{0(j,0)}$\tabularnewline
\hline
\end{tabular}
\par\end{centering}
\caption{\label{shortening}Shortening conditions
and short multiplets for the  $\NN=2$ superconformal algebra \cite{Dolan:2002zh}.}
\end{table}

A generic long multiplet $\AA_{R,r(j,\bar{j})}^{\Delta}$ of the $\NN=2$
superconformal algebra is generated by the action of the $8$ Poincar\'e supercharges
$\QQ$ and $\bar{\QQ}$ on a superconformal primary, which by definition is
 annihilated by all  conformal supercharges $\SS$. If  some combination of
the  $\QQ$'s  also annihilates the primary, the corresponding multiplet
is shorter and the conformal dimensions of all its members are protected against quantum corrections.
A comprehensive list of  the possible shortening 
conditions for the $\NN=2$ superconformal algebra was given in
 \cite{Dolan:2002zh} . Their findings are summarized in  Table \ref{shortening}.
We take a moment to explain the notation.\footnote{We  follow the conventions of
 \cite{Dolan:2002zh}, except that we have introduced the labels 
${\cal D}$, ${\cal F}$, ${\cal \hat F}$ and ${\cal G}$ to denote some shortening conditions that were left nameless in  \cite{Dolan:2002zh}.}
The state $|R,r\rangle^{h.w.}_{(j,\bar{j})}$ is the highest weight state with
$SU(2)_{R}$ spin $R >0$, $U(1)_{r}$ charge $r$,
which can have either sign, and Lorentz quantum numbers  $(j,\bar{j})$.
The multiplet  built on this state is  denoted as $\mathcal{X}_{R,r(j,\bar{j})}$,
where the letter $\mathcal{X}$ characterizes the shortening condition.
The left column of Table  \ref{shortening} labels
the condition. 
A superscript on the label  corresponds to the index $\II =1,2$ of the
supercharge that kills the primary:
or example ${\cal B}^1$ refers
to ${\cal Q}_\alpha^1$. Similarly a ``bar'' on the label refers to the conjugate condition: for example
$\bar{\BB}^{2}$ corresponds to $\bar Q_{2 \, \dot \alpha}$ annihilating the state;
this would result in the short anti-chiral multiplet $\bar{\BB}_{R,r(j,0)}$, obeying $\Delta = 2 R -r$.
Note that conjugation reverses the signs of $r$, $j$ and $\bar j$ in the expression of the conformal dimension.
We refer to \cite{Dolan:2002zh} for more details.

\section{ \label{chiralring}${\cal N} = 1$ Chiral Ring }

An important subset of the protected operators of a supersymmetry theory are the operators in the chiral ring. Chiral operators, by definition, are annihilated
by the supercharge of one chirality, $\bar \QQ^{\dot \alpha}$, and thus obey a ${\cal B}$-type shortening condition.
(If the theory has extended supersymmetry we focus on an ${\cal N} =1$ subalgebra.)
The product of two chiral operators is again  chiral. Chiral operators are normally
considered modulo $\bar \QQ^{\dot \alpha}$-exact  operators. The chiral cohomology classes 
can be specified by a set of generators and relations, which are easy to determine at weak (infinitesimal but non-zero) coupling.
At higher orders  the relations may get corrected, but the basic counting of chiral states is not expected to change \cite{Cachazo:2002ry, Kinney:2005ej}.

Let us first consider the case of pure $\NN=2$ SYM with gauge group $SU(N_c)$. 
Under an $\NN=1$ subalgebra the field content is decomposed as a chiral superfield $\Phi$ and a vector superfield $W_\alpha$,
both in the adjoint representation of the gauge group.. 
A generic chiral operator of the theory in the adjoint representation of the gauge group obeys
\be
 \left[ W_{\alpha} , \OO \right\}
= \left[ \bar{\mathcal{Q}}^{\dot{\alpha}} ,  D_{\alpha \dot{\alpha} } \OO \right\}   \,.\label{Qexact}
\ee
Substituting $\OO=\Phi$ and $\OO=W_\beta$ we see that, modulo $\bar{\mathcal{Q}}$ exact terms, $W_\alpha$ (anti-)commutes with $\Phi$ and $W_\beta$ respectively. Using these relations we can narrow down the single-trace  chiral operators to
\be
\Tr \, \Phi^{k+2} , \qquad \Tr \, \Phi^{k+1} W_{\alpha},   \qquad  \Tr\,  \Phi^k \epsilon^{\alpha\beta} W_{\alpha} W_{\beta} \, ,\qquad \mbox{for} \,\, k\geq0\,.
\label{Vchiral}
\ee
We have listed one representative from each cohomology class. For finite $N_c$ the operators are further related by trace relations. In the  large $N_c$ limit of $N=2$ supersymmetric Yang Mills, 
(\ref{Vchiral}) is  the complete and unconstrained list of single-trace chiral operators. Taking products we generate the whole chiral ring.
 In $\NN=2$ language the chiral operators are assembled in a single supermultiplet for each $k$,
 the  multiplet with primary $\Tr \, \phi^{k+2}$.

To obtain $\NN=2$ SCQCD we add $N_f$ fundamental hypermultiplets, equivalent to $N_f$ fundamental chiral multiplets $\q$ and $N_f$ antifundamental
chiral multiplets  $\tilde \q$, with the $\NN=2$ invariant superpotential $\tilde \q \Phi \q$. 
There are no chiral operators containing both $W_\alpha$ and $\q$ because $W_\alpha \q$ is $\bar \QQ$ exact.
Generally, in a  theory with superpotential, further relations are imposed by the equations of motion
\bea
\p_{A} W(A_i) = \bar{D}_{\dot{\alpha}}\bar{D}^{\dot{\alpha}}A \quad \Rightarrow   \quad    \p_{A} W(A_i)_{c.r.} = 0\, ,
\eea
where $\{ A_i \}$ is the set of  chiral superfields. The subscript \emph{c.r.} denotes that the relation is valid in the chiral ring. In our case this implies that 
operators containing both $\Phi$ and $\q$ are  constrained by
the equations of motion 
\bea
\Phi \q = 0 ,  \quad   \tilde{ \q}\Phi = 0    \quad  \mbox{and} \quad  \q^a\,_i \tilde{\q}^i\,_b -\frac{1}{N_c} \delta^a_b  \q^c\,_i \tilde{\q}^i\,_c =0 \,.
\eea
These relations set to zero all generalized single-trace operators\footnote{In the flavor non-singlet sector they also allow for $\q^{a}_{\,i}\tilde \q_{a}^{\, j}$.}
 containing $\q$, except for 
$ \Tr\, \q \tilde \q$. When expressed in  $SU(2)_R$ covariant fashion, this operator corresponds to the ${\cal N} = 2$ superconformal primary $\Tr \MM_{\bf 3}$. 
Note that for gauge group $U(N_c)$ instead of $SU(N_c)$ the third relation gets modified to $\q^a\,_i \tilde{\q}^i\,_b=0$ implying that even $\Tr \, \q\tilde \q$ is absent from the chiral ring. 
(For $U(N_c)$  we would have
to also {\it add} the  operator $\Tr \, \Phi$ to the list (\ref{Vchiral})). 
All in all, consideration of  the chiral ring for ${\cal N} = 2$ SCQCD has led to identify the following  protected  ${\cal N} = 2$   superconformal primaries:
\be
  \Tr \, {\cal M}_{\bf 3}\, , \quad \Tr \, \phi^{\ell + 2} \, ,\quad \ell \geq 0\,.
  \ee 
  Note that the multiplets $\{ \Tr \, T \phi^\ell \}$, as well as the extra exotic protected states discussed in section \ref{SCQCDindex}, are not part of the chiral ring.

It is  straightforward  to repeat this exercise for  the $\mathbb Z_{2}$ orbifold of $\NN=4$ SYM. 
 In $\NN=1$ language the field content of the orbifold theory consists of vector multiplets $(\Phi, W_\alpha)$ and $(\check \Phi, \check W_\alpha)$,  in the adjoint representation of  $SU(N_c)$ and $SU(N_{\check c})$ respectively. They are coupled to bifundamental chiral multiplets $(\q_{\IIh},\tilde \q^{\JJh})$ through the superpotential $\tilde \q^{\IIh} \Phi \q_{\IIh}+\q_{\IIh} \check \Phi \tilde \q^{\IIh}$. Here $\IIh, \JJh$ are  $SU(2)_L$ indices.  At large $N_c$, the 
 chiral ring of the orbifold is generated by  the operators (\ref{Vchiral}), by a second copy of  (\ref{Vchiral})  with $\Phi,W_\alpha\to\check \Phi, \check W_\alpha$ corresponding to the two vector multiplets,
 and by  single-trace operators involving the fields from hypermultiplets. The latter obey following constraints due to the superpotential:
 \bea 
\label{orbifoldrelations}
&&\tilde{\q}^{\IIh} \Phi  =- \check{\Phi}\tilde{\q}^{\IIh}  ,\qquad\qquad\qquad\qquad \qquad \Phi  \q_{\IIh} =-  \q_{\IIh} \check{\Phi}  
\\
&&   \q_{\IIh}^a\,_{\check{a}} \tilde{\q}^{\IIh\check{a}} \,_b -\frac{1}{N_c} \delta^a_b  \q_{\IIh}^c\,_{\check{a}}\,  \tilde{\q}^{\IIh\check{a}} \,_c =0, \qquad\,  \tilde{\q}^{\IIh\check{a}} \,_a {\q}_{\IIh}^a\,_{\check{b}}  -\frac{1}{N_{\check c}} \delta^{\check{a}} _{\check{b}}   \tilde{\q}^{\IIh \check{c}} \,_a {\q}_{\IIh}^a\,_{\check{c}}  =0
\nonumber
\eea
Using the first two equivalence relations we could always choose a class representative that doesn't contain any $\check \Phi$. %
 Then the relations in the second line 
  allow for highest $SU(2)_L$ spin  chiral operators of schematic form $\Tr \,(\q\tilde \q)_{{\bf 3_L}}^{\ell+1} \Phi^k$. %
 This operator is in the untwisted sector as it is invariant under quantum $\mathbb {Z}_2$  symmetry of the orbifold upto $\bar \QQ^{\dot \alpha}$ exact terms.
As before, the chiral ring of the $SU(N_c)$ theory (as opposed to $U(N_c)$), also contains the ``exceptional'' operator $\Tr \,(\q\tilde \q)_{{\bf 1_L}}$, which belongs to the twisted sector.
Assembling these ${\cal N} = 1$ chiral multiplets into full $\NN=2$ multiplets,  we find the following list of ${\cal N} = 2$ superconformal primaries:
\bea
&&\Tr\, (\phi^{k+2} + \check \phi^{k+2} ) \,, \qquad  \Tr\, (\MM_{{\bf 3_R}  {\bf 3_L}}^{\ell+1} \,\phi^k ) \, , \\
&& \Tr \, (\phi^{k+2} - \check\phi^{k+2})\, ,
  \qquad \Tr\,  \MM_{{\bf 3_R}  {\bf 1_L}}\,, \qquad    \qquad \mbox{for}\,\, \, k\geq0,\,\ell \geq0\,.
\eea
The primaries in the first line belong to the untwisted sector and the primaries in the second line belong to the twisted sector.
We know from inheritance from ${\cal N} = 4$ SYM that   in the untwisted sector  there are additional protected operators (see section \ref{sec:untwisted}).
On the other hand, in the twisted sector this is plausibly the complete list, as confirmed by the calculation of the superconformal index in appendix \ref{indexappendix}.

As we move away from the orbifold point by taking $\check g \neq g$, 
the calculation of the chiral ring is almost unchanged, we only need to perform the substitutions $\check \Phi, \check W_\alpha \to \kappa \check \Phi, \kappa \check W_\alpha$, 
with $\kappa \equiv \check g/g$ that take into account the deformation of the superpotential.
The quantum numbers of the chiral operators remain unchanged.

\section{\label{indexappendix}The Index of Some Short multiplets}

In this appendix we calculate the index of various short multiplets. A first  goal is to determine the index of the set $\{$ $\hat \BB_{1}$, $\EE_{\ell(0,0)}$, $\ell \geq 2$ $\}$
(the multiplets found by the analysis of the chiral ring in the twisted sector of the orbifold), and
show that it agrees with (\ref{twisted}).  A second goal is to calculate $\II_{naive}$, the index of the ``naive'' protected spectrum (\ref{list}) of ${\cal N} = 2$ SCQCD.

\newcommand{\bbf}{\underline}
\newcommand{\col}{\textcolor{Black}}
\subsection{$\EE_{\ell(0,0)}$ multiplet}

The chiral multiplet $\EE_{\ell(0,0)}$ \cite{Dolan:2002zh}
is defined to be the multiplet that descends from the operator with
$R=0$, that is annihilated by both $\mathcal{Q}^{1}$ and $\mathcal{Q}^{2}$.
The shortening condition is $\Delta=\ell$. We have arranged the operator content of the multiplet in the array below. We represent the action of the supercharge $\QQ$ to the left and  $\bar \QQ$ to the right. As 
$\EE_{\ell(0,0)}$ is annihilated by $\QQ$s, it only extends to the right.
\[
\begin{array}{l|cccccc}
\Delta\\
\ell & \bbf{0_{\left(0,0\right)}}\\
\ell+\frac{1}{2} &  & \bbf{\frac{1}{2}_{\left(0,\frac{1}{2}\right)}}\\
\ell+1 &  &  & 0_{\left(0,1\right)},\bbf{1_{\left(0,0\right)}}\\
\ell+\frac{3}{2} &  &  &  & \frac{1}{2}_{\left(0,\frac{1}{2}\right)}\\
\ell+2 &  &  &  &  &  & 0_{\left(0,0\right)}\\
\hline
r & \ell \quad& \ell-\frac{1}{2} & \ell-1 & \ell-\frac{3}{2} &  & \quad \ell-2\end{array}
\]
This multiplet contributes only to the left index $\II^{\ind L}$. The  operators with $\delta^{\ind L}=0$ are underlined and their contribution to the index is listed in table \ref{listE}.

\begin{table}[h]
\begin{centering}
\begin{tabular}{|l|c|l|}
\hline 
$\Delta$  & $R_{(j,\bar{j})}$  & $\II ^{\ind L}(t,y,v)$\tabularnewline
\hline
\hline 
$\ell$  & $0_{(0,0)}$  & $t^{2\ell}v^{\ell}$\tabularnewline
\hline 
$\ell+\frac{1}{2}$  & $\frac{1}{2}_{\left(0,\frac{1}{2}\right)}$  & $-t^{2\ell+1}v^{\ell-1}\left(y+\frac{1}{y}\right)$\tabularnewline
\hline 
$\ell+1$  & $1_{\left(0,0\right)}$  & $t^{2\ell+2}v^{\ell-2}$\tabularnewline
\hline
\end{tabular}
\par\end{centering}

\caption{\label{listE}Operators with $\delta^{\ind L}=0$ in $\mathcal{E}_{\ell(0,0)}$}

\end{table}

For $\ell>1$, we sum the contribution of the operators
from the above table and divide it by the contribution $\left(1-t^{3}y\right)\left(1-t^{3}y^{-1}\right)$
from the derivatives,
\begin{eqnarray*}
\sum_{\ell=2}^{\infty} \II^{\ind L}_{\EE_{\ell(0,0)}} & = & \frac{1}{\left(1-t^{3}y\right)\left(1-t^{3}y^{-1}\right)}\sum_{\ell=2}^{\infty}t^{2\ell}v^{\ell}(1-t^{1}v^{-1}(y+y^{-1})+t^{2}v^{-2})\\
 & = & \frac{t^{4}v^{2}(1-\frac{t}{vy})(1-\frac{ty}{v})}{(1-t^{2}v)\left(1-t^{3}y\right)\left(1-t^{3}y^{-1}\right)}\end{eqnarray*}
The conjugate multiplet $\bar \EE_{-\ell(0,0)}$
contributes exactly the same but to  $\II^{\ind R}$.

\subsection{ $\hat{\mathcal{B}}_{1}$ multiplet}

Next we consider the nonchiral multiplet $\hat{\mathcal{B}}_{1}$
\cite{Dolan:2002zh}, with the shortenning condition that the highest
weight state is anihilated by $\mathcal{Q}^{2},\bar{\mathcal{Q}}_{1}$.
This shortening condition requires $r=0$, $j=\bar{j}=0$ and $\Delta=2$
for the highest weight state.
\[ \label{B1multiplet}
\begin{array}{l|ccccc}
\Delta\\
2 &  &  & \bbf{1_{\left(0,0\right)}}\\
\frac{5}{2} &  & \bbf{\frac{1}{2}_{\left(\frac{1}{2},0\right)}} &  & \frac{1}{2}_{\left(0,\frac{1}{2}\right)}\\
3 & 0_{\left(0,0\right)} &  & 0_{\left(\frac{1}{2},\frac{1}{2}\right)} &  & 0_{\left(0,0\right)}\\
\frac{7}{2} &&&\\
4 &  &  & -0_{\left(0,0\right)}\\
\hline
r & 1 & \frac{1}{2} & 0 & -\frac{1}{2} & -1\end{array}
\]
The operator $-0_{(0,0)}$ at $\Delta=4$ stands for an equation of motion -- the negative sign in front of it means that its contribution to the index (partition function in general) has to be subtracted.
We have underlined the operators with $\delta^{\ind L}=0$ and their contribution to $\II^{\ind L}$ is listed in table \ref{listB}.

\begin{table}[h]
\begin{centering}
\begin{tabular}{|l|c|l|}
\hline 
$\Delta$  & $R_{(j,\bar{j})}$  & $\II^{\ind L}(t,y,v)$\tabularnewline
\hline
\hline 
$2$  & $1_{(0,0)}$$ $  & $\frac{t^{4}}{v}$\tabularnewline
\hline 
$\frac{5}{2}$  & $\frac{1}{2}_{\left(\frac{1}{2},0\right)}$  & $-t^{6}$\tabularnewline
\hline
\end{tabular}
\par\end{centering}

\caption{\label{listB}Operators with $\delta^{\ind L}=0$ in $\BB_{1}$}

\end{table}

Summing the individual contributions and dividing with the contribution from the derivatives,
we get the index for this multiplet as,
\begin{equation}
\II^{\ind L}_{\BB_1}=\frac{t^{4}\left(1-t^{2}v\right)}{v\left(1-t^{3}y\right)\left(1-t^{3}y^{-1}\right)}\,.
\end{equation}
\newpage

\subsection{$\hat{\mathcal{C}}_{0\left(0,0\right)}$ multiplet}
The stress tensor, supercurrents and R-symmetry currents of the $\NN=2$ theory are part of this multiplet. Its shortening condition $\hat \CC$ is explained in table \ref{shortening}. The operator content of this multiplet is displayed in the array below.
\begin{equation}
    \begin{array}{l|ccccc}
 \Delta      &                                                                              \\
 &&&&\\
 2          &   &  &0_{\left( 0,0 \right)}                                         &\\
 &&&&\\
 \frac{5}{2} &    &      \bbf{ \frac{1}{2}_{\left(  \frac{1}{2}, 0 \right)} }  & &  \frac{1}{2}_{\left( 0, \frac{1}{2} \right)}      \\
 &&&&\\
 3   &      \bbf{ 0_{\left( 1,0 \right)}}\quad &  &     {\bbf{1_{(\frac{1}{2},\frac{1}{2})}}}  ,\,  0_{\left(  \frac{1}{2}, \frac{1}{2} \right)}   &  &  \qquad 0_{\left( 0,1 \right)}  \\
 &&&&\\
\frac{7}{2}  &                &       \bbf{\frac{1}{2}_{\left( 1, \frac{1}{2} \right)} }  & &  \frac{1}{2}_{\left(  \frac{1}{2},1 \right)}      \\
&&&&\\
 4  &            &      &    0_{\left( 1,1 \right)} \\
 &&&-0_{(0,0)},\,-1_{(0,0)}&\\
 &&&&\\
 \frac{9}{2}&&-\frac{1}{2}_{(\frac{1}{2},0)}&&-\frac{1}{2}_{(0,\frac{1}{2})}\\
 &&&&\\
 10&&&-0_{(\frac{1}{2},\frac{1}{2})}&\\
 &&&&\\
 \hline
  r               &   1\qquad  & \frac{1}{2}       &   0    & -\frac{1}{2}  & \qquad-1
      \end{array}
\end{equation}
The operators with negative signs stand for equations of motion as before.
We have underlined the operators with $\delta^{\ind L}=0$ and their contribution is listed in the table below. 
\begin{table}[h]
\centering{}\begin{tabular}{|l|l|l|}
\hline 
$\Delta$ & $R_{(j,\bar{j})}$ & $\II^{\ind L}(t,y,v)$\tabularnewline
\hline
\hline 
$\frac{5}{2}$ & $\frac{1}{2}_{(\frac{1}{2},0)}$ & $-t^{6}$\tabularnewline
\hline 
$3$ & $0_{(1,0)}$ & $t^{8}v$\tabularnewline
\hline 
$3$ & $1_{(\frac{1}{2},\frac{1}{2})}$ & $\frac{t^{7}}{v}(y+\frac{1}{y})$\tabularnewline
\hline 
$\frac{7}{2}$ & $\frac{1}{2}_{(1,\frac{1}{2})}$ & $-t^{9}(y+\frac{1}{y})$\tabularnewline
\hline
\end{tabular}
\caption{\label{listChat}Operators with $\delta^{\ind L}=0$ in $\hat{\mathcal{C}}_{0(0,0)}$}
\end{table}
Summing the contributions, we get the left index of this multiplet to be
\begin{equation}
\II^{\ind L}_{\hat \CC_{(0,0)}} =    -t^6 ( 1-vt^2 ) (1-\frac{t}{v}(y+\frac{1}{y} ))\,.
\end{equation}
Being a nonchiral multipet, it contributes the same to the right index as well.

\subsection{$\CC_{\ell(0,0)}$ multiplet, $\ell \geq 1$}
This multiplet obeys the shortening condition $\mathcal{F}= \mathcal{C}_1\cap  \mathcal{C}_2$. The operator content of $\CC_{\ell(0,0)}$ is displayed below. 
\be
   \begin{array}{l|cccccccccc}
 \Delta      &                                                                              \\
  &&&&\\
 \ell+2          & & & 0_{\left( 0,0 \right)}                                         &\\
 &&  & & \\
 \ell+\frac{5}{2}  &   &    {\bbf {\frac{1}{2}_{\left(   \frac{1}{2}, 0 \right)}}}   & &  \frac{1}{2}_{\left( 0,  \frac{1}{2} \right)}      \\
   &&&&\\
\ell+ 3    &   {\bbf{ 0_{\left(  1,0 \right)}}}\quad\,\, &  &    {\bbf {1_{\left(  \frac{1}{2},  \frac{1}{2} \right)}}},\,  0_{\left(   \frac{1}{2},  \frac{1}{2} \right)}   &  &  0_{\left( 0, 1 \right)} \, , \, \col{ 1_{\left( 0,0 \right)} }   \\
   &&&&\\
\ell+\frac{7}{2}   &        & {\bbf {\frac{1}{2}_{\left( 1, \frac{1}{2} \right)}}}   & &    \frac{1}{2}_{\left(   \frac{1}{2}, 1 \right)}  \, , \, \col{  \frac{1}{2}_{\left(   \frac{1}{2}, 0 \right)},\,{\bbf {\frac{3}{2}_{\left(  \frac{1}{2}, 0 \right)}}}} & &  \col{   \frac{1}{2}_{\left( 0,  \frac{1}{2} \right)}   }\\
  &&&&\\
\ell+ 4   &            &      &   0_{\left(  1, 1 \right)}, \,\col{ {\bbf{   1_{\left(  1,0 \right)} }}} & &\col{  0_{\left(  \frac{1}{2},  \frac{1}{2} \right)}  \,,\,  1_{\left(  \frac{1}{2},  \frac{1}{2} \right)} }  & & \qquad\col{   0_{\left( 0,0 \right)} }\\
   &&&&\\
\ell+ \frac{9}{2}   &      &         &      & \col{  \frac{1}{2}_{\left( 1,  \frac{1}{2} \right)} } & &    \col{ \frac{1}{2}_{\left( \frac{1}{2},0 \right)} }  \\
   &&&&\\
\ell+5 & & &  & &   \col{  0_{\left( 1,0\right)}  }\quad \\
   &&&&\\
   \hline
  r               &  \ell+1\qquad  & \ell +\frac{1}{2}      &   \ell     & \ell -\frac{1}{2}  &  \ell  -1 \quad& \ell  -\frac{3}{2} & \qquad \ell -2
      \end{array}  \nonumber
\ee
The operators with $\delta^{\ind L}=0$ are underlined as usual. Table \ref{listC} lists their contribution to $\II^{\ind L}$. Summing the contribution to the left index from $\CC_{\ell(0,0)}$ with $\ell \geq 1$ we get,
\begin{equation}
\sum^{\infty}_{\ell=1} \II^{\ind L}_{\CC_{\ell(0,0)}} =    -t^8v ( 1-vt^2 ) (1-\frac{t}{v}(y+\frac{1}{y} ))-\frac{t^{10} }{v}\,.
\end{equation}

\begin{table}[h]
\centering{}\begin{tabular}{|l|l|l|}
\hline 
$\Delta$ & $R_{(j,\bar{j})}$ & $\II^{\ind L}(t,y,v)$\tabularnewline
\hline
\hline 
$\ell+\frac{5}{2}$ & $\frac{1}{2}_{(\frac{1}{2},0)}$ & $-t^{6+2\ell}v^{\ell}$\tabularnewline
\hline 
$\ell+3$ & $0_{(1,0)}$ & $t^{8+2\ell}v^{\ell+1}$\tabularnewline
\hline 
$\ell+3$ & $1_{(\frac{1}{2},\frac{1}{2})}$ & $t^{7+2\ell}v^{\ell-1}(y+\frac{1}{y})$\tabularnewline
\hline 
$\ell+\frac{7}{2}$ & $\frac{1}{2}_{(1,\frac{1}{2})}$ & $-t^{9+2\ell}v^{\ell}(y+\frac{1}{y})$\tabularnewline
\hline 
$\ell+\frac{7}{2}$ & $\frac{3}{2}_{(\frac{1}{2},0)}$ & $-t^{8+2\ell}v^{\ell-2}$\tabularnewline
\hline 
$\ell+4$ & $1_{(1,0)}$ & $t^{10+2\ell}v^{\ell-1}$\tabularnewline
\hline
\end{tabular}
\caption{\label{listC}Operators with $\delta^{\ind L}=0$ in $\mathcal{C}_{\ell(0,0)}$}
\end{table}

\subsection{The $\II_{twist}$ of the orbifold and $\II_{naive}$ of SCQCD}
The protected operators in the twisted sector of the orbifold are listed in Table \ref{twistedtable}. The conjugates, which contribute to $\II^{\ind L}$,
 are of the type:
\be
 \hat \BB_1, \qquad  \EE_{\ell(0,0)} \qquad\qquad \mbox{for} \quad \ell \geq 2 \,. 
 \ee
So we get,
\bea
\II_{twist}&=&\II_{\hat \BB_1}+\sum_{\ell=2}^{\infty} \II_{\EE_{\ell(0,0)}}\\
&=&\frac{t^{4}\left(1-t^{2}v\right)}{v\left(1-t^{3}y\right)\left(1-t^{3}y^{-1}\right)}+\frac{t^{4}v^{2}(1-\frac{t}{vy})(1-\frac{ty}{v})}{(1-t^{2}v)\left(1-t^{3}y\right)\left(1-t^{3}y^{-1}\right)}\\
&=& \frac{t^{2}v}{1-t^{2}v}-\frac{t^{3}y}{1-t^{3}y}-\frac{t^{3}y^{-1}}{1-t^{3}y^{-1}} - f_V(t,y,v)\,.
\eea
This precisely matches with (\ref{twisted}), confirming the protected operators in the twisted sector of the orbifold.
Let us now compute the $\II_{naive}$ of SCQCD that follows from the preliminary list \ref{list} of protected operators. Their conjugates, which contribute to $\II^{\ind L}$, are of the type:
\be
\hat \BB_1,\qquad \EE_{\ell+2(0,0)},\qquad \hat \CC_{0,0}, \qquad \CC_{\ell+1(0,0)} \qquad \qquad\mbox{for} \quad \ell\geq0 \,.
\ee
The $\II_{naive}$ then is
\bea
\II_{naive}&=& \II_{\hat \BB_1}+\sum_{\ell=2}^{\infty} \II_{\EE_{\ell(0,0)}}+\II_{\hat \CC_{0,0}}+\sum_{\ell=1}^{\infty} \II_{\CC_{\ell(0,0)}}\\
&=& \frac{-t^6(1-\frac{t}{v}(y+\frac{1}{y}))-\frac{t^{10}}{v}+\frac{t^4v^2(1-\frac{t}{vy})(1-\frac{ty}{v})}{1-t^2v}+\frac{t^4}{v}(1-t^2v)}{(1-t^3y)(1-\frac{t^3}{y})}\,.
\eea

\section{\label{tensorKK}KK Reduction of the $6d$ Tensor Multiplet on $AdS_5 \times S^1$}

In this appendix we discuss the Kaluza-Klein reduction of the $6d$ tensor multiplet
on $AdS_5 \times S^1$, and its matching with the twisted spectrum of the orbifold theory.

The tensor multiplet of maximal chiral  supersymmetry in six dimensions (we will refer to it as (2,0) susy)
has the following field content  \[
B_{ \mu \nu}^{-} \,,  \quad \la^{\usp J}_{\alpha}\,, \quad \Phi^{[\usp J\usp K]} \,.\]
The indices $\usp J,\usp K$ are the $USp(4)$ indices which is the
R-symmetry group of the chiral supergravity. The  spinors $\la^{\usp J}_{\alpha}$ are  in the ${\bf 4}$ (complex) representation of $USp(4)$
and the scalars  $\Phi^{[\usp J\usp K]}$ in the ${\bf 5}$ (real) representation.
The $\la^{\usp J}_{\alpha}$
are Weyl, symplectic Majorana spinors. The symplectic Majorana condition
is a psuedo-reality condition,
$\bar \la_{\usp I} = \Omega_{\usp I \usp K} \la^{\usp K}$,
where $\Omega$ is the symplectic form.

Consider now the background $AdS_5 \times S^1$. 
The natural embedding of the  $SU(2)_R \times U(1)_r$ R-symmetry of the $\NN = 4$ $AdS_5$ superalgebra 
(or equivalently of the $\NN = 2$ $4d$  superconformal algebra)
into $USp(4)$ is
 \[ \label{choice}
\left(\begin{array}{cc|cc} SU(2)_R \times U(1)_r & & &
\\  & & & \\
\hline   & & &
\\  & & & SU(2)_R \times  U(1)_r^*
\end{array}\right) \,
\]
The five scalars decompose as 
\bea \label{u1r}
 \Phi^{[\usp J\usp K]}  & \longrightarrow& \Phi^i +\; \Phi \;\;+\; \;\bar \Phi  \\
{\bf 5} &\longrightarrow& {\bf 3}_0 + \bf{1}_{-1} +{\bf 1}_{+1} \, ,\nonumber
\eea
where the subscripts denote $U(1)_r$ charges. The spinors decompose as two (conjugate) $SU(2)_R$ doublets, with opposite $U(1)_r$ charges $r = \pm \frac{1}{2}$.

We are interested in the Kaluza-Klein reduction of the tensor multiplet on the $S^1$. We  borrow the 
 results of \cite{Gukov:1998kk} (see also \cite{Berenstein:2000hy}), where all the  KK modes with non-zero momentum
were matched with the   multiplets $\{ \bar {\cal E}_{2+\ell (0,0)} \, \ell \geq 0 \}$, corresponding to the twisted
 primaries $\{ \Tr \phi^{2+\ell} - \Tr \check \phi^{\ell+2} \, \}$ of the orbifold theory. 
 We will add  the zero modes to the analysis of \cite{Gukov:1998kk}.

Let us indeed start with the zero modes on $S^1$. The bosonic zero modes comprise the following $AdS_5$ fields \cite{Gukov:1998kk}:
a complex scalar $\Phi$, with $m^2 = -3$ (in $AdS$ units)\footnote{The complex scalar $\Phi$ corresponds to the $k=-1$ real scalar  in Family 2  and the $k=1$ real scalar
in Family 3 of \cite{Gukov:1998kk}.  We have just relabeled them as $n=0$ modes. 
};  a triplet of scalars $\Phi^i$, with $m^2 = -4$; a massless two form $B_{\hat m \hat n}$, {\it or} equivalently a massless gauge field $A_{\hat m}$.
The massless two-form $B_{\hat m \hat n}$ arises from the $6d$ anti-selfdual two-form $B_{\mu \nu}^-$ when both indices
are taken to be along $AdS_5$, while the gauge field $A_{\hat m}$ arises from  $B_{\mu \nu}^-$ when one index
is taken to be along $AdS_5$ and the other along $S^1$. Because of the anti-selfduality of   $B_{\mu \nu}^-$, the two possibilities
are  not independent: $B_{\hat m \hat n}$ and $A_{\hat m}$ are dual to each other as $5d$ fields, and we must
pick one {\it or} the other. This ambiguity translates into
 two alternative ways to fit the zero modes into supermultiplets
of the $\NN  =2$ $4d$ superconformal algebra. 
Let us look at  them in turn:
\begin{itemize}
\item   Choosing $B_{\hat m \hat n}$.

The massless two-form $B_{\hat m \hat n}$ is dual to a boundary two-form operator $F'_{m n}$ of dimension $\Delta =2$.
We claim that the full supermultiplet of boundary operators is $\{ \phi' \, , \lambda_\alpha^{'\cal I}\, , F'_{m n}\, D'_i \}$, which
is the the familiar off-shell ${\cal N} = 2$ vector multiplet (or ${\cal N} = 2$ ``supersingleton'' multiplet).
 Here $\phi'$ is a complex scalar with $r = \pm 1$ and $\Delta=1$, dual to the bulk scalar $\Phi$ of $m^2 = -3$. The mass of $\Phi$ is in the range
that allows both the $\Delta_+$ and the $\Delta_-$ quantization schemes \cite{Breitenlohner:1982bm, Klebanov:1999tb}, and supersymmetry forces the choice of $\Delta_- = 2 - \sqrt{m^2+4}=1$.
Since  $\phi'$ saturates the unitarity bound, it must  be a free scalar field. We recognize
  $F'_{m n}$ as the Maxwell field strength and $D'_i$, $i=1,2,3$,  which form $SU(2)_R$ triplet with $\Delta =2$ and are  dual to the bulk fields $\Phi^i$, as the auxiliary fields.
  Finally $\lambda_\alpha^{'\cal I}$ are the free fermionic fields with $\Delta =\frac{3}{2}$. The AdS/CFT relation for spin $\frac{1}{2}$
  fields is usually quoted as $\Delta = 2 + |m|$, but this is evidently a case where we must pick instead $\Delta_- = 2 -|m|$, with $m=\frac{1}{2}$.
  We are not aware of an explicit discussion of the $\Delta_\pm$ quantization ambiguity for spinors, but it must be there because of supersymmetry.
  (Incidentally, similar issues arise  in the familiar IIB on $AdS_5 \times S^5$ background if one looks at the zero modes, which can be organized in the ${\cal N} =4$
  supersingleton multiplet. Again both the scalars in the {\bf 6} of $SU(4)$ {\it and} the spinors in the ${\bf 4}$ must be quantized in the $\Delta_-$ scheme.)

\item  Choosing $A_{\hat \mu}$.

The boundary dual to  $A_{\hat m}$ is a conserved current $J_m$ ($\Delta=3$). In this case
we claim that supersymmetry forces the usual $\Delta_+$ quantization scheme for $\Phi$ and $\la^{\usp J}_{\alpha}$.
 It is easy to check that the zero modes can be precisely organized into the $\hat {\cal B}_1$ multiplet (summarized in (\ref{B1multiplet})).

\end{itemize}
The two possibilities have a nice physical interpretation. The first
alternative corresponds to keeping the $U(1)$ degree of freedom in the twisted
sector (this is the ``relative'' $U(1)$ in the product gauge recall the discussion after equ.(\ref{survive})) --
 in other terms we should identify $\phi' = {\rm Tr} (\phi - \hat \phi)$.
The second possibility corresponds instead to {\it removing} the relative $U(1)$. Then clearly the multiplet
built on  ${\rm Tr} (\phi - \hat \phi)$ is lost, but
  as we have emphasized in section \ref{twistedsubsection} and appendix B,
  an {\it additional}  protected multiplet appears, the  $\hat {\cal B}_1$ multiplet built on the primary
${\rm Tr}\, {\cal M}_{\bf 3}$. The AdS/CFT dictionary handles this subtle
ambiguity in a very elegant way. For our purposes,  the second alternative is the relevant one,
since we must remove the relative $U(1)$ in order to have a truly conformal field theory.

The matching of the higher Kaluza-Klein modes was
discussed in \cite{Gukov:1998kk}, we summarize the results
 in Table \ref{tensormatching}.

\begin{table}
\begin{centering}
\begin{tabular}{|l|l|l|l|l|}
\hline 
\multicolumn{3}{|c|}{Field Theory} & \multicolumn{2}{c|}{Gravity}\tabularnewline
\hline 
Operator &  $U(1)_{r}$ & $\Delta$ & Mass & Field \tabularnewline
\hline
\hline 
Tr$[\bar{\f}^{n+1}]$ $-$ Tr$[\bar{\check \f}^{n+1}]$&  $n+1$ & $n+1$ & $(n+1)(n-3)$ & $\bar{\Phi}$ \tabularnewline
\hline 
Tr$[F\fbar^{n}]$ $-$  Tr$[\check F \bar{\check{\f}}^{n}]$&  $n$ & $n+2$ & $n^{2}$ & $B_{\hat{m}\hat{n}}$ \tabularnewline
\hline
Tr$[\la\la\fbar^{n-1}]$ $-$  Tr$[\check \la\check \la\bar {\check \phi}^{n-1}]$  & $n$ & $n+2$ & $n^{2}-4$ & $\Phi^{i}$ \tabularnewline
\hline 
Tr$[F^{2}\fbar^{n-1}]$ $-$ Tr$[\check F^{2}\bar {\check  \phi}^{n-1}]$&  $n-1$ & $n+3$ & $(n-1)(n+3)$ & $\Phi$ \tabularnewline
\hline 
\end{tabular}
\par\end{centering}

\caption{\label{tensormatching}Matching of the positive KK modes ($n\geq 1$) \cite{Gukov:1998kk}. The negative
KK modes $(n \leq -1)$ correspond to the conjugate operators.}

\end{table}

\global\long\def\m{\mu}

\global\long\def\n{\nu}

\global\long\def\La{\Lambda}

\global\long\def\s{\sigma}

\global\long\def\f{\phi}

\global\long\def\e{\epsilon}

\global\long\def\del{\partial}

\global\long\def\D{\Delta}

\global\long\def\al{\alpha}

\global\long\def\ad{\dot{\alpha}}

\global\long\def\bd{\dot{\beta}}

\global\long\def\la{\lambda}

\global\long\def\ra{\rightarrow}

\global\long\def\fbar{\bar{\phi}}

\global\long\def\p{\partial}

\global\long\def\bA{{\bf A}}

\global\long\def\OO{\mathcal{O}}

\global\long\def\II{\mathcal{I}}

\global\long\def\JJ{\mathcal{J}}

\global\long\def\KK{\mathcal{K}}

\global\long\def\LL{\mathcal{L}}

\global\long\def\TT{\mathcal{T}}

\global\long\def\NN{\mathcal{N}}

\global\long\def\MM{\mathcal{M}}

\global\long\def\PP{\mathcal{P}}

\global\long\def\nn{ \mathfrak{n} }

\global\long\def\qq{ \mathfrak{q} }

\global\long\def\mm{ \mathfrak{m} }

\global\long\def\pp{ \mathfrak{p}}

\global\long\def\Tr{\mbox{Tr}}

\global\long\def\Q{\mathcal{Q}}

\global\long\def\TT{\mathcal{T}}

\global\long\def\SS{\mathcal{S}}

\global\long\def\RR{\mathcal{R}}

\global\long\def\TpT{\mathcal{T}^{\prime}}

\global\long\def\IIh{\hat{\mathcal{I}}}

\global\long\def\JJh{\hat{\mathcal{J}}}

\global\long\def\KKh{\hat{\mathcal{K}}}

\global\long\def\LLh{\hat{\mathcal{L}}}

\global\long\def\SSh{\hat{\mathcal{S}}}

\global\long\def\RRh{\hat{\mathcal{R}}}

\global\long\def\dprime{\prime\prime}

\global\long\def\topp#1{\check{#1}}

\global\long\def\fh{\topp{\f}}

\global\long\def\QQ{\mathcal{Q}}

\global\long\def\AA{\mathcal{A}}

\global\long\def\BB{\mathcal{B}}

\global\long\def\CC{\mathcal{C}}

\global\long\def\DD{\mathcal{D}}

\global\long\def\EE{\mathcal{E}}

\global\long\def\vf{\varphi}

\section{ The Cigar Background 
 and 7d Gauged Sugra}

This appendix collects some facts about  the non-critical
string theory obtained in the double-scaling limit of two colliding NS branes \cite{Giveon:1999px, Giveon:1999tq},
namely  IIB on $\mathbb{R}^{5,1} \times SL(2)_2/U(1)$. We start by reviewing well-known results,
see {\it e.g.} 
\cite{Aharony:1998ub, Giveon:1999zm, Giveon:1999zm, Giveon:1999px, Giveon:1999tq, Aharony:2003vk, Aharony:2004xn, Murthy:2003es}, 
and then make a new claim about a space-time ``effective action'' description. We are going to argue that
the ``lighest''  delta-function normalizable modes in the continuum are described by a $7d$ maximally supersymmetric
supergravity with non-standard gauging, recently constructed in \cite{Samtleben:2005bp, Weidner:2006rp}.

\subsection{Preliminaries and Worldsheet Symmetries}

A class of  ``non-critical'' supersymmetric string backgrounds can defined
in the RNS formalism  by  taking the tensor product of  $\mathbb{R}^{d-1,1}$ 
with the Kazama Suzuki supercoset $SL_{2}(\mathbb{R})_k/U(1)$. The $\mathbb{R}^{d-1,1}$ part
is described as usual by $d$ free bosons $X^\mu$ and $d$ free fermions $\psi^\mu$.
The coset $SL_{2}(\mathbb{R})_k/U(1)$
has a sigma-model description with target space the ``cigar'' background (setting $\alpha' = 2$)
\be
ds^{2}  = d\rho^{2}+\tanh^{2}(\frac{Q\rho}{2})d\theta^{2} \qquad\rho \geq 0\, \quad \theta\sim\theta+\frac{4\pi}{Q}
\ee
with vanishing $B$ field and dilaton varying as
\be
\Phi  =  -\ln\cosh(\frac{Q\rho}{2})\,.
\ee
The level $k$ of the coset is related to the parameter $Q$ as $k = 2/Q^2$. The central charge is
 \be
 c_{cig} = 3 +  \frac{6}{k} = 3 + 3 Q^2\,. 
 \ee
Adding the usual superconformal ghost system $\{ b\,, c\,,  \beta\,, \gamma \}$ of central charge -15 and requiring cancellation
of the total conformal anomaly, one finds  $Q=\sqrt{\frac{1}{2}(8-d)}$.
 In the asymptotic region $\rho\rightarrow\infty$ the cigar
becomes a cylinder  of radius $\frac{2}{Q}$, with the dilaton varying linearly with $\rho$,
and the theory is thus a free CFT. We will soon restrict to the $d=6$ case, implying $c_{cig} =6$, $Q=1$ and $k=2$.

For generic level $k$ the Kazama-Susuki coset $SL(2)_k/U(1)$ has  $(2,2)$ supersymmetry. 
In the asymptotic linear-dilaton region the holomorphic currents of ${\cal N} = 2$ susy take the form 
\begin{eqnarray}
T_{{\rm cig}} & = & -\frac{1}{2}(\del\rho)^{2}-\frac{1}{2}(\del\theta)^{2}-\frac{1}{2}(\psi_{\rho}\del\psi_{\rho}+\psi_{\theta}\del\psi_{\theta})-\frac{1}{2}Q\del^{2}\rho\\
J_{{\rm cig}} & = & \label{Jcig}
 -i\psi_{\rho}\psi_{\theta}+iQ\del\theta\equiv i\del H+iQ\del\theta\equiv i  \del\f \\
G_{{\rm cig}}^{\pm} & = & \frac{i}{2}(\psi_{\rho}\pm i\psi_{\theta})\del(\rho\mp i\theta)+\frac{i}{2}Q\del(\psi_{\rho}\pm i\psi_{\theta})\, ,
\end{eqnarray}
with  analogous expressions for the anti-holomorphic currents. For $k=2$, which is the case of interest for us,  
worldsheet supersymmetry is enhanced to $(4,4)$. This is the generic enhancement of worldsheet susy from ${\cal N} = 2$ to ${\cal N} = 4$
that takes place when $c=6$. Indeed for this value of the central charge the currents
 $J_{\rm cig}^i = \{ e^{\pm \int J_{\rm cig}}\, , J_{\rm cig} \}$, $i = \pm, 3$, generate a left-moving $SU(2)$ current algebra,  the R subalgebra
 of the left-moving  ${\cal N} = 4$ worldsheet superconformal algebra. The two extra odd currents  $\hat G_{{\rm cig}}^{\pm}$ are
 generated in the OPE of $G_{{\rm cig}}^{\pm}$ with $J_{\rm cig}^i$. Similarly for the right-movers.
In the full cigar background the wordsheet superconformal currents 
 have more complicated expressions  but the theory still has exact $(2,2)$ susy, enhanced to $(4,4)$ for $k=2$.

In the free linear dilaton theory,  $i \partial \theta$ and $i \partial H$ defined in (\ref{Jcig})
  are separately holomorphic, but only their linear combination  $J_{\rm cig}$ is holomorphic in the full cigar background.  
    This reflects the non-conservation of winding around the cigar  (strings can unwrap at the tip).
Momentum $P^\theta$ around the cigar is still conserved, and there is a corresponding Noether current
 with both holomorphic and anti-holomorphic components, which asymptotically takes the form $\frac{1}{Q}(i \partial \theta\, , i \bar \partial \theta)$.
 For $k=2$,  the field $\theta$ is asymptotically at  the free fermion radius.  Thus in the linear dilaton theory 
  the left-moving susy $U(1)$ generated by $(i \partial \theta\,, \psi_\theta)$ is enhanced to a left-moving $SU(2)_2$ current algebra,
  which can be  represented by three free fermions $\psi_i$, with $\psi_3 \equiv \psi_\theta$ and $\psi_\pm \equiv e^{\pm i \theta}$.
  To avoid confusions with other $SU(2)$ symmetries will refer to this algebra as $SU(2)_{\psi_i}$.
  Similarly in the right-moving sector we have the analogous $SU(2)_{\tilde \psi_i}$.
   In the full cigar background the $SU(2)_{\psi_i}$ and $SU(2)_{\tilde \psi_i}$ current algebras are {\it not} symmetries, and
   only a {\it global} diagonal  $SU(2)$  survives, 
  whose Cartan generator is the momentum $P^\theta$. This  is interpreted as the $SU(2)_R$ {\it spacetime} R-symmetry.

\subsection{Cigar Vertex Operators}

To characterize the primary vertex operators of the cigar it is sufficient to give their asymptotic form 
in the linear-dilaton region. While the exact expressions are more complicated, their quantum numbers (including conformal dimensions)
remain the same and can thus be evaluated in the asymptotic region. 
Splitting the vertex operators in left-moving and right-moving parts,
we have the asymptotic left-moving expressions 
\begin{eqnarray}
\label{primaryL}
V_{j,m}^{NS} & = & e^{iQm\theta} e^{Qj\rho} \nonumber\\
V_{j,m}^{R} & = & e^{\pm\frac{i}{2}\f}e^{iQm\theta}e^{Qj\rho}
\end{eqnarray}
and the asymptotic  anti-holomorphic expressions 
\begin{eqnarray}
\label{primaryR}
\tilde{V}_{j,\tilde{m}}^{NS} & = & e^{-iQ\tilde{m}\bar{\theta}}  e^{Qj \bar \rho} \nonumber\\
\tilde{V}_{j,\tilde{m}}^{R} & = & e^{\pm\frac{i}{2}\tilde{\f}}e^{-iQ\tilde{m}\bar{\theta}}e^{Qj \bar \rho}\,.
\end{eqnarray}
Left-moving and right-moving terms can be glued together provided they have the same
value of the quantum number $j$. We will sometimes re-express $j$ in terms of $p$, the momentum in the radial direction,
as
\be
{j}=-\frac{Q}{2}+i p \,.
\ee
The quantum numbers $m$ and $\tilde{m}$ are related to the integer winding
$w$ and the integer momentum $n$ in the angular direction of the cylinder as
\begin{equation} 
m=\frac{1}{2}(n+w k)\qquad\tilde{m}=-\frac{1}{2}(n-wk)\,.
\label{eq:relation}\end{equation}
Recall however that winding is not a conserved quantum number in the cigar background.
Conformal dimensions 
 of the primary operators (\ref{primaryL},\ref{primaryR}) are
\begin{eqnarray*}
\Delta_{j,m}^{NS} & = & \frac{m^{2}-j(j+1)}{k}\qquad\qquad\qquad\:\:\: 
\\
\bar{\Delta}_{j,\tilde{m}}^{NS} & = & \frac{\tilde{m}^{2}-j(j+1)}{k}\qquad\qquad\qquad\:\:\:\ 
\\
\Delta_{j,m}^{R \pm} & = & \frac{1}{8}+\frac{(m\pm\frac{1}{2})^{2}-j(j+1)}{k}\qquad 
\\
\bar{\Delta}_{j,\tilde{m}}^{R \pm} & = & \frac{1}{8}+\frac{(\tilde{m}\mp\frac{1}{2})^{2}-j(j+1)}{k} 
\end{eqnarray*}
.

\subsection{Spacetime Supersymmetry}

From now on we restrict to the case of interest, $d=6$.
The RNS vertex operators for $\mathbb{R}^{5,1}$ are familiar. 
To describe the Ramond sector, we bosonize the fermions in the usual fashion,
\begin{eqnarray*}
\pm\psi_{0}+\psi_{1} & = & e^{\pm\f_{0}}\\
\psi_{2}\pm i\psi_{3} & = & e^{\pm i\phi_{1}}\\
\psi_{4}\pm i\psi_{5} & = & e^{\pm i\f_{2}}
\end{eqnarray*}
Spinors of $\mathbb{R}^{5,1}$ are then written
\[
V_{\alpha}=e^{\frac{1}{2}(\e_{0}\f_{0}+i\e_{1}\f_{1}+i\e_{2}\f_{2})}\]
with $\e_a = \pm1$. 
With these notations at hand, the
BRST invariant vertex operators for the spacetime supercharges for the IIB theory read
 \begin{eqnarray*}
S_{\alpha} & = & e^{-\vf/2}e^{+\frac{i}{2}\f}V_{\alpha}^{+} \qquad \bar S_{\alpha}  =  e^{-\vf/2}e^{-\frac{i}{2}\f}V_{\alpha}^{+} 
\\
\tilde{S}_{\alpha} & = & e^{-\tilde{\vf}/2}e^{+\frac{i}{2}\tilde{\f}}\tilde{V}_{\alpha}^{+} \qquad \bar {\tilde S}_{\alpha}  = e^{-\tilde{\vf}/2}e^{-\frac{i}{2}\tilde{\f}}\tilde{V}_{\alpha}^{+} 
\end{eqnarray*}
where $\varphi$ is the usual chiral boson arising in the  bosonization of the $\beta \gamma$ system. We use a  bar  to denote conjugation, and a tilde to distinguish the right-movers.
By $V_{\alpha}^{+}$ we mean the positive chirality spinor, {\it i.e.} we impose  $\e_{0}\e_{1}\e_{2}=1$.
Choosing the same chirality  in the left and right-moving  sectors is the
statement of the type IIB GSO projection. The supercharges obey the supersymmetry algebra
\be
\{ S_\alpha\, , \bar S_\beta \} = 2 \gamma^\mu_{\alpha \beta} P_\mu \, \qquad \{ \tilde S_\alpha\, , \bar {\tilde S}_\beta \} = 2 \gamma^\mu_{\alpha \beta} P_\mu \, ,
\ee
where $P_\mu$ is the momentum in $\mathbb{R}^{5,1}$. Thus the theory has $(2,0)$ supersymmetry in the six Minkowski directions.
Note that
\be
[P^\theta, S_\alpha\,  (\tilde S_\alpha) ] = \frac{1}{2} S_\alpha\,  (\tilde S_\alpha) \, , \quad [P^\theta, {\tilde S}_\alpha\, (  \bar {\tilde S}_\alpha)] = -\frac{1}{2} S_\alpha\,  (  \bar {\tilde S}_\alpha) \, ,
\ee
confirming the interpretation of $P^\theta$ as a spacetime R-symmetry. 

Physical vertex operators 
are constrained to be local with the spacetime supercharges. Locality implies the GSO condition
\begin{eqnarray*}
m+F_{L} & \in & 2\mathbb{Z}+1\qquad(NS)\\
m+F_{L} & \in & 2\mathbb{Z}\quad\qquad\quad(R)\end{eqnarray*}
where $F_L$ is the left-moving worldsheet fermion number.
 The analogous condition holds for the right-movers. In the asymptotic region we may fermionize the field $\theta$ into $\psi^{\pm}$.
 Then the quantum number $m$, instead of denoting left-moving momentum
in the $\theta$ direction, gets re-interpreted as $\psi^\pm$ fermion number. Denoting by $F'_L = F_L + m$ the  new
 total left-moving fermion number, the
GSO projection becomes simply\begin{eqnarray*}
F_{L}^{\prime} & \in & 2\mathbb{Z}+1\qquad(NS)\\
F_{L}^{\prime} & \in & 2\mathbb{Z}\qquad\qquad(R) 
\end{eqnarray*}
and analogously for the right-movers.

\subsection{Spectrum: generalities}

The physical spectrum of the theory comprises:
\begin{enumerate}
\item[(i)] A discrete set of truly normalizable states, localized at the tip of the cigar. \\($j < - Q/2$)
\item[(ii)] A continuum of delta-function normalizable states, corresponding to incoming and outgoing waves in the $\rho$ direction. \\($j = -Q/2 + i \mathbb{R}$, {\it i.e.} $p \in \mathbb{R}$)
\item[(iii)]  Non-normalizable vertex operators, supported in the asymptotic large $\rho$ region.\\  {$(j > -Q/2)$}
\end{enumerate}
States of type (i) live in $\mathbb{R}^{5,1}$ at $\rho  \sim 0$ and they fill in a massless tensor multiplet of the 6d $(2,0)$ supersymmetry.
More precisely they are: 
\begin{itemize}
\item[NSNS:] four scalars, in the {\bf 3} + {\bf 1} of $SU(2)_R$; 
\item[RR:] one scalar and one anti-selfdual antisymmetric tensor, both $SU(2)_R$ singlets;
\item[RNS:] one left-handed Weyl spinor, which can be thought of an $SU(2)_R$ doublet of left-handed Majorana-Weyl spinors;
\item[NSR:]  same as RNS.
\end{itemize}
See \cite{Mizoguchi:2008mk} for a detailed analysis.

In the rest of this appendix we will focus on the states of type (ii). These are the states relevant
for the determination of a spacetime ``effective action'' for the non-critical string. Recall that our philosophy is to use  the  $\mathbb{R}^{5,1} \times$ cigar background  as an intermediate step
 towards the $AdS$ background dual to ${\cal N} =2$ SCQCD. Both backgrounds should arise as solutions of the same non-critical string field theory.
 We would like to use the cigar background, for which we have a solvable worldsheet  CFT, to derive an ``effective action'' description.
The ``effective action'' is expected to be background independent and should admit both the cigar background and the $AdS$ background
as different classical solutions. We will restrict to the lowest level in a ``Kaluza-Klein expansion'' on the cigar circle (to be defined more precisely below). The states will then propagate
in seven dimensions, $\mathbb{R}^{5,1}$ times the radial direction $\rho$. Because of the linear dilaton, they obey  massive field equations
in 7d, but they are in another sense ``massless'' -- they are closely related to the massless states of the critical IIB 10d theory and
possess the gauge invariances expected for massless 7d fields. We should emphasize 
from the outset that  the linear dilaton varies with a string-scale gradient, so there is no real separation of scales between
the ``massless'' level that we are keeping and the higher levels.
This is why we are using ``effective action''  in quotation marks.
Nevertheless the distinction between the lowest level obeying massless gauge-invariances and the higher genuinely massive levels
is a meaningful one, and we still expect such an ``effective action'' to contain useful
information. Remarkably, we will see that it is a $7d$ gauged supergravity with non-standard gauging.

 Finally we should mention the operators of type (iii). They  have an interesting holographic interpretation as
``off-shell'' observables of  little string theory, which ``lives'' on the $\mathbb{R}^{5,1}$ boundary at $\rho = \infty$. 
However we are not interested in the cigar background per se and we are after a different incarnation of holography,
so it is not immediately clear what the significance of these operators is for our story. In analogy with $c=1$ non-critical
string, our non-critical superstring background is expected to possess a rich spectrum of ``discrete states'', with Liouville dressing
of type (iii). A closely related phenomenon is the existence of a chiral ring, which has been demonstrated in \cite{Konechny:2005df} (see also \cite{Rastelli:2005ph}).
This infinite tower of discrete states may be related to the exotic extra protected states of ${\cal N} = 2$ SCQCD.

\subsection{Delta-function normalizable states: the lowest  mass level}

We are now going to exhibit in detail the  physical states of type (ii) at the lowest
mass level.  We first organize the states according their symmetries in the asymptotic linear dilaton
region, and later discuss the symmetry breaking induced by the cigar interaction.
The asymptotic cylinder is at free-fermion radius, and we wish to  work covariantly in the
enhanced $SU(2)_{\psi_i} \times SU(2)_{\tilde \psi_i}$ symmetry.

After fermionizing $\theta$ into $\psi^\pm$, we have in total ten worldsheet fermions: $\psi_\mu$, $\mu = 0,\dots 5$
associated with $\mathbb{R}^{5,1}$, $\psi_\rho$ associated to the radial direction  and $\psi_i$, $i=3, \pm$ associated to  the stringy circle.
It is then clear from outset that the lowest mass level of our theory will be formally similar
 to the massless spectrum of  10d {\it critical} IIB string theory, but of course the states will propagate only in the seven 
 dimensions $x_{\hat \mu}=(x_\mu, \rho)$.

\subsubsection{NS sector}

In the left-moving NS sector the lowest states are the three 7d scalars
\be
V^{{\rm NS}}_i = 
\psi_i e^{-\varphi}e^{j\rho}e^{ik\cdot X} \, ,
\ee
in a triplet of $SU(2)_{\psi_i}$, and the 7d vector
\be
V^{{\rm NS}}_{\hat \mu} = 
\psi_{\hat \mu} e^{-\varphi}e^{j\rho}e^{ik\cdot X} \, ,
\ee
where $\hat \mu = \mu \, ,\rho$.
The mass-shell condition $L_{0}=1$ gives, for both the scalar and the vector,
\begin{equation}
\frac{1}{2}k^{2}-\frac{1}{2}j(j+1)=0\, ,\label{eq:massless}
\end{equation}
which using $j = -1/2 + i p$  we may write as
\be
\label{massless6d}
-k^2 - p^2 = k_0^2 - \vec k^2 -p^2=  \frac{1}{4} \,.
\ee
Because of the linear dilaton, the wave equations appear to be ``massive'' with $m^2 =\frac{1}{4}$. 
Introducing a polarization vector $e^{\hat \mu} = (e_\mu\, e_\rho)$, 
the superconformal invariance condition  $G_{\frac{1}{2}}  e^{\hat \mu} V^{{\rm NS}}_{\hat \mu} = 0$ gives a modified
transversality equation for the vector\footnote{Apologies for  the $\sqrt{-1}$, but here the symbol $i$ would look  confusing next to the momentum $j$.} 
\[
k\cdot e- \sqrt{-1}(j+1)e_{\rho}=0 \,.
\]
A short calculation shows that the polarization
\[
e=k\quad\mbox{and}\quad e_{\rho}= -\sqrt{-1} \, { j  }\]
corresponds to a null state. Thus despite the mass term in the wave equation,
$V^{{\rm NS}}_{\hat \mu}$  the 7-2 = 5 physical degrees of freedom of a massless 7d vector.

The theory is super-Poincar\'e invariant in $\mathbb{R}^{5,1}$, and we 
may label the states in terms of 6d quantum numbers. In assigning 6d Lorentz
quantum numbers, we may  focus  for convenience on the states with  radial momentum $p=\frac{1}{2}$, which obey a massless
6d wave-equation (see \ref{massless6d}). We can then label them according to the 6d little group $SO(4) = SU(2) \times SU(2)$.
It must kept in mind that this is  just a notational device, since the states are really part of a 7d continuum with arbitrary real $p$.
We use the notation $| j_1 \, , j_2 \rangle^{2I+1}$ for a state with spins $(j_1, j_2)$ under the 6d little group, and in the $2I+1$-dimensional representation of $SU(2)_{\psi_i}$.
All in all,  in this 6d notation we may summarize the lowest NS states as
\[
|\frac{1}{2},\frac{1}{2}\rangle^1  \oplus|0,0\rangle^1   \oplus|0,0\rangle^{3} \, .\]

\subsubsection{R sector}

The construction of vertex operators in the Ramond sector proceeds just as in
to the familiar critical (10d) case, except of course that momenta are only seven-dimensional,
\be
V^R = e^{-\varphi/2} e^{\frac{i}{2} (\epsilon_0 \phi_0 + \epsilon_1 \phi_1 + \epsilon_2 \phi_2 \epsilon_\theta \theta + \epsilon_H H)} e^{j \rho} e^{i p \cdot X}\, , \quad \epsilon_0 \epsilon_1 \epsilon_2 \epsilon_\theta \epsilon_H = 1 \, ,
\ee
which we may write as
\be
\Psi_\alpha (p_\mu) e^{\pm \frac{i}{2} (\theta+H)} \, e^{j \rho}\,, \qquad \quad \Psi_{\dot \alpha}(p_\mu)  e^{\pm \frac{i}{2} (\theta- H)} \, e^{j \rho} \, .
\ee
Here $\Psi_\alpha$ and $\Psi_{\dot \alpha}$ are 6d pseudo-real (Majorana-Weyl) spinors, respectively  left-handed and right-handed.
Choosing the 7d momentum as $p = \frac{1}{2}$ the spinors obey a massless 6d wave equation, but as above
we should keep in mind that they are really part of 7d continuum. 
For each chirality we have an $SU(2)$ doublet of 6d Majorana-Weyl spinors (equivalently, one complex Weyl spinor) 
so in ``massless 6d notation'' we write the spectrum as
\[
|\frac{1}{2},0\rangle^{2}\oplus|0,\frac{1}{2}\rangle^{2} \,.\]
 In 7d the wave-equation looks ``massive'', but the counting
of degrees of freedom is again the one for massless states.

\subsubsection{Gluing}

Table \ref{NSNS}--\ref{RNS} show the result of gluing the left- and right-moving sectors.
 In the first column of each table we list the  $(m, \tilde m)$ quantum numbers, recall (\ref{eq:relation}). In the second
 and  third columns the
 Lorentz quantum numbers are specified in the  the 6d ``massless'' notation, that is we label states by
  their spins $(j_1, j_2)$ of the little group $SO(4) = SU(2)_1 \times SU(2)_2$.
  The superscripts $2I+1$ and $2 \tilde I+1$ in the second column denote  the dimensions of the representations
  under $SU(2)_{\psi_i}$ and $SU(2)_{\tilde \psi_i}$, respectively (the superscript is omitted for singlets). Finally 
  the superscript $2R+1$ in the third column denotes the dimension of the $SU(2)_R$ representation, with $SU(2)_R$
  defined as the {\it diagonal} combination   of $SU(2)_{\psi_i}$ and $SU(2)_{\tilde \psi_i}$ which is preserved
  by the cigar interaction. 

\begin{table}
\centering{}\begin{tabular}{|c|l|l|l|}
\hline 
$(\{m\},\{\tilde{m}\})$ & $|j_{1},j_{2}\rangle^{2 I +1}\otimes| {j}_{1}, {j}_{2}\rangle^{2 \tilde I +1}$ & Decomposition: $| j_1, j_2 \rangle^{2R+1}$ & $6d$ Fields
\tabularnewline
\hline
\hline 
$(\{0\},\{0\})$ & $|\frac{1}{2},\frac{1}{2}\rangle\otimes|\frac{1}{2},\frac{1}{2}\rangle$ & $|1,1\rangle\oplus|1,0\rangle\oplus|0,1\rangle\oplus|0,0\rangle$ & $G_{\mu\n},B_{\m\n},\f$\tabularnewline
\cline{2-4} 
 & $|\frac{1}{2},\frac{1}{2}\rangle\otimes|0,0\rangle$ & $|\frac{1}{2},\frac{1}{2}\rangle$ & $V_{\m}$\tabularnewline
\cline{2-4} 
 & $|0,0\rangle\otimes|\frac{1}{2},\frac{1}{2}\rangle$ & $|\frac{1}{2},\frac{1}{2}\rangle$ & $\tilde{V}_{\m}$\tabularnewline
\cline{2-4} 
 & $|0,0\rangle\otimes|0,0\rangle$ & $|0,0\rangle$ & $\rho$\tabularnewline
\hline 
$(\{\pm1,0\},\{0\})$ & $|0,0\rangle^{3}\otimes|\frac{1}{2},\frac{1}{2}\rangle$ & $|\frac{1}{2},\frac{1}{2}\rangle^{3}$ & $\tilde{V}_{\mu}^{3}$\tabularnewline
\cline{2-4} 
 & $|0,0\rangle^{3}\otimes|0,0\rangle$ & $|0,0\rangle^{3}$ & $\rho^{3}$\tabularnewline
\hline 
$(\{0\},\{\pm1,0\})$ & $|\frac{1}{2},\frac{1}{2}\rangle\otimes|0,0\rangle^{3}$ & $|\frac{1}{2},\frac{1}{2}\rangle^{3}$ & $V_{\m}^{3}$\tabularnewline
\cline{2-4} 
 & $|0,0\rangle\otimes|0,0\rangle^{3}$ & $|0,0\rangle^{3}$ & $\tilde{\rho}^{3}$\tabularnewline
\hline 
$(\{\pm1,0\},\{\pm1,0\})$ & $|0,0\rangle^{3}\otimes|0,0\rangle^{3}$ & $|0,0\rangle^{5}\oplus|0,0\rangle^{3}\oplus|0,0\rangle$ & $T^{5},T^{3},T$\tabularnewline
\hline
\end{tabular}\caption{\label{NSNS}Field Content in NSNS sector. }

\end{table}
\begin{table}
\centering{}\begin{tabular}{|c|l|l|l|}
\hline 
$(\{m\},\{\tilde{m}\})$ & $|j_{1},j_{2}\rangle^{2 I +1}\otimes| {j}_{1}, {j}_{2}\rangle^{2 \tilde I +1}$ & Decomposition: $| j_1, j_2 \rangle^{2R+1}$ & $6d$ Fields
\tabularnewline
\hline
\hline 
$(\{0\},\{0\})$ & $|\frac{1}{2},0\rangle^{2}\otimes|\frac{1}{2},0\rangle^{2}$ & $|1,0\rangle^{3}\oplus|1,0\rangle\oplus|0,0\rangle^{3}\oplus|0,0\rangle$ & $A_{\m\n}^{3+},A_{\m\n}^{+},A^{3},A$\tabularnewline
\hline 
$(\{\pm1\},\{0\})$ & $|0,\frac{1}{2}\rangle^{2}\otimes|\frac{1}{2},0\rangle^{2}$ & $|\frac{1}{2},\frac{1}{2}\rangle^{3}\oplus|\frac{1}{2},\frac{1}{2}\rangle$ & $A_{\m}^{3},A_{\m}$\tabularnewline
\hline 
$(\{0\},\{\pm1\})$ & $|\frac{1}{2},0\rangle^{2}\otimes|0,\frac{1}{2}\rangle^{2}$ & $|\frac{1}{2},\frac{1}{2}\rangle^{3}\oplus|\frac{1}{2},\frac{1}{2}\rangle$ & $\tilde{A}_{\m}^{3},\tilde{A}_{\m}$\tabularnewline
\hline 
$(\{\pm1\},\{\pm1\})$ & $|0,\frac{1}{2}\rangle^{2}\otimes|0,\frac{1}{2}\rangle^{2}$ & $|0,1\rangle^{3}\oplus|0,1\rangle\oplus|0,0\rangle^{3}\oplus|0,0\rangle$ & $A_{\m\n}^{3-},A_{\m\n}^{-},A^{\prime3},A^{\prime}$\tabularnewline
\hline
\end{tabular}\caption{\label{RR}Field Content in RR sector}

\end{table}
\begin{table}
\centering{}\begin{tabular}{|c|l|l|l|}
\hline 
$(\{m\},\{\tilde{m}\})$ & $|j_{1},j_{2}\rangle^{2 I +1}\otimes| {j}_{1}, {j}_{2}\rangle^{2 \tilde I +1}$ & Decomposition: $| j_1, j_2 \rangle^{2R+1}$ & $6d$ Fields
\tabularnewline
\hline
\hline 
$(\{0\},\{0\})$ & $|\frac{1}{2},\frac{1}{2}\rangle\otimes|\frac{1}{2},0\rangle^{2}$ & $|1,\frac{1}{2}\rangle^{2}\oplus|0,\frac{1}{2}\rangle^{2}$ & $\Psi_{\m\ad}^{2},\Psi_{\ad}^{2}$\tabularnewline
\cline{2-4} 
 & $|0,0\rangle\otimes|\frac{1}{2},0\rangle^{2}$ & $|\frac{1}{2},0\rangle^{2}$ & $\Psi_{\al}^{2}$\tabularnewline
\hline 
$(\{\pm1,0\},\{0\})$ & $|0,0\rangle^{3}\otimes|\frac{1}{2},0\rangle^{2}$ & $|\frac{1}{2},0\rangle^{4}\oplus|\frac{1}{2},0\rangle^{2}$ & $\Psi_{\alpha}^{4},\Psi_{\al}^{2}$\tabularnewline
\hline 
$(\{0\},\{\pm1\})$ & $|\frac{1}{2},\frac{1}{2}\rangle\otimes|0,\frac{1}{2}\rangle^{2}$ & $|\frac{1}{2},1\rangle^{2}\oplus|\frac{1}{2},0\rangle^{2}$ & $\Psi_{\m\al}^{2},\Psi_{\al}^{2}$\tabularnewline
\cline{2-4} 
 & $|0,0\rangle\otimes|0,\frac{1}{2}\rangle^{2}$ & $|0,\frac{1}{2}\rangle^{2}$ & $\Psi_{\ad}^{2}$\tabularnewline
\hline 
$(\{\pm1,0\},\{\pm1\})$ & $|0,0\rangle^{3}\otimes|0,\frac{1}{2}\rangle^{2}$ & $|0,\frac{1}{2}\rangle^{4}\oplus|0,\frac{1}{2}\rangle^{2}$ & $\Psi_{\ad}^{4},\Psi_{\ad}^{2}$\tabularnewline
\hline
\end{tabular}\caption{\label{NSR}Field Content in NSR sector}

\end{table}
\begin{table}
\centering{}\begin{tabular}{|c|l|l|l|}
\hline 
$(\{m\},\{\tilde{m}\})$ & $|j_{1},j_{2}\rangle^{2 I +1}\otimes| {j}_{1}, {j}_{2}\rangle^{2 \tilde I +1}$ & Decomposition: $| j_1, j_2 \rangle^{2R+1}$ & $6d$ Fields
\tabularnewline
\hline
\hline 
$(\{0\},\{0\})$ & $|\frac{1}{2},0\rangle^{2}\otimes|\frac{1}{2},\frac{1}{2}\rangle$ & $|1,\frac{1}{2}\rangle^{2}\oplus|0,\frac{1}{2}\rangle^{2}$ & $\Psi_{\m\ad}^{2},\Psi_{\ad}^{2}$\tabularnewline
\cline{2-4} 
 & $|\frac{1}{2},0\rangle^{2}\otimes|0,0\rangle$ & $|\frac{1}{2},0\rangle^{2}$ & $\Psi_{\al}^{2}$\tabularnewline
\hline 
$(\{\pm1\},\{0\})$ & $|0,\frac{1}{2}\rangle^{2}\otimes|\frac{1}{2},\frac{1}{2}\rangle$ & $|\frac{1}{2},1\rangle^{2}\oplus|\frac{1}{2},0\rangle^{2}$ & $\Psi_{\m\al}^{2},\Psi_{\al}^{2}$\tabularnewline
\cline{2-4} 
$(\{0\},\{\pm1,0\})$ & $|0,\frac{1}{2}\rangle^{2}\otimes|0,0\rangle$ & $|0,\frac{1}{2}\rangle^{2}$ & $\Psi_{\ad}^{2}$\tabularnewline
\hline 
 & $|\frac{1}{2},0\rangle^{2}\otimes|0,0\rangle^{3}$ & $|\frac{1}{2},0\rangle^{4}\oplus|\frac{1}{2},0\rangle^{2}$ & $\Psi_{\alpha}^{4},\Psi_{\al}^{2}$\tabularnewline
\hline 
$(\{\pm1\},\{\pm1,0\})$ & $|0,\frac{1}{2}\rangle^{2}\otimes|0,0\rangle^{3}$ & $|0,\frac{1}{2}\rangle^{4}\oplus|0,\frac{1}{2}\rangle^{2}$ & $\Psi_{\ad}^{4},\Psi_{\ad}^{2}$\tabularnewline
\hline
\end{tabular}\caption{\label{RNS}Field Content in RNS sector}

\end{table}

It is interesting to organize the spectrum according to massless supermultiplets
of $6d$ supersymmetry (again, we may pretend that the states are massless in $6d$ by focussing
on the value $p = \frac{1}{2}$ of the momentum along $\rho$).
Massless supermultiplets  are constructed by taking the direct
product of a primary $|j_{1},j_{2}\rangle^{2R+1}$ with a set
${\cal R}$ of  raising operators. For $(2,0)$ susy
in six dimensions,\begin{equation}
{\cal R} =(1,0)+2 (\frac{1}{2},0)^{2}+(0,0)^{3}+2 (0,0)\label{eq:supercharges}\end{equation}
 For example the graviton multiplet is obtained acting with ${\cal R}$ on the primary $|0,1\rangle$,
while the tensor multiplet is obtained starting with 
the primary $|0,0\rangle$.  
The complete field content of (the lowest level of) the cigar theory is obtained
by action of ${\cal R}$ on the set of primaries,\[
|0,1\rangle+2|0,\frac{1}{2}\rangle^{2}+|0,0\rangle^{3}+2|0,0\rangle\]
 Comparison with (\ref{eq:supercharges}) suggests us that there are two
 other hidden supercharges at work, of opposite chirality, namely $(0,2)$,
which relate the primaries of all the $(2,0)$ supermultiplets. 
In other words, we might conclude that
we have obtained the maximally supersymmetric non-chiral $(2,2)$
supergravity in six dimensions. This is  correct as the counting of states with $7d$ momentum $p=\frac{1}{2}$ 
goes, but the right-handed supersymmetries are broken by interactions.
Nevertheless this is a useful hint: we should regard  the effective theory for the lowest level
as a spontaneously broken version of a maximally supersymmetric theory. And since the $7d$ momentum can be arbitrary, the candidate
theory before symmetry breaking is maximally supersymmetry {\it seven}-dimensional supergravity.

\subsection{Maximal 7d   Supergravity with $SO(4)$ Gauging}

To pursue this hint, in  Table \ref{spectrum7d} we have organized
the lowest level of the linear-dilaton theory (before turning on the cigar interaction)
according to $7d$ quantum numbers. The little group in $7d$ is $SO(5) \cong USp(4)$
and we label $ USp(4)$ representations  by their dimension. In the linear dilaton theory the full
$SU(2)_{\psi_i} \otimes  SU(2)_{\tilde \psi_i} \cong SO(4)$ is unbroken and we label
states with superscripts $(2I +1, 2\tilde I+1)$ indicating the representation dimensions
of the two $SU(2)$s.
\begin{table}
\begin{centering}
\begin{tabular}{|l|l|l|l|}
\hline 
Sector & $|USp(4) \rangle^{2 I + 1} \otimes  |USp(4) \rangle^{2 \tilde I + 1} $ &   Decomposition:  $|USp(4) \rangle^{({2 I + 1}, 2 \tilde I +1 )} $ &  $7d$ Fields\tabularnewline
\hline
\hline 
NSNS & $|5\rangle\otimes|5\rangle$ & $|14\rangle\oplus|10\rangle\oplus|1\rangle$ & $G_{\hat \mu \hat \nu},\: B_{\hat \mu \hat \nu},\:\f$\tabularnewline
\cline{2-4} 
 & $|5\rangle\otimes|1\rangle^{3}$ & $|5\rangle^{ (3,1) \oplus (1,3)}$ & $V_{\hat \mu}^{(3,1) \oplus (1,3)}$\tabularnewline
\cline{2-2} 
 & $|1\rangle^{3}\otimes|5\rangle$ &  & \tabularnewline
\cline{2-4} 
 & $|1\rangle^{3 }\otimes|1\rangle^{3}$ & $|1\rangle^{ (3,3)}$ & $T^{ (3,3)}$\tabularnewline
\hline 
RR & $|4\rangle^{2 }\otimes|4\rangle^{2 }$ & $|10\rangle^{(2,2)}\oplus|5\rangle^{ (2,2)}\oplus|1\rangle^{(2,2)}$ & $C_{\hat \mu \hat \nu}^{(2,2)},\: C_{\hat \mu}^{(2,2)},\: C^{(2,2)}$\tabularnewline
\hline
\hline 
RNS & $|4\rangle^{2 }\otimes|5\rangle$ & $|16\rangle^{(2,1)\oplus (1,2) }\oplus|4\rangle^{(2,1) \oplus (1,2) }$ & $\Psi_{\hat \mu}^{(2,1) \oplus (1,2)},\:\Psi^{(2,1)\oplus (1,2)}$\tabularnewline
\cline{2-2} 
NSR & $|5\rangle\otimes|4\rangle^{2}$ &  & \tabularnewline
\cline{2-4} 
 & $|4\rangle^{2}\otimes|1\rangle^{3}$ & $ |4\rangle^{(2,3)\oplus (3,2)}$ & $ \Psi^{(2,3) \oplus (3,2)}$\tabularnewline
\cline{2-2} 
 & $|1\rangle^{3 }\otimes|4\rangle^{2}$ &  & \tabularnewline
\hline
\end{tabular}
\par\end{centering}
\caption{\label{spectrum7d}Seven-dimensional labeling of the spectrum
of the linear-dilaton theory}
\end{table}
Remarkably, the resulting spectrum is precisely the field content of maximal
$7d$ supergravity with $SO(4)$ gauging, a theory that has been fully constructed 
only quite recently \cite{Samtleben:2005bp, Weidner:2006rp}.   The massless vector $V_{\hat \mu}^{(3,1)+(1,3)}$ are the $SO(4)$ gauge fields.
On the other hand the vectors $C_{\hat \mu}^{4}$
are eaten by the two forms $C_{\hat \m \hat \n}^{4}$, which become massive through a vectorial
Higgs mechanism \cite{Samtleben:2005bp, Weidner:2006rp}.

Recall that the standard gauging of maximal $7d$ sugra
is of the full $SO(5)$ R-symmetry --  this is the famous supergravity that arises
by consistent truncation of $11d$ supergravity compactified on $S^4$ and that admits  a maximally supersymmetric $AdS_7$ vacuum.
By contrast, the scalar potential of the $SO(4)$ theory 
does not allow for  a stationary solution, but only for a domain wall solution \cite{Samtleben:2005bp, Weidner:2006rp}, that is,
our linear-dilaton background. A closely related interpretation of the $SO(4)$ gauged
supergravity was given in \cite{Boonstra:1998mp} (before its explicit construction!) 
as the effective $7d$ supergravity arising from a ``warped compactification'' of IIB supergravity on the near-horizon NS5 brane background
 $\mathbb{R}^{5,1} \times$ linear dilaton $\times S^3$. 
 
 The cigar background
 is obtained by further turning on a ``tachyon'' perturbation, a profile for  the NSNS scalar fields $T^{(3,3)}$ that decays for large $\rho$ and acts as a wall for $\rho \sim 0$.
 Note that the scalars are in the symmetric traceless tensor of $SO(4)$, and choosing a vev for them breaks $SO(4) \to SO(3) \cong SU(2)_R$, the diagonal combination of $SU(2)_{\psi_i} \times SU(2)_{\tilde \psi_i}$, as expected. In the IIA set-up of colliding NS5 branes, this breaking corresponds to choosing an angular direction in the transverse $S^3$ to the coincident NS5 brane --
  the direction along which the branes are separated (we called it $\tau$ in Figure \ref{HWfig}). Under the preserved diagonal $SU(2)_R$, the nine NSNS scalars $T^{(3,3)}$ decompose
  as ${\bf 5 + 3+1}$. The ${\bf 1}$ and the ${\bf 3}$ are associated to moduli, corresponding respectively  (in the T-dual picture) to the radial and angular separations
  of the two NS5 branes; together with an extra $SU(2)_R$-singlet scalar from the RR sector they comprise
  the five scalars of the $6d$ tensor multiplet localized at the tip of the cigar.

In the application of the  $SO(4)$-gauged $7d$ supergravity to our problem of finding the dual
$\NN=2$ SCQCD, we are not interested in turning on a background for the NSNS scalars,
but rather for the RR fields corresponding to
 $N_{c}$ D3 branes and $N_{f}$ D5
branes.
  D3 branes are magnetically
charged unde the RR one-form $C_{\mu}^{(2,2)}$ and D5 branes are magnetically
charged under the RR zero-form $C^{(2,2)}$. As the superscripts indicate
both of the RR one-form and zero-form  transform as vectors of $SO(4)$. It is possible to choose a common direction in $SO(4)$ space for both forms,
so that again we break $SO(4) \to SO(3) \cong SU(2)_R$. This is  again consistent with the IIA Hanany-Witten picture.
 Separating the NS5 branes in
breaks $SO(4)$ to $SO(3)$, and it is clear that both the compact and the non-compact $D4$-branes 
are extended in the same direction along which the NS5 branes are separated, so that their fluxes are oriented coherently in $SO(4)$ space.
 The surviving $SO(3) \cong SU(2)_R$ is interpreted as the $SU(2)_R$ R-symmetry of the $\NN = 2$ gauge theory.

\bibliographystyle{JHEP}
\bibliography{Orbifoldbib}

\end{document}